\documentclass[a4paper,fleqn,usenatbib]{mnras}  
\usepackage{newtxtext,newtxmath}
\usepackage[T1]{fontenc}
\usepackage{ae,aecompl}
\usepackage{graphics,graphicx}
\usepackage{amsmath}	

\title[Shear-selected clusters]{The uncommon intracluster medium features of the first massive clusters selected independently of their baryon content}
\author[Andreon et al.]{S. Andreon,$^{1}$\thanks{E-mail:stefano.andreon@inaf.it} M. Radovich,$^2$ A. Moretti,$^{1}$  F.-X. Desert$^3$, T. Hamana$^4$, M. Pizzardo$^5$, \and C. Romero$^6$,  H. Roussel$^7$, G. Trinchieri$^{1}$   
\\
$^1$ INAF--Osservatorio Astronomico di Brera, via Brera 28, 20121, Milano, Italy\\
$^2$ INAF--Osservatorio Astronomico di Padova, Vicolo Osservatorio 5, 35122, Padova, Italy\\
$^3$ Univ. Grenoble Alpes, CNRS, IPAG, 38000 Grenoble, France\\
$^4$ National Astronomical Observatory of Japan, Mitaka, Tokyo 181-8588, Japan\\
$^5$ Department of Astronomy and Physics, Saint Mary’s University, 923 Robie Street, Halifax, NS-B3H3C3, Canada\\
$^6$ Center for Astrophysics $\vert$ Harvard \& Smithsonian, 60 Garden Street, Cambridge, MA 02138, USA \\
$^7$ Institut d'Astrophysique de Paris, Sorbonne Universit\'es, UPMC
Univ. Paris 06, CNRS UMR 7095, 75014 Paris, France\\
}
\date{Accepted XXX. Received YYY; in original form ZZZ}
\pubyear{2021}

\begin{document}
\label{firstpage}
\pagerange{\pageref{firstpage}--\pageref{lastpage}}
\maketitle

\begin{abstract}
Our current knowledge of the thermodynamic properties of galaxy clusters comes primarily from detailed studies of clusters selected by their minority components: hot baryons. 
Most of these studies select the clusters using the component that is being investigated, the intracluster medium (ICM), making the sample choice prone to selection effects. Weak-gravitational lensing allows us to select clusters by the total mass component and,  being independent of the type of matter, makes the sample choice unbiased with respect to the baryon content. In this paper, we study four galaxy clusters at intermediate redshift ($0.25<z<0.61$), selected from the weak-lensing survey of Miyazaki et al. (2018). 
We derive core-excised X-ray luminosities, richness-based masses, Compton parameters, and profiles of mass, pressure and electron densities. 
These quantities are derived from shear data, Compton maps, and our own X-ray and SZ follow-up.
When compared to ICM-selected clusters of the same mass, in the range $2$ to $5 \ 10^{14}$ M$_\odot$, our small sample of four clusters 
is expected to have on average 0.2 rare ($>2\sigma$) features, while we  
observed on average two rare features in each one of the seven explored properties:
richness, core-excised luminosity, Compton parameter, pressure and electron pressure profiles, and central values of them.
The abundance of rare and unique features in such a small sample indicates a fundamental bias in our knowledge of the thermodynamic properties of clusters when derived from ICM-selected samples.
\end{abstract}

\maketitle

\section{Introduction}

It is well-known that all surveys miss objects below their detection threshold. When a sample is selected by a given observable,  it is biased for that selection quantity and for quantities with covariance with it, as it lacks objects below the detection threshold. The bias also exists when the sample selection is probabilistic (the ``threshold" is fuzzy and some objects below the threshold enter in the sample, while some above the threshold are missed, see Andreon \& Weaver 2015 about how to deal with this case) in place of deterministic. The sample bias often becomes evident when the selection quantity, or any other quantity showing covariance with it, is plotted against another observable. Biased samples typically show preferential distribution on the side where objects easier to detect are located. There are many examples in the literature illustrating these biases, for example in the X-ray luminosity - temperature scaling of X-ray selected samples (e.g., Pacaud et al. 2007; Andreon et al. 2011),  the X-ray luminosity - mass scaling of X-ray selected samples (Vikhlinin et al. 2009; Mantz et al. 2010; Andreon \& Berge 2012; Giles et al. 2016), the Compton Y vs mass scaling of X-ray selected samples (Nagarajan et al. 2019; Andreon 2016).
To summarize, X-ray and SZ- selected samples (Compton Y shows covariance with X-ray luminosity) are biased samples for, e.g.,  X-ray luminosity vs mass scaling relations and Compton Y vs mass scaling relations . For the same argument, optically-selected samples are biased samples for the richness vs mass scaling relation (e.g. Andreon \& Hurn 2010, see the appendix in particular).

Instead, when the probability of inclusion of an object in a sample is independent of the quantity being investigated (at fixed value plotted in the abscissa if a scaling relation is considered), the sample is unbiased for that quantity. 
Those samples are rare because they cannot be built from selection in a survey in the investigated quantity (X-ray, or SZ, for scaling involving ICM-related quantities such as X-ray luminosity, count rate, Compton Y, SZ detection significance, etc.). 
Furthermore, investigation in the interested quantity requires a follow-up that can be  ``expensive" because there will likely be objects that are 
below the detection threshold in surveys.  One such a sample is the X-ray Unbiased Cluster Sample (XUCS, hereafter), selected by velocity dispersion (of the galaxies) and for which the probability of inclusion of a cluster in the sample is independent of the cluster X-ray or SZ properties at fixed mass  (Andreon et al. 2016, 2017a, 2017b).

In biased samples, the amplitude of the selection bias cannot be estimated from the survey itself (e.g., Vikhlinin et al. 2009) and inferring the property of the whole population requires assumptions about the size and properties of the population not entering in the sample. For example: do X-ray and SZ surveys miss a minority of clusters or do they miss most of them (at a given fixed mass when scaling relation with mass are considered)? Given that this information is absent in the data, a strong prior is usually taken.  A common one is assuming that
the scatter in the selection quantity at fixed mass is mass- and redshift- independent (e.g. Pacaud et al. 2007, Ghirardini et al. 2024), i.e. that the value measured on the most massive objects at low reshifts applies to objects of all masses at all redshifts. This is a very risky assumption. 
For example, at $\log M/M_{\odot}<14.8$,
the $Y-M$ scaling of the excellent analysis by Nagarajan et al. (2019, see Fig.~\ref{fig:Y_M}) is only fit to clusters drawn from half of the population of clusters brighter-than-the-average for their mass. The fit assumes that the scatter around the mean relation of the fainter-than-the-average population is identical to the one of their brighter-than-the-average analogs. However, this is  
hard to know with no, or almost no, example of a fainter-than-the-average population in the observational sample. Second, is the size of absent
population close to the 
value supposed to be in the analysis?  
In the case of the X-ray scaling relations, while it has been known for at least 30 years  that the X-ray selection misses some objects of low surface brightness (e.g. Rosati et al. 1995), the size of the missed fraction is still under discussion (e.g. Andreon et al. 2024) and basically parametrized by the scatter around the mean relation. The scatter is found to be $0.5$ dex in optically or dynamical selected samples (Andreon \& Moretti 2011; Andreon et al. 2016). In ICM selected samples, the values stayed in the range 0.02-0.17 (see Pratt et al. 2009) and only recently they have become as large as 0.4 dex based on the latest eROSITA analysis (Ghirardini et al. 2024).

In addition to the above statistical issues, astrophysical challenges arise from our reliance on baryon tracers to select the sample to be studied. Our current knowledge of the thermodynamic properties of galaxy clusters comes almost exclusively from studying samples that are selected using baryon tracers, either the galaxies or, very often, the intracluster medium. However, these are minority components compared to the cluster's main ingredient, dark matter. In principle at least, clusters with much lower fraction of gas, galaxies, or baryons than observed in clusters selected by baryon tracers may exist. Or perhaps the amount is the same but the baryons are differently distributed, say less concentrated (as in CL2015, Andreon et al. 2019).
Are the objects absent in baryon-selected samples similar to those in the observed samples or different? Sometimes they are, and sometimes they are not:
for example, clusters missed in X-ray because of their low surface brightness obey the same $L_X-T$ relation of those entering in X-ray samples  whereas they fall in a different part of the $T-M$ plane because they have low temperature for their mass (Andreon et al. 2024), showing that the behaviour of clusters absent in X-ray samples depends on the considered scaling relation. It also seems that the whole population of clusters includes objects with lower pressure profiles than found in ICM-selected samples (Andreon et al. 2019, 2021, 2023;  Di Mascolo et al. 2020; Dicker et al. 2020; Sayers et al. 2023; Hilton et al. 2018).

To address these challenges,
an unbiased sample of galaxy clusters must be selected independently of the ICM itself, and possibly selected from a non-minority mass component. Weak gravitational lensing (Bartelmann \& Schneider 2001) allows it because the technique measures the distortions (shear) of background galaxies due to the gravitational potential, regardless of the type/proportion of matter by which it is generated.  The release of the first catalog (Miyazaki et al. 2018) of high signal-to-noise shear detections in the Hyper Suprime-Cam Subaru Survey (HSC, Aihara et al.  2018) allows us to select clusters independently of their baryon content and to investigate the bias of the baryon selection.
As far as we know, previous studies (e.g., Giles et al. 2015 and Deshpande et al. 2017; Ramos-Ceja et al. 2021) did not have 
data of sufficient quality to derive a radial profile (of, say, shear, pressure or electron density) and sometimes were confronted with identification problems at such a point that there was a large mismatch between the number of shear peaks and clusters (e.g. 8 vs 17 in Deshpande et al. 2017).

Throughout this paper, we assume $\Omega_M=0.3$, $\Omega_\Lambda=0.7$, 
and $H_0=70$ km s$^{-1}$ Mpc$^{-1}$. 
Results of stochastic computations are given
in the form $x\pm y$, where $x$ and $y$ are 
the posterior mean and standard deviation. The latter also
corresponds to 68\% uncertainties because we only summarize
posteriors close to Gaussian in this way. All logarithms are in base 10.

\begin{table*}
\caption{Cluster sample, available observations, and derived quantities}
\footnotesize
\begin{tabular}{l r r r r}
\hline
Quantity & id5 & id17 & id34 & id48 \\
\hline
RA (deg)  & 179.0448 & 178.0613 & 223.0791 & 220.7880 \\
Dec (deg) & -0.3516 & 0.5218 & 0.1677 & 1.0370 \\
$z_{spec}$   & 0.256 & 0.463 & 0.601 & 0.529 \\
X-ray  data & Swift & Swift & Swift & Chandra \\
SZ data & ACT & ACT & NIKA2 \& ACT & NIKA2 \& ACT \\
X-shear fit type & disjoint & joint & disjoint & joint \\ 
$\log$ M$_{500}$/M$_{\odot}$ & $14.68\pm0.10$ & $14.45\pm0.10$ & $14.80\pm0.26$ & $14.38\pm0.14$  \\
$\log$ M$_{200}$/M$_{\odot}$ & $14.85\pm0.10$ & $14.62\pm0.14$ & $14.97\pm0.25$ & $14.47\pm0.17$ \\
$\log$ M$_{200,{\rm rich}}$/M$_{\odot}$ & $14.14\pm0.16$ & $14.52\pm0.16$ & $14.87\pm0.16$ & $14.50\pm0.16$ \\
$\log$ M$_{200,{\rm caustics}}$/M$_{\odot}$ & $14.79\pm0.08$ &  \\
$\log$ L$_{X,{\rm ce,500}}$/(erg s$^{-1})$ & $43.72\pm0.03$ & $44.16\pm0.03$ & $44.50\pm0.02$ & $44.29\pm0.04$ \\
$\log$ Y$_{\rm sph,500}$/(Mpc$^{2})$ & $-4.62\pm0.07$ & $-4.58\pm0.04$ & $-4.58\pm0.11$ & $-4.80\pm0.12$ \\
Other name & & ACT-CL J1152.2+0031 & ACT-CL J1452.3+0009 & ACT-CL J1443.1+0102 \\
         & & &  & eMACSJ1443.2+0102 \\
\hline      
\end{tabular} 
\hfill\break
X-ray luminosities are in the [0.5-2] keV band. Y$_{\rm sph,500}$ needs a 0.06 dex reduction if they need to be compared with values in Nagarajan et al. (2019). 
\label{tab1}
\end{table*}

\begin{figure}
\centerline{\includegraphics[width=9truecm]{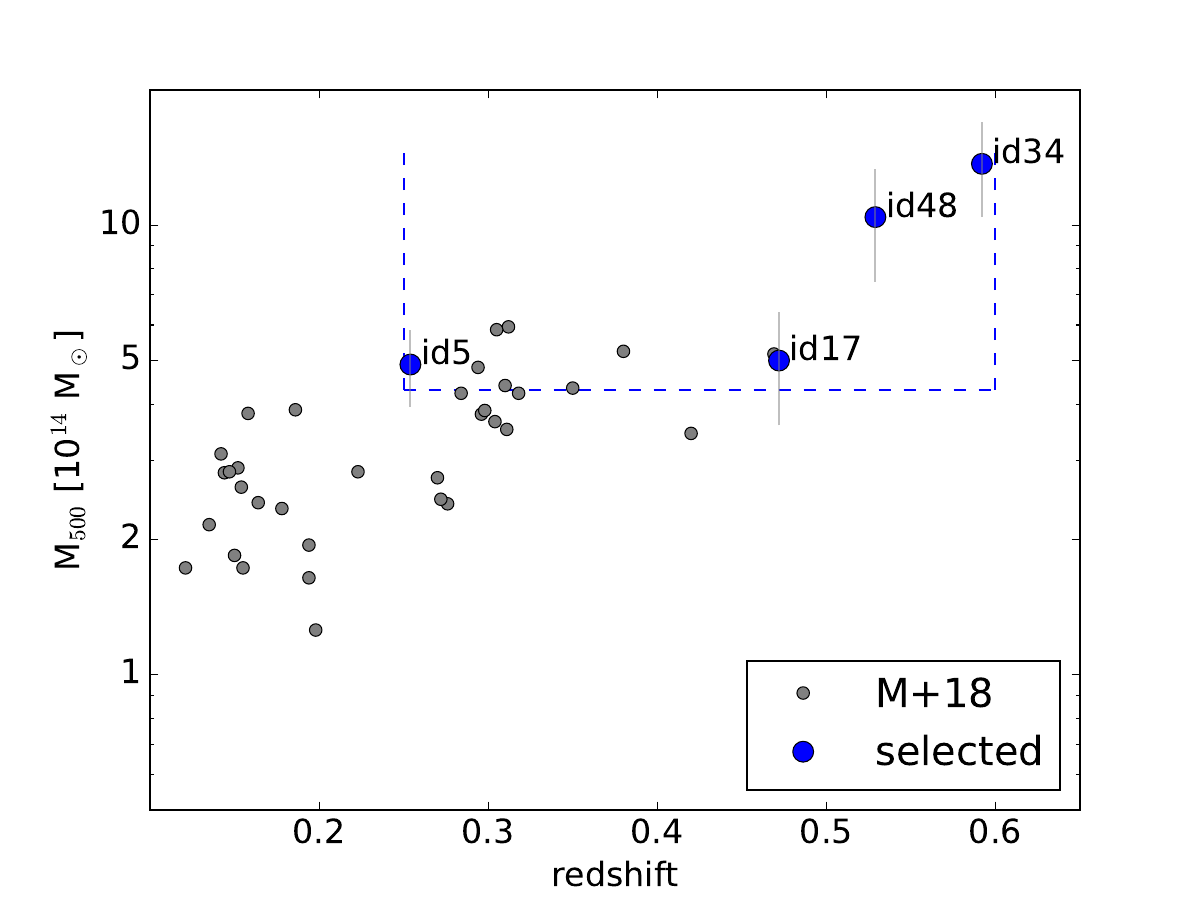}}
\centerline{\includegraphics[width=9truecm]{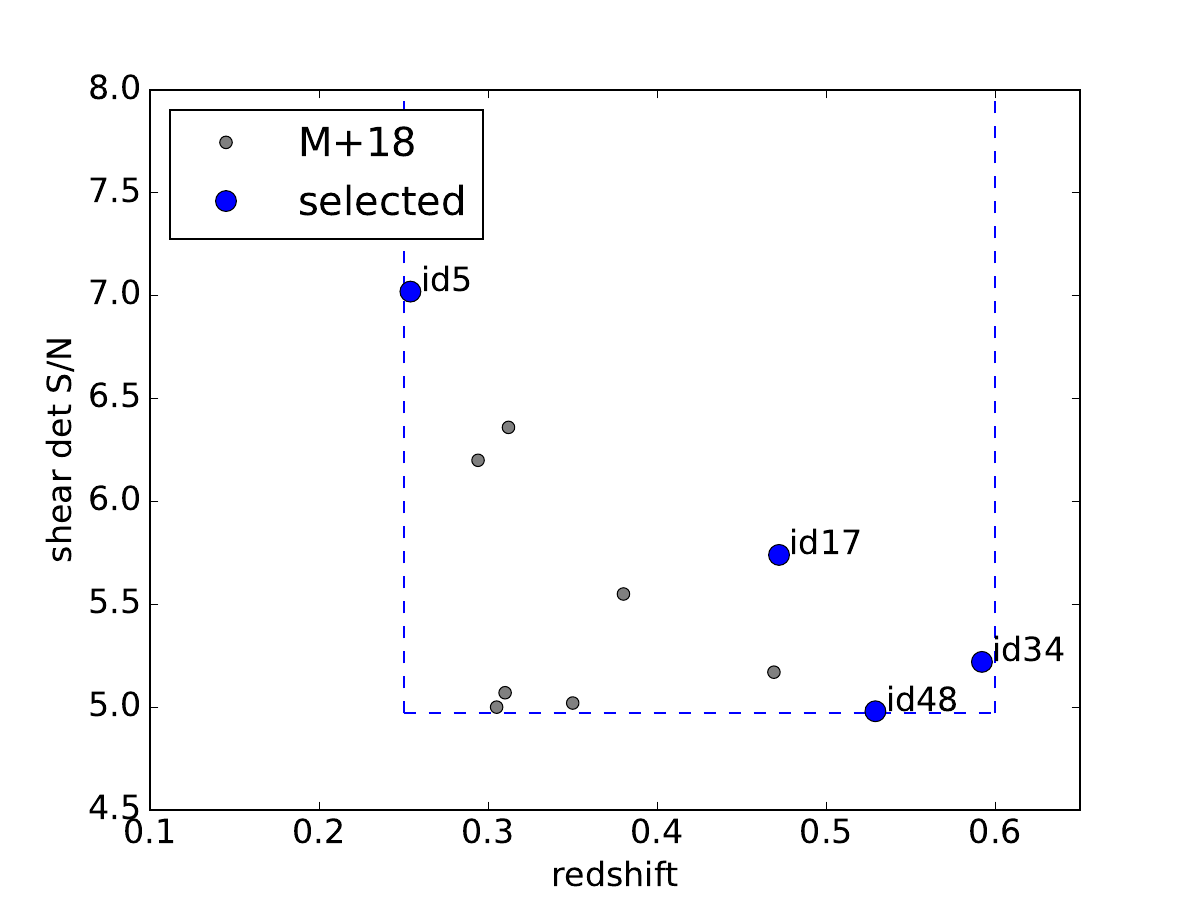}}
\caption[]{Top panel: Mass-redshift plot of the $\geq 4.98$ sample with available masses.  Bottom panel: detection (shear) S/N - redshift plot of the sub-sample with $M_{500}>4.3 \ 10^{14} $ M$_{\odot}$. In both panels the followed up clusters are indicated by a solid circle. Masses in this figure are as in Miyazaki et al. (2018).}
\label{fig:sel}
\end{figure}

\section{Data, Analysis, and Results}

\subsection{Sample and selection}

For this pilot investigation we considered the shear-selected sample in the HSC catalog (Miyazaki et al. 2018) and specifically four of the 11 clusters with $ 0.25<z<0.60$, signal-to-noise $\geq 4.98$, and $M_{500}>4.3 \ 10^{14} $ M$_{\odot}$ (Fig.~\ref{fig:sel}). The Miyazaki et al. (2018) cluster catalog has been built using very deep observations with excellent seeing performed on the 160 deg$^{2}$ area deeply observed by the Hyper Suprime-Cam (HSC) as a Subaru Strategic Program. We decided to observe in X-ray and/or SZ four of the above 11 clusters, specifically we selected two pairs of clusters with two different criteria: the two most massive clusters (id34 and id48) and randomly two, out of three clusters with largest signal-to-noise visible in the spring nights (id5 and id17). Id5 is also the cluster with the largest S/N among all those that have mass above the considered threshold at any right ascension.  None of the four clusters had pointed X-ray observations with XMM-Newton, Chandra, or Swift at the time of the target selection. We followed three of them with Swift. Id48  serendipitously falls in a Chandra pointing of an unrelated target and we use this observation in our work. 
None of the clusters had pointed SZ observations and therefore we followed up the two most massive (id34 and id48) with NIKA2. 

After the execution of our observations, all targets but id5 were listed in the ACT (SZ) catalog of Hilton et al. (2021). Near the completion of our analysis, id48 was listed as eMACSJ1443.2+0102 in the X-ray catalog of Ebeling et al. (2024). All four clusters are in the western half of the eROSITA sky (Merloni et al. 2024), but these data, that have become available close to the completion of this paper, are very shallow compared to the ones we use. For example, there are 5 net eROSITA photons at the sky position of id5 (vs 560 in our X-ray follow-up).

With the use of SDSS DR18 spectroscopic data (Almeida et al. 2023), we confirm and refine the photometric redshift reported in Miyazaki et al. (2018) of all but the highest redshift cluster (id48), for which SDSS data are inconclusive. This latter, together with id17 and id34, have spectroscopic redshifts reported in the Hilton et al. (2021) catalog. id48 has also been spectroscopically confirmed by Ebeling et al. (2024). DESI reshifts (DESI collaboration, 2023) confirm the id5 redshift and allow us to perform further analysis as detailed below. Table~\ref{tab1} summarizes some of the properties of the cluster sample.

\subsection{Optical: richness of all clusters and a closer look at id5}
\label{richness_section}

Inspection of the HSC images of the four clusters shows that id17 and id48 have gravitational arcs and that id34 is clearly bimodal (has a second galaxy excess about 2 arcmin South of the main one). Fig.~\ref{fig:optima} shows the true color images of the clusters.

Cluster richnesses $n_{200}$ and richness-based masses $M_{200,\rm rich}$ are derived using photometry from the third HSC data release (Aihara et al. 2022) following Andreon (2016) with minor updates. Briefly, galaxies on the red sequence and brighter than the evolved $M_V=-20$ mag are counted within a radius $r_{200}$ iteratively chosen to follow the richness-radius relation observed in nearby clusters (Andreon 2015). The contribution of background/foreground galaxies is estimated from an annulus of 3 to 7 Mpc radius, accounting for contamination by other clusters/groups by dividing the annulus in octants and discarding the two octants with larger counts and the two with lowest counts. $M_{200, \rm rich}$ is derived from richness via a scaling calibrated on clusters of known masses (Andreon 2015) accounting for evolution and with errors that account for the scatter between mass and the used observable (richness). We slightly update Andreon (2016): a) in the way the rest-frame $g-r$ color is computed (interpolating across adjacent filters in place of taking the closest filter pair); b) in the precise values of the background radii, and c) we automatize the self-calibration of the color of the red sequence which uses, in our case, clusters in the HSC survey with spectroscopic redshifts. Table~\ref{tab1} lists the derived masses. Three clusters turn out to have $\log M_{200,\rm rich}/M_\odot>14.4$ , while the remaining (id5) has $\log M_{200,\rm rich}/M_\odot\sim14.2$.

\begin{figure}
\centerline{\includegraphics[width=9truecm]{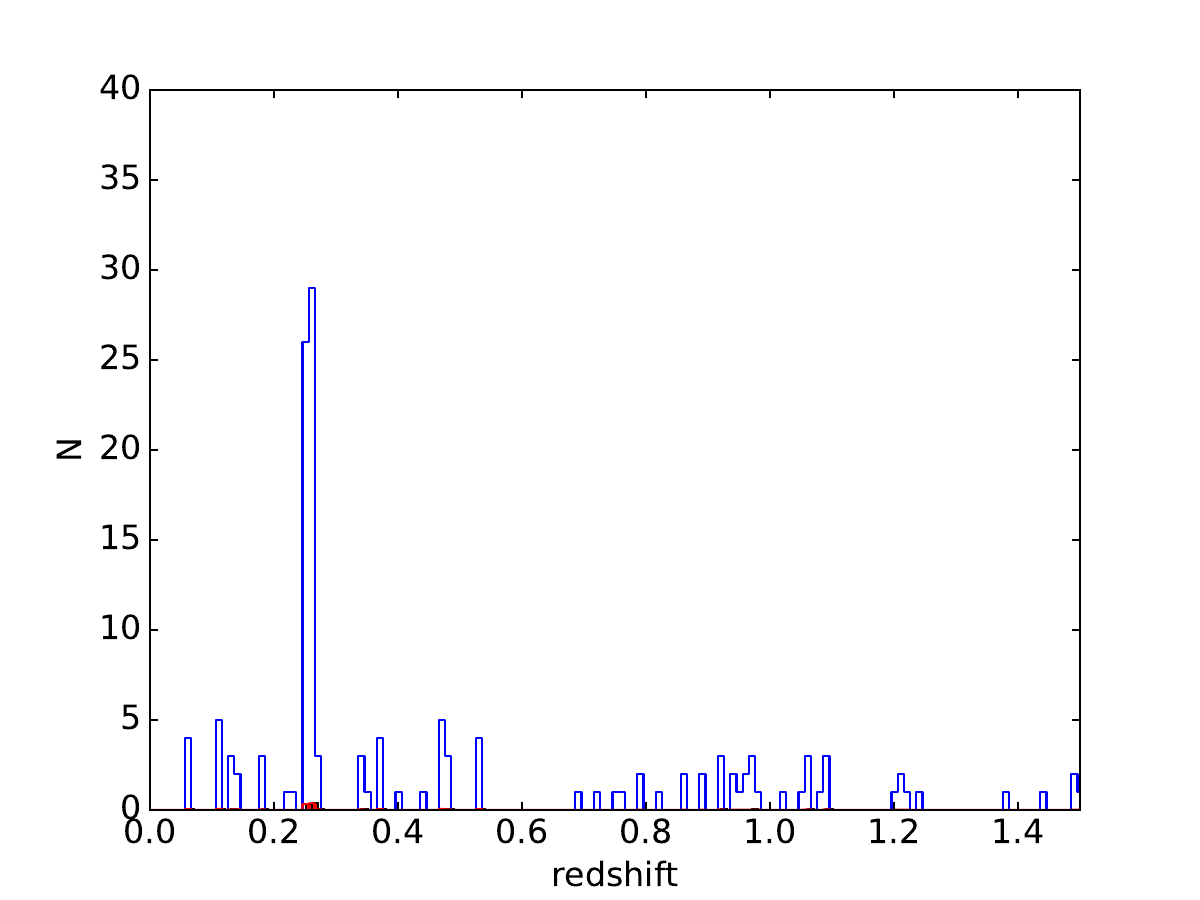}} 
\caption[h]{Redshift distribution in the id5 line of sight (within $r_{500}$, blue histogram, the cluster is the peak at $z=0.256$) and average background distribution around id5 (red filled histogram, barely visible because it is very close to zero at all redshifts), normalized to the cluster solid angle. There is only one cluster in the id5 line of sight.}
\label{fig:zdistrib}
\end{figure}

As will become apparent in our analysis of id5, described in the following sections, the cluster has anomalous properties for its mass. We therefore decided to scrutinize further the data. 
At 9 and 18 arcmin from id5 there are Abell 1419 and Abell 1411, respectively.  These clusters are both at $z_{\rm spec}\sim0.11$. They are both too far in redshift and angular distance to affect the id5 richness measurement but since we consider them in a later section of the paper, we decided to estimate their masses as well. Because of their low redshift, their brighter galaxies risk to be saturated in HSC images. Therefore, for those two clusters and for id5 we used shallower SDSS (Almeida et al. 2023) photometry. Id5 was added to test whether HSC photometry could be faulty (for unknown causes). Furthermore, to test the code we use here, we retrieved  the code originally used in Andreon (2016) to estimate the cluster richness, and from it, its mass. The richness-based id5 mass derived is entirely consistent with that obtained from HSC photometry (less than $0.5\sigma$ away).  The estimated masses of the two Abell clusters are more than 1 dex lower than our weak lensing estimate of the id5 mass in Sec.~\ref{shear_section}.

\begin{figure}
\centerline{\includegraphics[width=9truecm]{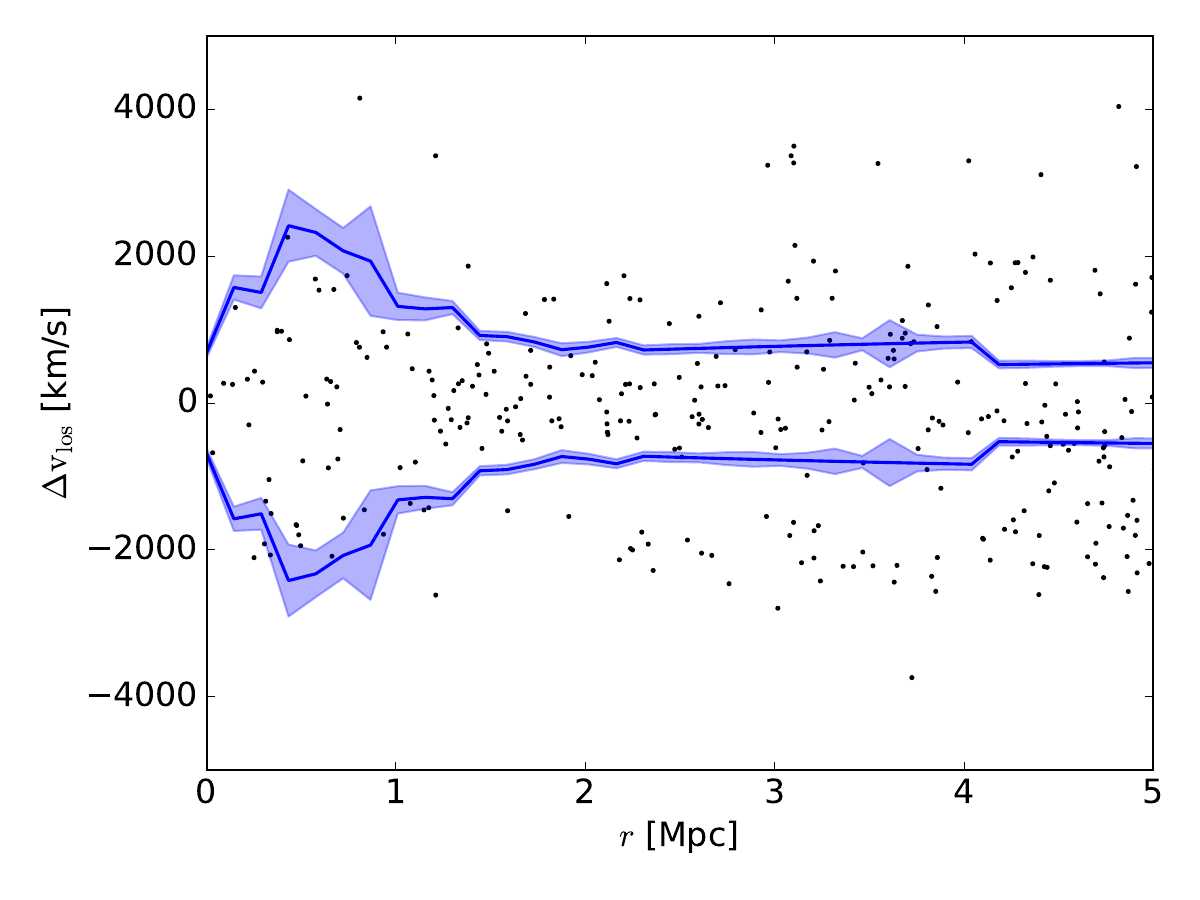}} 
\caption[h]{Velocity differences $\Delta {\rm v_{ los}}$ as a  function of clustercentric distance $r$ for individual galaxies in the field-of-view of 
the id5 cluster. Curves show the caustic profile estimated by the caustic technique. Shaded areas report the uncertainty of the caustics. Id5 cluster is a massive cluster, with $\log$ M$_{200,{\rm caustics}}$/M$_{\odot} = 14.8$.
}
\label{fig:caustics}
\end{figure}

\begin{figure*}
\centerline{\includegraphics[width=8truecm]{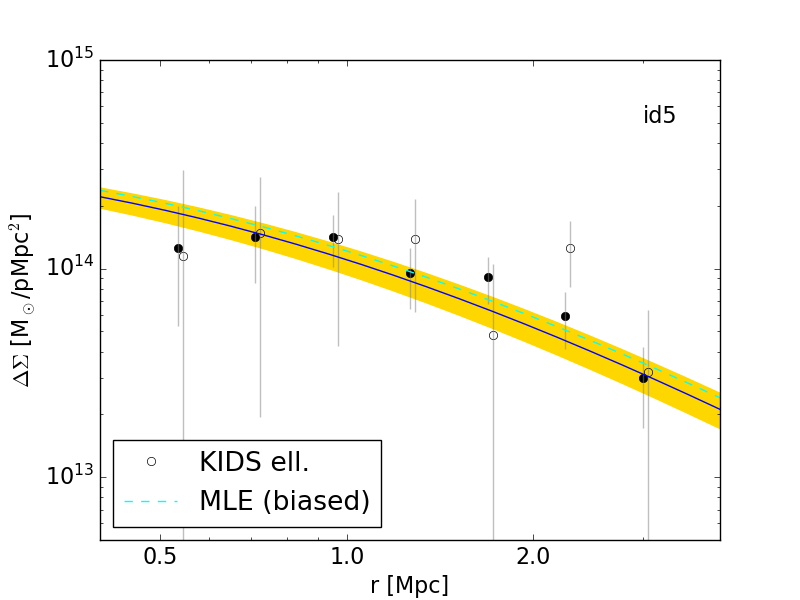} 
\includegraphics[width=8truecm]{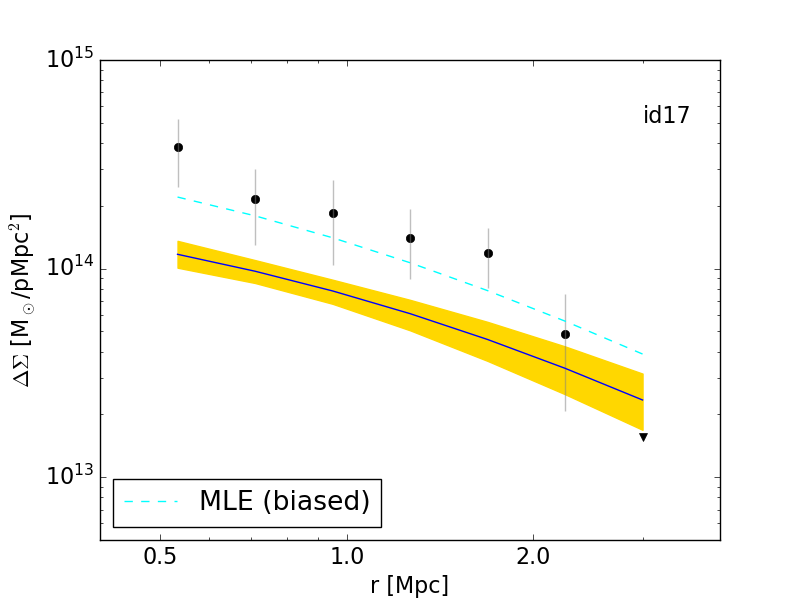}}
\centerline{\includegraphics[width=8truecm]{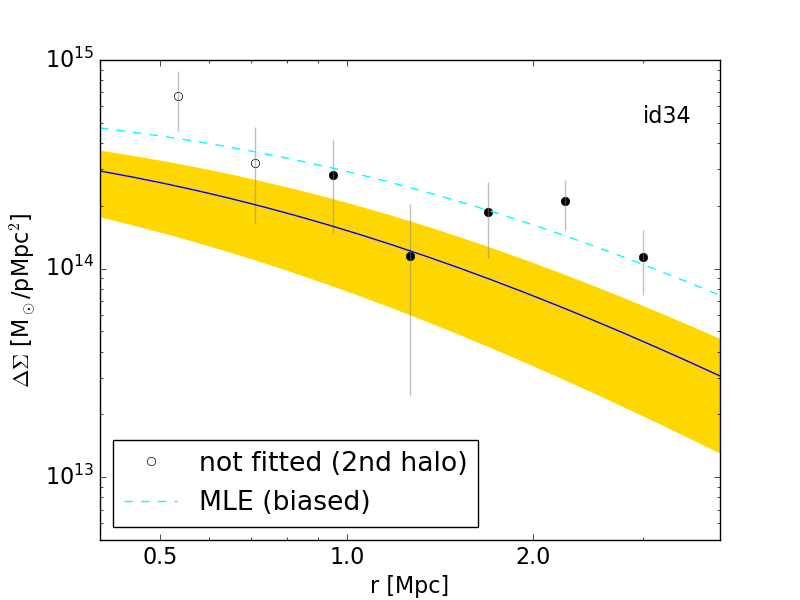} 
\includegraphics[width=8truecm]{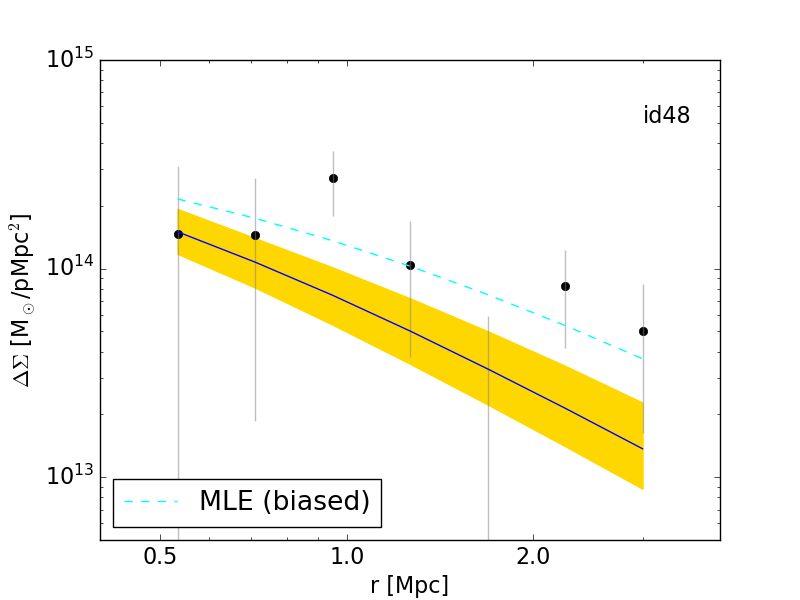}}
\caption[h]{Binned tangential shear profile of the four clusters. The solid line with yellow shading indicates the mean model and 68\% uncertainty fitted to shear and X-ray data (id17 and id48) or shear only (id5 and id34) data. Uncertainty in the model also accounts for intrinsic scatter, whereas the plotted error bars only account for shape noise and large scale structure. The model fit (solid line) avoids the region occupied by the data because of the Eddington bias. The dashed cyan line shows the maximum likelihood estimate (MLE), which is biased (by a negligible amount in the case of id5 cluster given its high signal-to-noise). For id5 we also report observed binned tangential shear profile derived from KiDS data (we plot them slightly offset in radii to avoid cluttering), see text for details.}
\label{fig:WL}
\end{figure*}

Id5 is also in the field of the DESI spectroscopic early data release (DESI collaboration, 2023), which provides 133 spectroscopic redshift within 8 arcmin of the cluster and within $|\Delta v|<3000$ km/s, in addition to many other redshifts in the area. We found two groups that are infalling onto id5, which suggest that the cluster should be a very massive one (such a situation is atypical in low mass clusters). These are: a) a group with more than 10 spectroscopic confirmed members centered 2 arcmin East of the brightest cluster galaxy (BCG, hereafter), formed by many of the galaxies visible in the East half of the top-left panel of Fig.~\ref{fig:optima}. The group has $\Delta v\sim-1000$ km/s from the BCG and from the velocity barycenter of the cluster; and b) a group with at least 6 spectroscopic members centered about 6 arcmin West of the BCG with $\Delta v\sim1000$ km/s spatially coincident with an extended X-ray emission visible in our X-ray Swift data. 
Spectroscopic data do not reveal any foreground or background massive cluster aligned with id5 (see Fig.~\ref{fig:zdistrib}), in agreement with the photometric data (Fig.~\ref{fig:id5CM}), that shows no indication of a second red sequence at a different color. The latter plot also rules out a large population of massive blue members. Furthermore, the large spectroscopic coverage of id5 allows us to check the statistical background subtraction performed in the richness estimation: an almost identical richness is obtained considering red sequence galaxies with $|\Delta v|<3000$ km/s. 

Using DESI spectroscopy, we derive the caustic mass of id5 
following Diaferio \& Geller (1997), Diaferio (1999), and Serra et al. (2011). Basically, the caustic technique estimates the three-dimensional cumulative mass profile of a cluster from the line-of-sight escape velocity profile of the cluster members. It is important to note that this technique does not assume dynamical equilibrium, so it can be applied in this complex dynamical case.
Diaferio (1997) showed that the caustic amplitude approximates the escape velocity 
from a cluster, 
which is in turn is directly related to the enclosed mass
(Diaferio \& Geller 1997, Diaferio 1999).
We use ${\cal F}_\beta=0.5$ 
(Diaferio \& Geller 1997, Diaferio 1999) and the X-ray center determined in Sec.~\ref{xray_data_reduction}.

Figure~\ref{fig:caustics} shows the redshift diagram of id5. Curves and shaded areas show the caustics and their $1\sigma$ uncertainty. 
There are 163 galaxies within $3R_{200c}$ and previous studies (e.g., Serra et. al 2011) show that this sampling is good for robust location of the cluster mass profile. We found
$\log$ M$_{200,{\rm caustics}}$/M$_{\odot} = 14.79\pm0.08$ and the cumulative mass profile shown in Fig.~\ref{fig:massprof}.

\begin{figure*}
\centerline{\includegraphics[trim={5 203 30 30}, clip,height=3.5truecm]{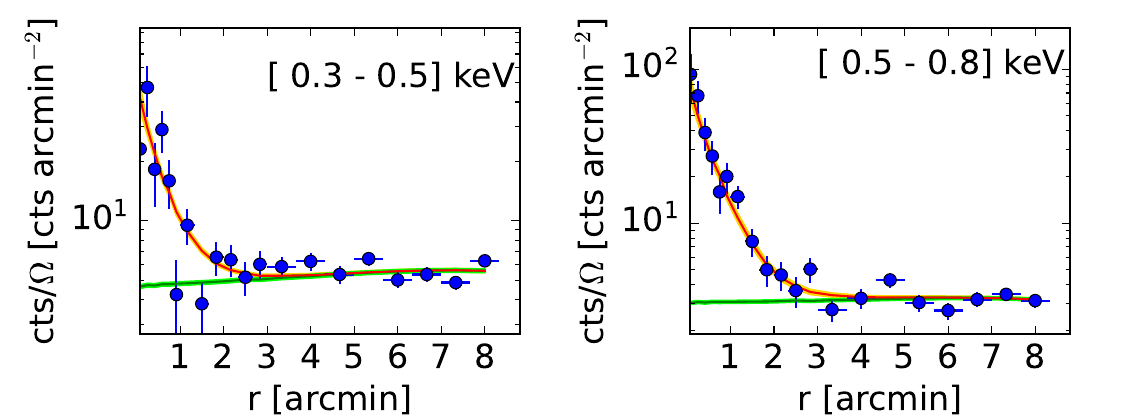}%
\includegraphics[trim={5 5 310 230}, clip,height=3.5truecm]{fitwithmod_1_wlHE_id17.pdf}}
\centerline{\includegraphics[trim={255 5 40 230}, clip,height=3.5truecm]{fitwithmod_1_wlHE_id17.pdf}%
\includegraphics[trim={5 203 30 30}, clip,height=3.5truecm]{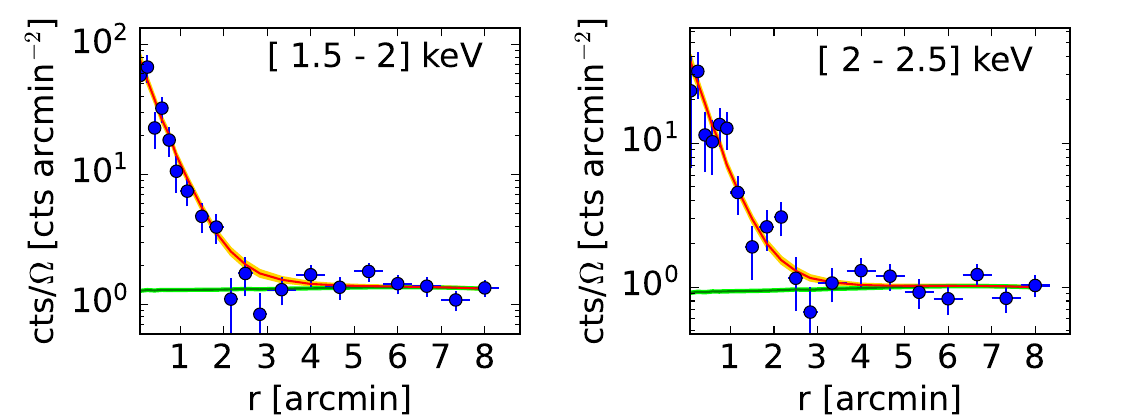}}
\centerline{\includegraphics[trim={5 5 30 230}, clip,height=3.5truecm]{fitwithmod_2_wlHE_id17.pdf}%
\includegraphics[trim={5 203 310 30}, clip,height=3.5truecm]{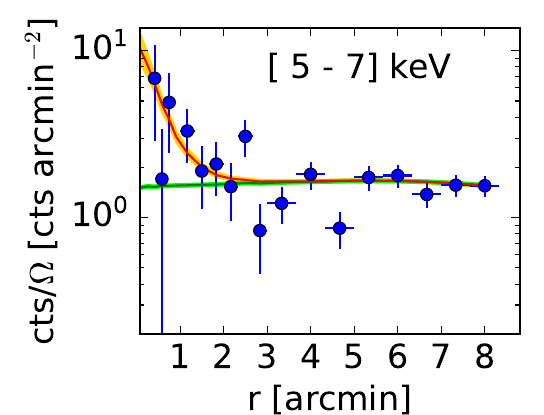}}
\caption[h]{id17 surface brightness profiles (points with error bars) in the X-ray bands with 68\% uncertainties
on the fitted model (red line and yellow shading). 
The green line with lime shading (barely visible) is the background radial
profile and its 68\% uncertainty. The analysis
uses Poisson errors, not the $\sqrt{n}$ plotted as 
errorbars. The assumed X-ray cluster model (red line)
represents well the observed X-ray profiles at all energies. Profiles of the other three clusters are shown in appendix~\ref{other_profs}.}
\label{fig:xray_obsprof_id34}
\end{figure*}

\subsection{Shear profiles and a closer look at id5}
\label{shear_section}

The shear data used in this work are those
those available in the Hyper Suprime-Cam Strategic program first-year shape catalog \citep{Mandelbaum2018}. The galaxy selection criteria are those presented in \citet[][Sect. 5.1]{Mandelbaum2018}: in particular, only galaxies with a signal-to-noise ratio larger than $10$ and $i<24.5$ mag are used for the shape analysis. 
Photometric redshifts are based on the subsample of galaxies with a SNR $\ge 5$ in all of the other ($grzy$) photometric bands.
Background galaxies used for the lensing analysis were selected adopting the {\em P-cut} method described by \citet{Oguri2014} and \citet{Umetsu2020}.
We used  the Ephor\_AB photometric redshifts \citep{Tanaka2018}, $p_{\rm cut}=0.95$, $z_{\rm min} = z_{\rm clus} + 0.1$, and with $z_{\rm MC} < 3$, where $z_{\rm MC}$ is the Monte-Carlo estimate of the redshift derived from the photometric redshift probability
distribution.
The surface density of background galaxies after these cuts  is
12.1, 7.6, 4.1, and 5.9 galaxies arcmin$^{-2}$ for id5, id17, id34, and id48, respectively.
We used radial bins with equal logarithmic spacing ($\Delta \log{R}=0.125)$ and we computed the tangential shear profile as usual \citep[see e.g.][]{Umetsu2020} accounting for shape measurement uncertainty, rms ellipticity, calibration bias and the responsivity factor \citep[e.g.,][]{Mandelbaum2018}, as detailed in Appendix~\ref{appendix_shear}. We use the
X-ray center determined in Sec.~\ref{xray_data_reduction}. The derived radial tangential profiles are shown in Fig.~\ref{fig:WL}.

In all fits performed in Sec.~\ref{fit}, we account for the galaxy shape noise ($C^{\rm stat}$ in the notation in  Umetsu et al. 2014) and we allow for a 20\% (Gruen et al. 2015, Chen et al. 2020) intrinsic scatter of the cluster lensing signal, excess surface mass to be precise, at fixed mass, due, for example, to deviations from spherical and the presence of correlated halos ($C^{\rm sys}$ in the Umetsu et al. 2014 notation). We do not use galaxies within 500 kpc to mitigate possible departures from the weak lensing regime and inaccuracies in shape measurements due to the dense cluster environment in this area. Similarly, we ignore galaxies at radii larger than 3.5 Mpc, where the effect of other clusters and large scale structures could require a modelling (e.g., Gruen et al. 2011). We do not account for the cosmic shear covariance due to uncorrelated large-scale structures projected along the line of sight (Hoekstra 2003, $C^{\rm lss}$ in the Umetsu et al. 2014 notation), which is negligible (its median is 5\% of the shape noise term).

To check the id5 mass, we compute the tangential shear profile from independent data, using using ellipticities from the Kilo-Degree Survey (KiDS) shape catalog \citep[DR4:][]{Kuijken2019,Wright2020,Hildebrandt2021,Giblin2021} and HSC photometric redshifts. Although the signal to noise of the profile is lower for KiDS ellipticities (because KiDS is shallower), the derived profile, shown in Fig.~\ref{fig:WL}, agrees with the one based on HSC. To further stress-test our mass derivation, we evaluate the impact of other clusters in the line of sights adjacent to id5. Specifically, we computed the expected change of the observed tangential shear profile of id5 in a number of configurations of increasing complexity: a) id5 alone; b) id5 plus a second halo with id5 mass at about  17 arcmin north and $z=0.11$ (to mimic Abell 1419); c) the previous case plus an additional halo with id5 mass at about 18' SWW at $z=0.13$ (to mimic Abell 1411); d) the previous case plus an additional halo with a quarter of the id5 mass at about 5' W at $z=0.254$ (to mimic the observed infalling group). In all four simulated cases,  the mass of the additional halos have been boosted for precaution by a large factor compared to the richness-based masses derived in Sec.~\ref{richness_section}.  In spite of the margin taken, the observed tangential shear profile within 2 Mpc is unaffected, and the point at 3 Mpc (the largest radius used) changes by less than $1\sigma$. These clusters are therefore not affecting our derivation of id5 tangential shear profile.

\subsection{X-ray data reduction}
\label{xray_data_reduction}

We observed id5, id17, and id34 for 38, 57, and 90
ks respectively (after time filtering) with the X-ray telescope on Swift, XRT, between January 2019 and December 2022.
The XRT background is low and stable because of the Swift low Earth orbit, making this instrument preferable to other X-ray telescopes for extended objects at fixed exposure time (Mushotzky et al. 2019; Walker et al. 2019). In fact, for id5, which is the cluster with the lowest signal in the [0.5-2] keV band within 1 arcmin, the background represents a minor (10\%) fraction of the cluster signal.

The X-ray data were reduced as already done in 
Andreon et al. 
(2019)
going through the usual steps of flare filtering, point source detection and flagging, accounting for vignetting, computing energy-dependent exposure maps, etc. At the date of the observations of our targets, the calibration radioactive sources onboard Swift have significantly decayed and, therefore, as in Andreon et al. (2019) we used the full XRT field of view at all energies. For background estimation, we use 18 high latitude background fields observed after 2016 with comparable exposure times and devoid of bright extended sources selected for this purpose from the XRT archive. These fields were processed as the cluster data.

Fig.~\ref{fig:image} shows the 0.5-2 keV band image of the three clusters. There are 560, 580, and 950 net photons within 2.5 arcmin of the cluster centre of id5, id17, and id34, respectively in the [0.5-2] keV band. Id34 shows a second extended X-ray emission ($\sim 2.6$ arcmin South, 200 net photons within 1 arcmin in the [0.5-2] keV band). Spectroscopic information is scarce for this cluster: in addition to cluster redshift reported in literature (Hilton et al. 2021), in the SDSS DR18 spectroscopic sample there is one galaxy only with spectroscopic redshift in each region where X-ray emission is detected and the two velocities differ by about 1700 km/s rest-frame. The asymmetric extension of the central emission toward the Southern one suggests that we are observing two parts of the same object and that the cluster is not in equilibrium. 

Nine arcmin north of id5 there is a second evident extended X-ray emission (outside the field of view of the Fig.~\ref{fig:image}), spatially coincident with the already mentioned Abell 1419. Because of its large distance and modest flux, the contamination by Abell 1419 is negligible and we can simply flag out the region. 

For the spatial-spectral analysis, XRT photons are partitioned into nine bands: [0.3-0.5], [0.5-0.8], [0.8-1.1], [1.1-1.5], [1.5-2], [2.0-2.5],[2.5-3.5], [3.5-5.0], and [5.0-7.0] keV. We then measured counts and effective exposure time in the nine energy bands in circular annuli with width increasing with radius to counterbalance the decreasing intensity of the cluster. The minimal width is taken to be 10 arcsec, comparable to the XRT PSF. We only considered annuli where vignetting is less than 50\% and we truncate the profile when less than two third of the area of the annulus is within the 50\% vignetted region. The cluster centres, reported in Table~\ref{tab1}, are iteratively computed as the centroid of X-ray emission within the inner 300 kpc radius. These centers closely correspond to the center inferred from the gravitational arcs for the two clusters showing them, id17 and id48. Derived surface brightness profiles of id17 are shown in Fig.~\ref{fig:xray_obsprof_id34} whereas those of id5 and id34 are shown in Fig.~\ref{fig:x-ray_id5} and Fig.~\ref{fig:x-ray_id34}.

The radial profile of the background is derived, for each background field, in the same way as for the cluster and using the cluster center in pixels coordinates as extraction center (so each cluster has its own background profile even if the same 18 background fields are used) to account for the background spatial pattern. We then measure the shape of the background radial profile in each energy band stacking the profiles derived in each field after normalizing them at $6<r<9$ arcmin. Finally, the background profile is fitted to the cluster radial profile at radii where the cluster contamination is minimal to derive a preliminary background normalization that we use as prior to our X-ray fit.

The XRT PSF is modelled following Moretti et al. (2007) in each of the nine energy bands. A posteriori, accounting for the PSF proved to be unnecessary for our purposes. 

Id48 was serendipitously observed for clean 8 ks in two ACIS-I chips switched on in an ACIS-S observation targeting a very distant quasar (OBSID=3960). The analysis of these data are pretty similar to those of Swift data, and identical to the analysis of Chandra data in Andreon et al. (2021). Differences from the Swift analysis are: we used the [0.7-1.0], [1.0-2.0], [2.0-3.0], [3.0-5.0], and [5.0-7.0] keV bands, and the background radial profile uses blank field images computed with CIAO BLANKSKY (Fruscione et al. 2006), normalized to the count rate in the hard band [9-13] keV, in place of empty fields. The [0.5-2] keV image of id48 is shown in Fig.~\ref{fig:image}. Derived surface brightness profiles are shown in Fig.~\ref{fig:x-ray_id48}. There are 340 net photons within 2.5 arcmin of the id48 in the [0.5-2] keV band.

\begin{figure*}
\centerline{\includegraphics[width=8truecm]{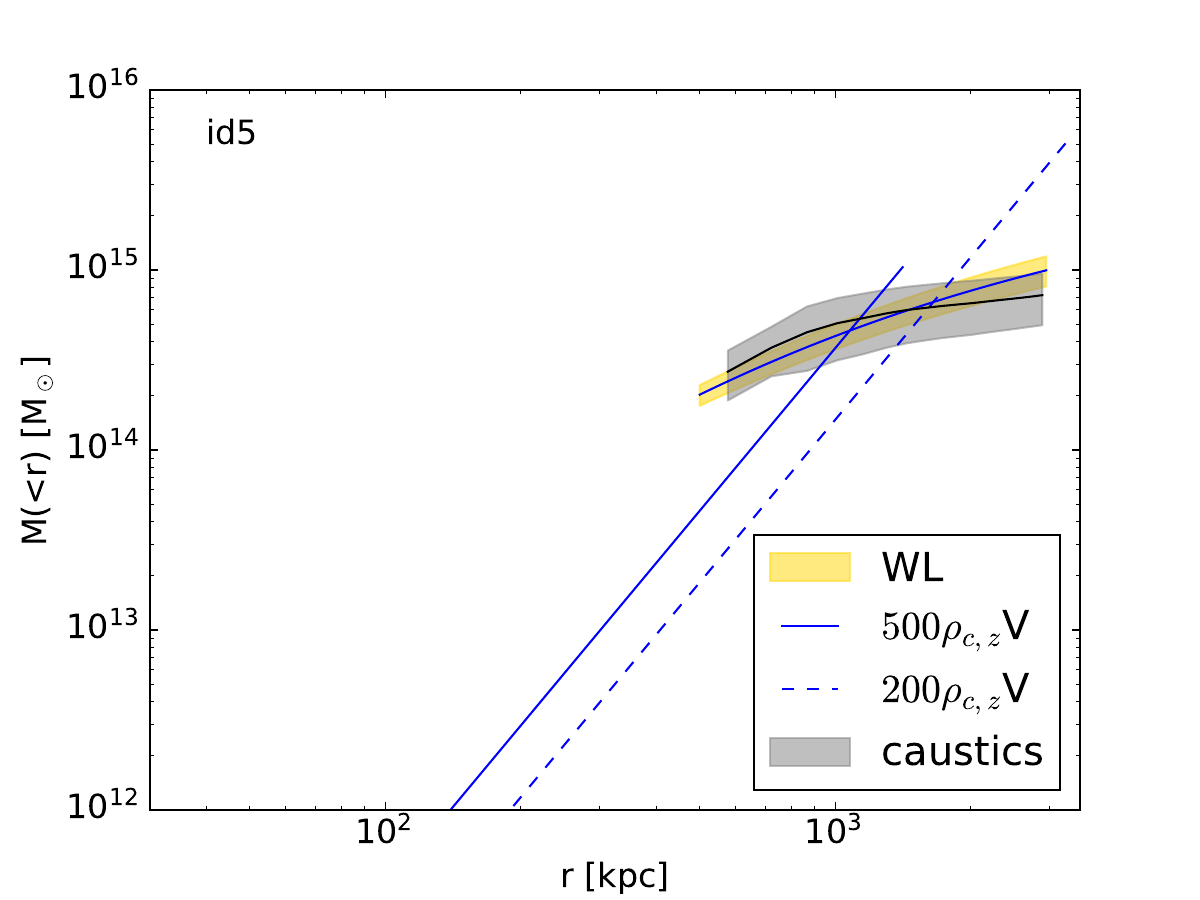}\includegraphics[width=8truecm]{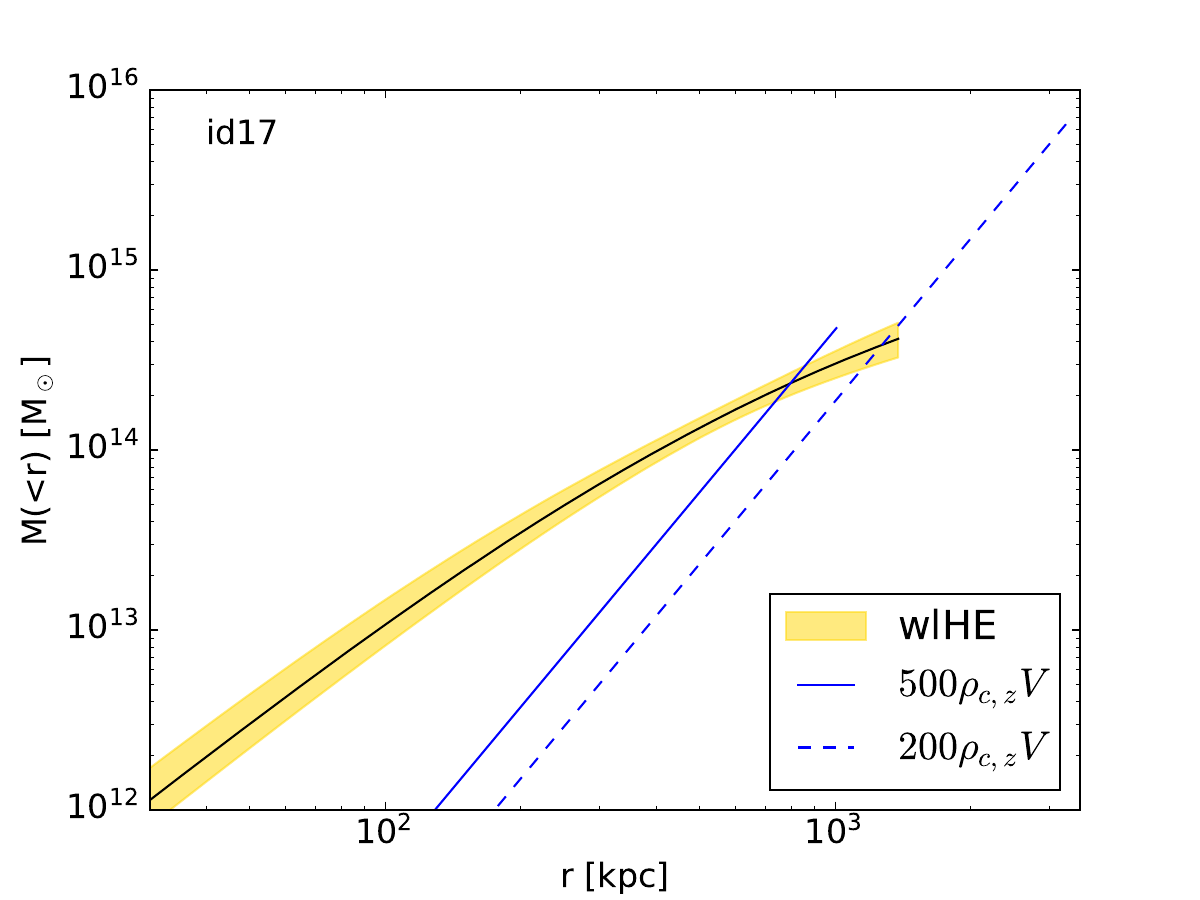}}
\centerline{\includegraphics[width=8truecm]{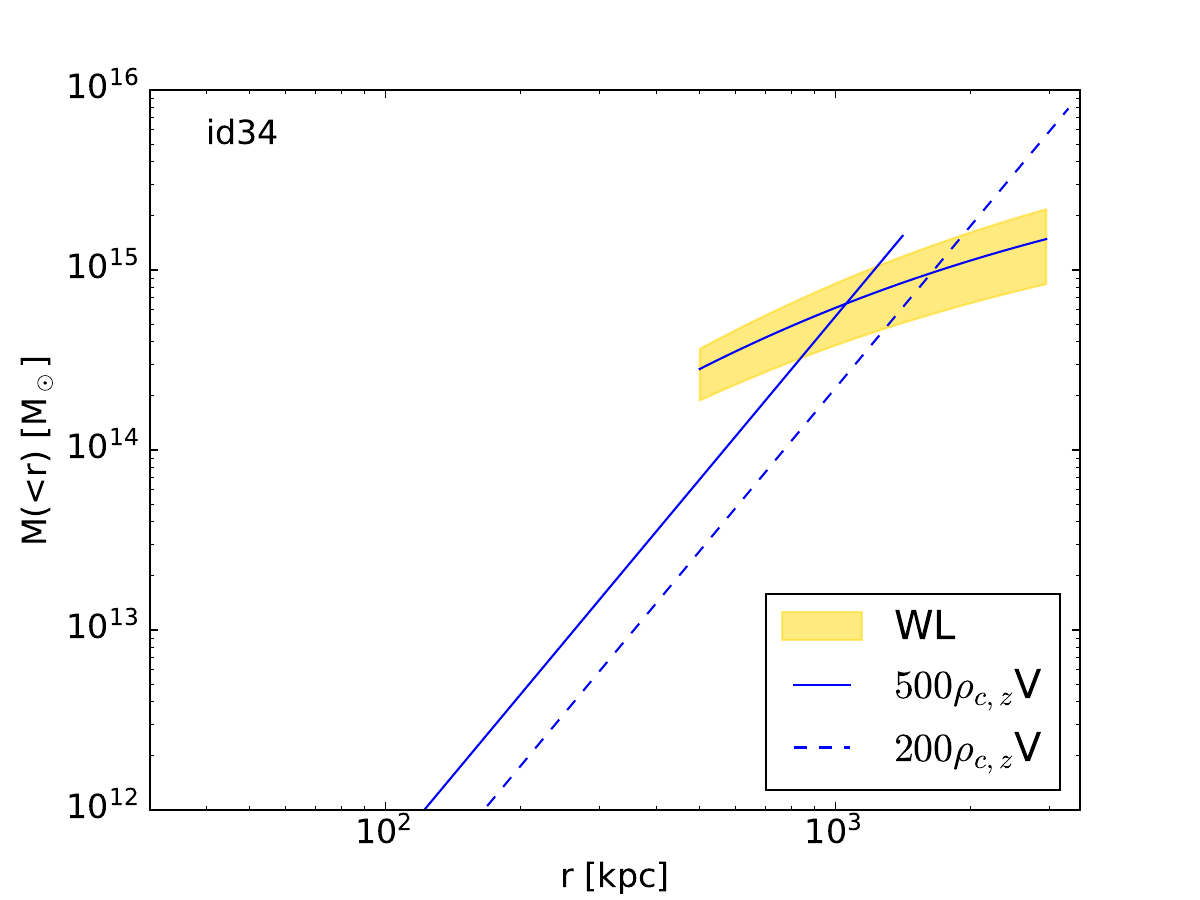}\includegraphics[width=8truecm]{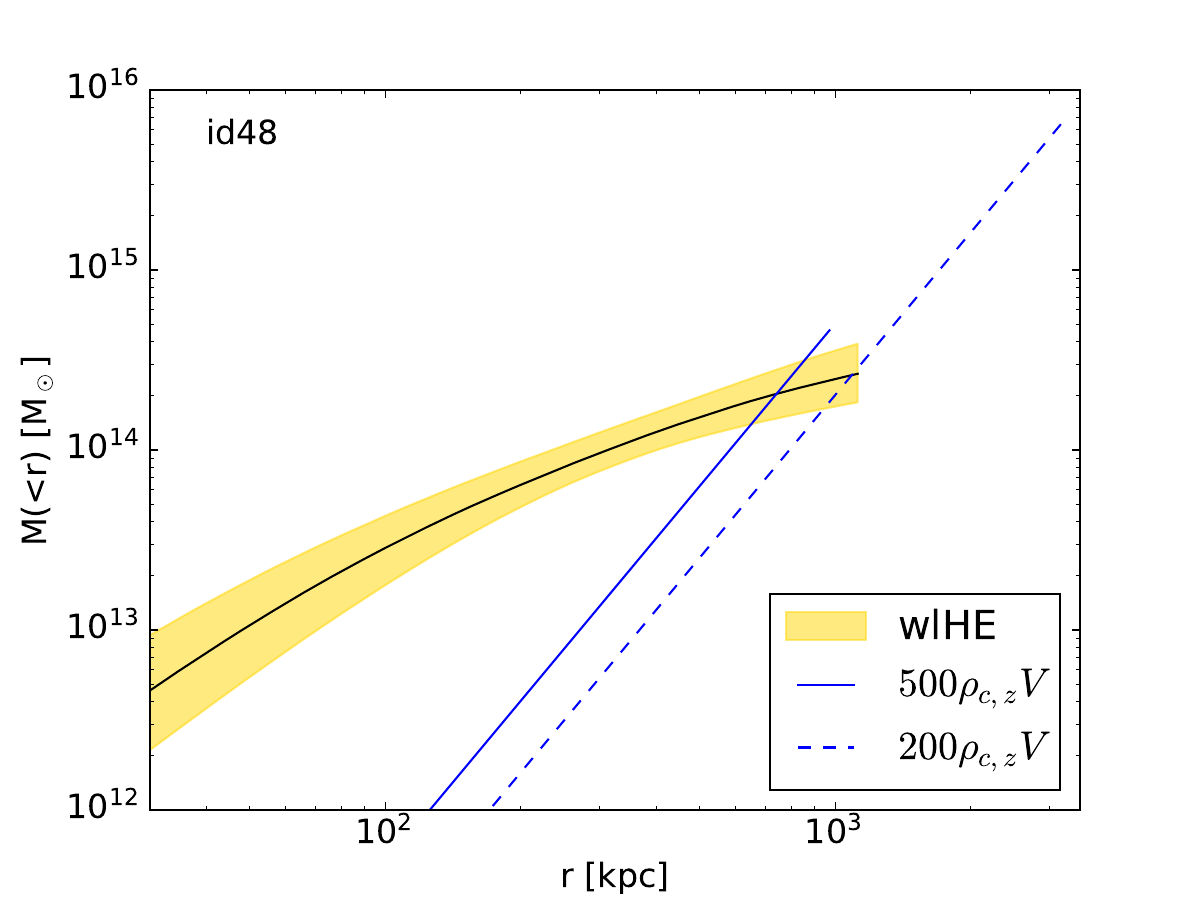}}
\caption[h]{Cumulative mass profile of the four clusters (solid line with 68\% uncertainty in yellow). The mass profile is constrained by weak-lensing data at large radii and, in the case of joint fits, by X-ray data at small radii (roughly, $r<500$ kpc). Id5 cumulative mass profile derived from the caustic method (solid line with 68\% uncertainty in gray) agrees well with the profile derived from the shear.}
\label{fig:massprof}
\end{figure*}

\subsection{Joint/disjoint Analyses}
\label{fit}

We want to
derive core-excised X-ray luminosities, and profiles of mass, pressure and electron densities of the four clusters. Based on the available data, it is reasonable to assume spherical symmetry and hydrostatic equilibrium for id17 and id48. We therefore do a joint fit of shear and X-ray data under these hypotheses. Id34 is bimodal both in the optical and in X-rays and therefore we need to proceed in a different way: 1) we do not consider X-ray and shear data jointly (it would require hydrostatic equilibrium); 2) we consider a reduced radial range,  where the assumption of spherical symmetry is more likely to hold, namely, the inner 2 arcmin for the X-ray analysis (after generously flagging the second clump) and radii larger than 0.9 Mpc for the shear one; and 3) we do not assume hydrostatic equilibrium in the analysis of X-ray data. The assumption of hydrostatic equilibrium for cluster id5  is risky because of its dynamical complexity (Sec.~\ref{richness_section}), and, as illustrated below, it has anomalous optical and, to lower extent, X-ray properties for its mass. Therefore, to be conservative, we do not assume hydrostatic equilibrium in the analysis of this object as well,  and we do not perform a joint fit to shear and X-ray data. Nevertheless, we checked that the electron density and pressure profiles of id5 derived using a joint fit are comparable (identical) to what we find here. 

The shear data are fitted with a Navarro et al. (1997) radial profile for the total matter with an uniform prior on $\log$ concentration between ($\log$) 0.1 and 30 following Hamana et al. (2023). The fit to the shear data only (the disjoint fits) of id5 and id34 assumes the Dutton \& Macci\`o \ (2014) concentration-mass relation, whereas there is no such prior for joint fits of the other two clusters (because the X-ray data at small radii have a large constraining power). The shear fit to id34 further assumes a minimal mass $\log M_{200}=14.2$ inferred from an hydrostatic analysis of the X-ray data only of the main clump. The latter prior is needed to remove fits with very small concentration allowing large amounts of mass at the center (made possible because we flag off all the inner $\sim1$ Mpc radial profile) that are clearly ruled out by the X-ray data. The larger signal-to-noise of id5 data and the reduced flagging at small radii make this prior unnecessary for it. The log-uniform prior on concentration, although used in all fits, can be safely ignored for three of the objects (the joint fits and id5).

\begin{figure}
\centerline{\includegraphics[width=9truecm]{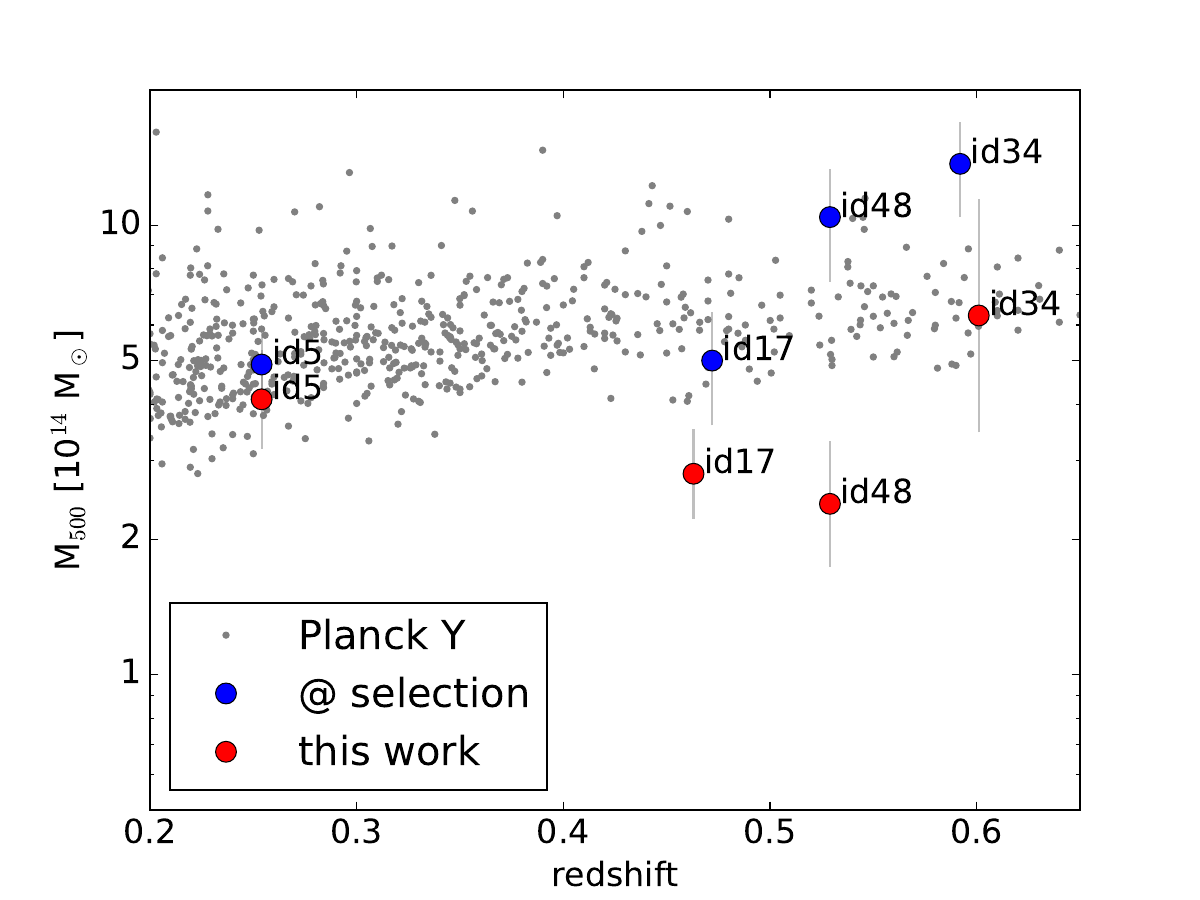}}
\caption[]{Mass-redshift plot. We plot the SZ-selected Planck sample of clusters (Planck collab. 2016, gray points) and the studied, shear selected, sample with both the masses at the time of the selection 
(large blue circles) and those derived in this work (red large circles). Planck masses are, in practice, the
Compton $Y$ signal converted in mass assuming a linear relation on log quantities. 
None of the four shear-selected clusters are detected by Planck, in line with the mass values derived in this paper.}
\label{fig:Mz_prepost}
\end{figure}

Both the observed tangential shear profile  and a fit to it with parameters taken at maximum likelihood estimate (i.e., the minimum $\chi^2$) are biased because of the Eddington (1913) bias induced by the steep mass function, 
as already remarked by Andreon et al. (2009), Andreon \& Congdon (2014), and Andreon \& Weaver (2015) for cluster masses, and by Hamana et al. (2023) for the mass of these very same clusters. The mass prior, taken to be a Tinker et al. (2008) mass function, corrects for the Eddington bias. The prior has the effect of lowering the estimated mass with the rationale that the detected objects likely belong to the population of low mass clusters scattered up. As a consequence of the bias, the fit (to the shear data alone or to the joint X-ray and shear data) with parameters taken at the posterior mean will have a non-minimal $\chi^2$ and, in extreme circumstances, a large $\chi^2$, mimicking a bad fit to the shear data, as best visible in the top-right panel of Fig.~\ref{fig:WL} (the solid line avoids the data points). Of course, when the signal-to-noise of the shear data is high (as for id5, top-left panel), the bias is small. This Eddington correction can be safely ignored when our X-ray data are jointly fit because they tightly constrain the cluster mass, but we applied it anyway. To sum up, the mass prior is used for all fits, but could be safely ignored for three of the objects (the joint fits and id5). For illustration purposes only,  we also performed a shear-only fit with an uniform prior on mass to all clusters to visualize the amplitude of the Eddington correction in Fig.~\ref{fig:WL}.

In all our fits, we model the electron density profile following Vikhlinin et al. (2006) with a flexible function constrained by the data to impose regularity and smoothness. We use a modified single-$\beta$ profile for the electron density with six parameters ($n_0$,$\alpha$,$\beta$,$\epsilon$,$r_c$, $r_s$):
\begin{equation}
 n_\mathrm{e}^2 = n_0^2
 \frac{(r/r_\mathrm{c})^{-\alpha}}{(1+r^2 / r_\mathrm{c}^2)^{3\beta-\alpha/2}}
 \frac{1}{(1 + r^\gamma / r_\mathrm{s}^\gamma)^{\epsilon / \gamma}}.
\label{eq:dens} 
\end{equation}

We adopt weak priors for the parameters fixing $\gamma=3$ as in  McDonald et al. (2013), positive priors for the slopes and $\epsilon<10$. In all our fits, metallicity is a free parameter, absorption is fixed at the total Galactic $N_H$ value in the direction of the cluster (Willingale et al. 2013), and the results are marginalized over a background scaling parameter to account for systematics (differences in the background level between the cluster and control fields), taken to have a Gaussian prior centered on one with 10\% sigma which is close to the observed background scatter across fields (Moretti et al. 2009; Andreon et al. 2023). The model is integrated on the same energy and radial bins as the observations, so that the results do not depend on the binning choice.

For the joint fit of id17 and id48, 
we assume hydrostatic equilibrium, and we derive the other thermodynamic profiles combining the mass profile and the electron density assuming hydrostatic equilibrium and the ideal gas law. For the disjoint fits of id5 and id34, instead, we model the temperature profile with a Vikhlinin et al. (2006) model with parameters $a=0$ and $a_{cool}=0$:
\begin{equation}
T = \frac{T_{m}}{(1+(r/r_t)^b)^{c/b}} \, .
\end{equation}
and we use weak priors for the parameters (zeroed at negative, unphysical, values when these are not allowed by the considered parameter, such as the parameters b and c. We also adopted $T_m>0.1$, $b<6$ and $c<4$. 

Finally, in all our fit, metallicity is left free, and we adopt an uniform prior for it (zeroed at negative, unphysical, values).

These fits are performed by extending \texttt{MBProj2} (Sanders et al. 2018) to also fit the binned tangential shear profile. In short, \texttt{MBProj2} is a Bayesian forward-modeling projection code that fits the X-ray data cube accounting for the background and computes the log-likelihood summing over all the radial-energy bins. Our joint fit adds the log-likelihood coming from the weak-lensing data to the \texttt{MBProj2} X-ray log-likelihood. In the case of disjoint fits (id5 and id34), we use the same software twice: once with the shear likelihood set to one, effectively coming back to (our own implementation of) \texttt{MBProj2} to derive the gas profiles, and a second time setting to one the X-ray likelihood to derive the mass profile from the shear data only. The Tinker et al. mass prior is computed using \texttt{hmf} (Murray et al. 2013),
whereas the surface mass density uses \texttt{cluster-lensing} (Ford \& Vanderplas 2016).

\begin{figure}
\centerline{\includegraphics[width=9truecm]{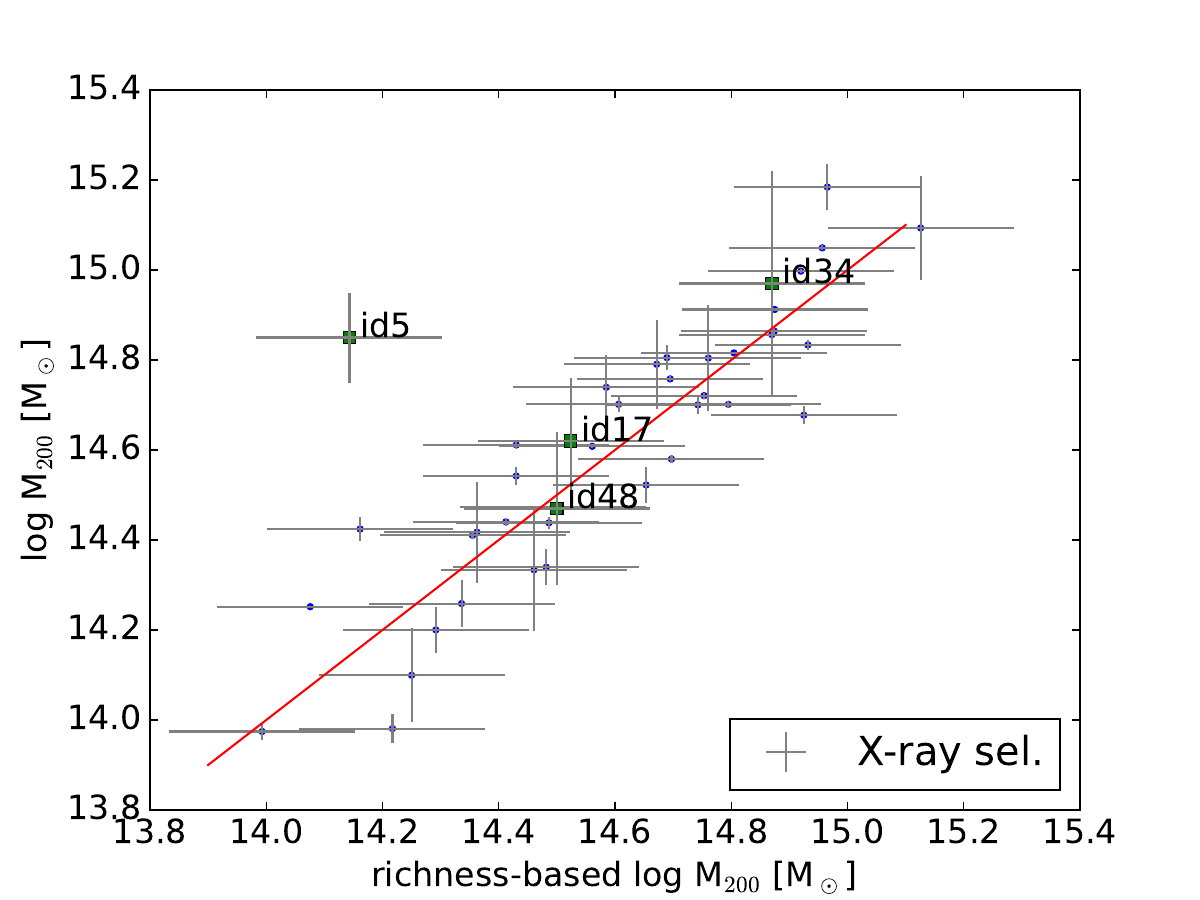}}
\caption[]{Mass-Mass plot of our studied sample (green squares) and of the X-ray selected comparison sample in Andreon (2016). The abscissa is the mass estimated from the richness, the ordinate is our fitted (mostly weak-lensing based) mass for our sample and the caustic mass for the comparison sample. There is a good agreement between mass estimates, except for id5, whose richness is too low by a factor of about three (see text for details). Errorbars also include intrinsic scatter between observable and mass (due to, for example, cluster elongation).}
\label{fig:M_rich}
\end{figure}

\subsection{Fit results}
\label{results}

Table~\ref{tab1} shows the type of fit that we perform, the derived $\log M_{500}$ and $\log M_{200}$ of our shear-selected sample. Table~\ref{tab1} also lists [0.5-2] keV core-excised ($0.15<r/r_{500}<1$) luminosities that we derived by integrating the electron-temperature radial profiles, which account for variation in sensitivity, exposure time, vignetting and flagging at the pixel level. For the three unimodal clusters, we verified that identical point estimates of luminosity are obtained counting [0.5-2] keV photons within the corona as usually done in literature. Id34 is bimodal and therefore we add back the photons of the second clump to the integral of the model of the main clump.

The weak-lensing data have extremely low constraining power at small radii, where instead X-ray data are most sensitive. Identical electron density and pressure profiles would be derived with disjoint fits for the two clusters with joint fits (i.e. for id17 and id48). X-ray data offer little constraints to mass at radii where the weak-lensing data are sensitive, and identical mass profiles at large radii would be derived in absence of X-ray data (we obtain $\log$ M$_{200}$/M$_{\odot}=14.69\pm0.20$ and $14.25\pm0.4$ for id17 and id48, respectively). In the case of id17 and id48, the joint fit of X-ray and shear data largely breaks the mass-concentration degeneracy allowing us to derive reliable masses in a large radial range.

The electron density radial profile, with its six parameters, overfits the data making it just a regularizing kernel. In particular, $\alpha$ and $\epsilon$ parameters are poorly determined. Since we are not interested in the parameters values, but only in the profiles, these degeneracies do not  affect strongly our results. The backscale parameter, that measures the amplitude of a potential (multiplicative) offset between the X-ray backgrounds in the line of sight of the cluster and empty fields, is extremely well determined and very close to one. The global gas metallicity is poorly determined. 

The bottom-left panel of Fig.~\ref{fig:WL} shows the shear-only fit to the tangential shear profile of the bimodal id34 cluster. The derived mass values turn out to be largely independent of the minimal mass taken and also whether we only fit larger radii only, for example $r>1.7$ Mpc (the second clump is at 1 Mpc from the center). As mentioned, for this shear-only fit the Eddington correction is important (compare solid blue and dashed cyan lines). The top-left panel of Fig.~\ref{fig:WL} also shows the id5 shear-only fit. Given the high signal-to-noise of the shear data, the Eddington correction is negligible for this object. Figures~\ref{fig:WL} also shows the model fits to the shear data of the id17 and id48. For the latter two clusters, shear and X-ray data are jointly fit, as mentioned.

Fig.~\ref{fig:massprof} shows the derived cumulative mass radial profiles of the four clusters. Clusters with joint fits have, of course, cumulative mass radial profiles extending over a wider range. Id5 cumulative caustic mass profile agrees well with the shear one.

Mostly as a consequence of accounting for Eddington (1913) bias, our masses are smaller than those used at the time of the cluster selection, as visible in Fig.~\ref{fig:Mz_prepost}, in agreement with later weak-lensing analysis of HSC data (Hamana et al. 2020, 2023; Oguri et al. 2021; all works appeared after the start of our observational program). In particular, our mass estimates also agree with Hamana et al. (2023), who use a different photo-z estimator, and conceptually similar, but not identical, correction for Eddington bias. None of our four shear-selected clusters is detected by Planck, although they have the mass (as reported Miyazaki et al. 2018) and redshift appropriate to be included in the Planck cluster catalog (Planck collab. 2016), see blue and gray points in Fig.~\ref{fig:Mz_prepost}.
With the revised mass, the non-detection by Planck of these four shear-selected cluster is expected, because the clusters are below, or near the bottom of, the cloud of Planck-selected clusters in Fig.~\ref{fig:Mz_prepost}, unlike the case in which the Eddington correction is not applied (red points in Fig.~\ref{fig:Mz_prepost}).

Fig.~\ref{fig:xray_obsprof_id34} illustrates how well the model (red line with 68\% uncertainty shaded) fits the id17 X-ray data, whereas the fit to the other clusters is shown in Fig.~\ref{fig:x-ray_id5}.

\begin{figure}
\centerline{\includegraphics[width=9truecm]{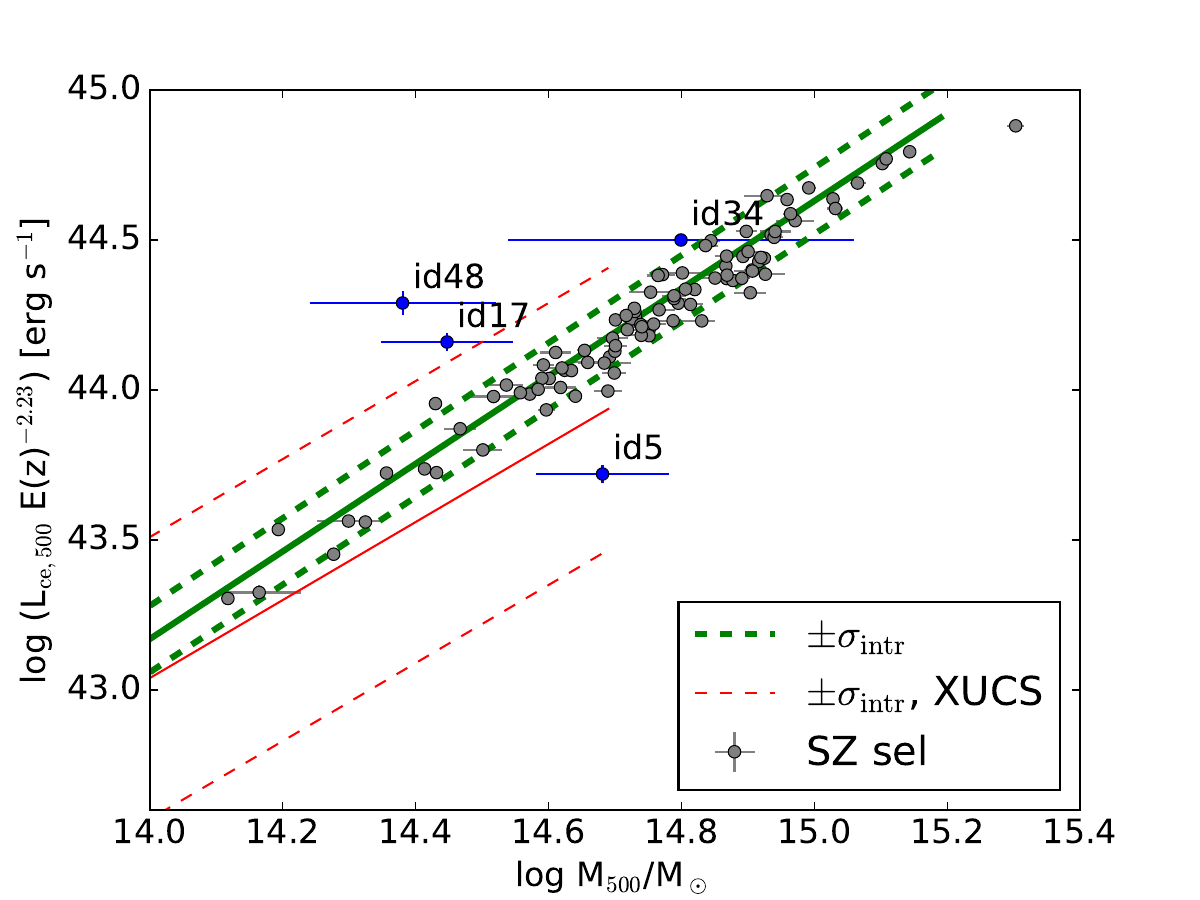}}
\caption[]{[0.5-2] keV core-excised ($0.15<r/r_{500}<1$) luminosity vs mass of our sample (squares) and and of the (biased) SZ-selected sample in Pratt et al. (2022). Masses in the latter sample are based on $Y_X$. The green solid line and the dashed corridor show the mean relation and the $\pm 1$ intrinsic scatter corrected for underestimation (from Pratt et al. 2022, see text) for the biased SZ-selected sample. The red solid line and the dashed corridor show the same, but for an X-ray and SZ-unbiased sample (Andreon et al. 2016). 
Id5 has an atypical (low) X-ray luminosity compared to the plotted comparison, SZ-selected, sample for its mass. The same is true using X-ray selected samples as comparison samples.
However, when compared to an X-ray unbiased sample (in red) is well within the $\pm 1\sigma_{\rm intr}$, i.e. it is typical.}
\label{fig:Lx_M}
\end{figure}

\begin{figure*}
\centerline{\includegraphics[width=8truecm]{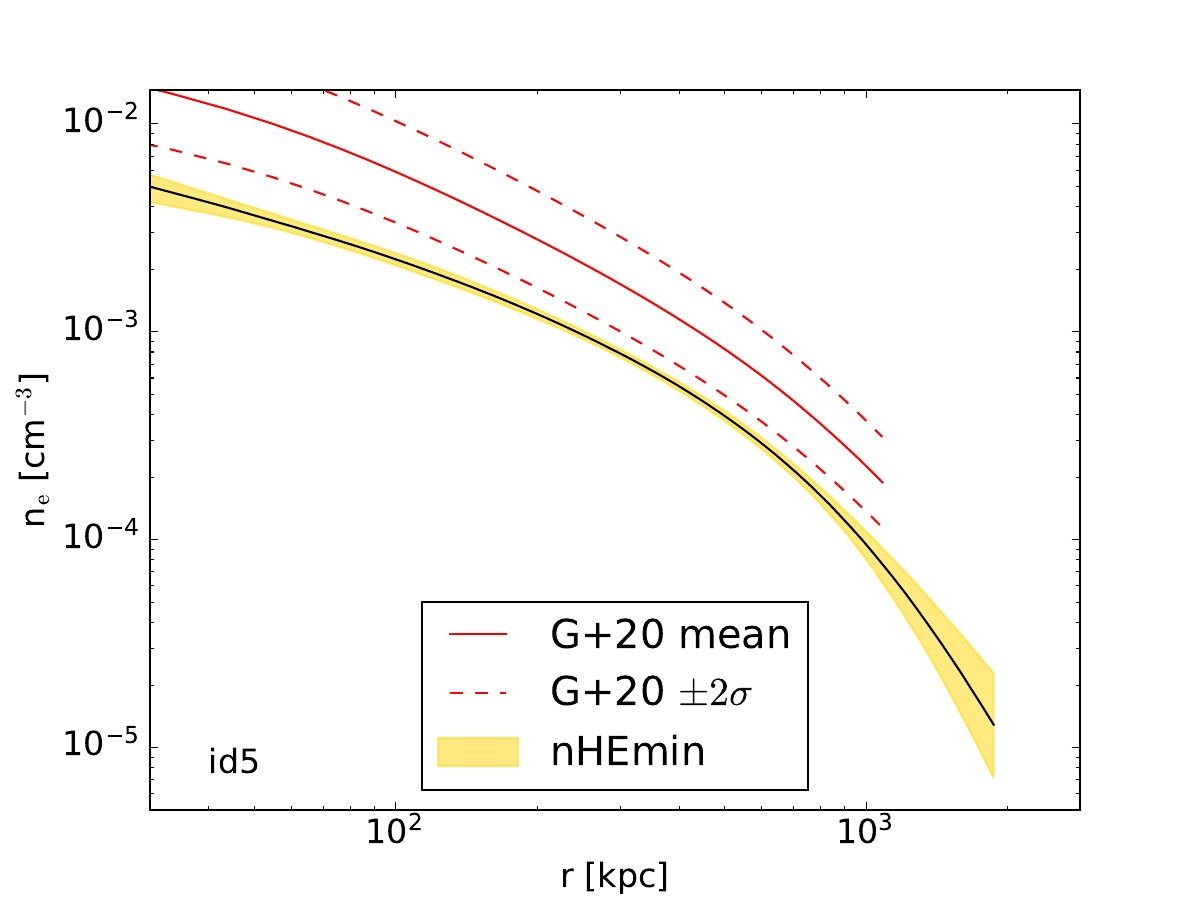}\includegraphics[width=8truecm]{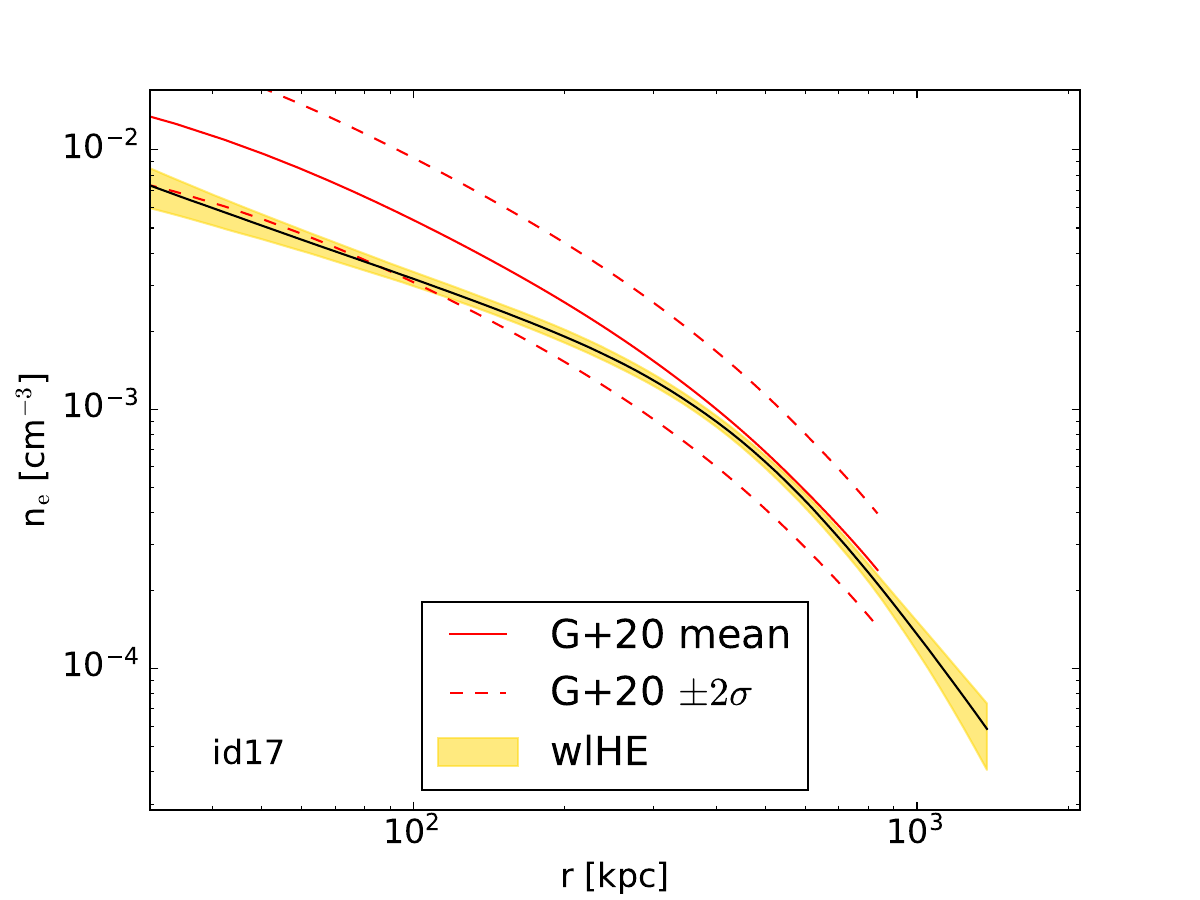}}
\centerline{\includegraphics[width=8truecm]{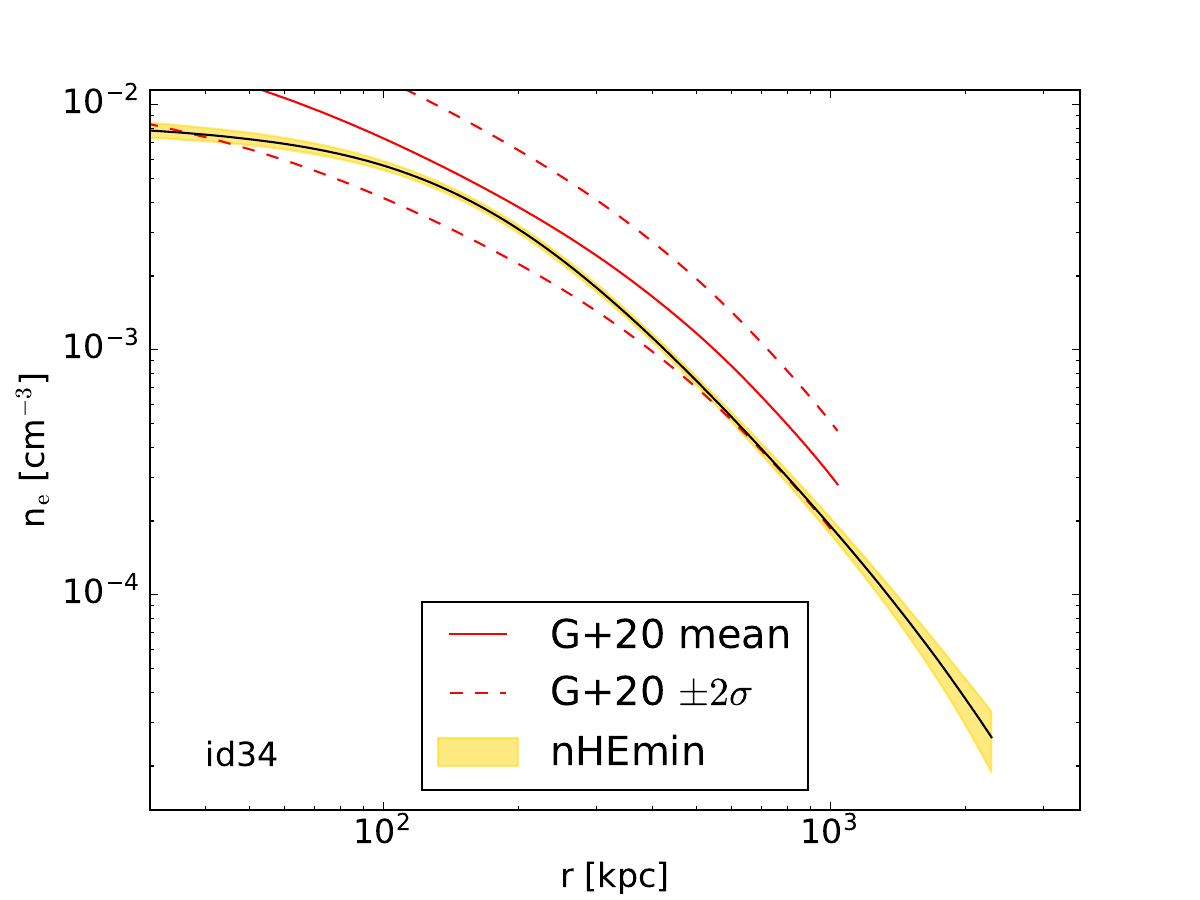}\includegraphics[width=8truecm]{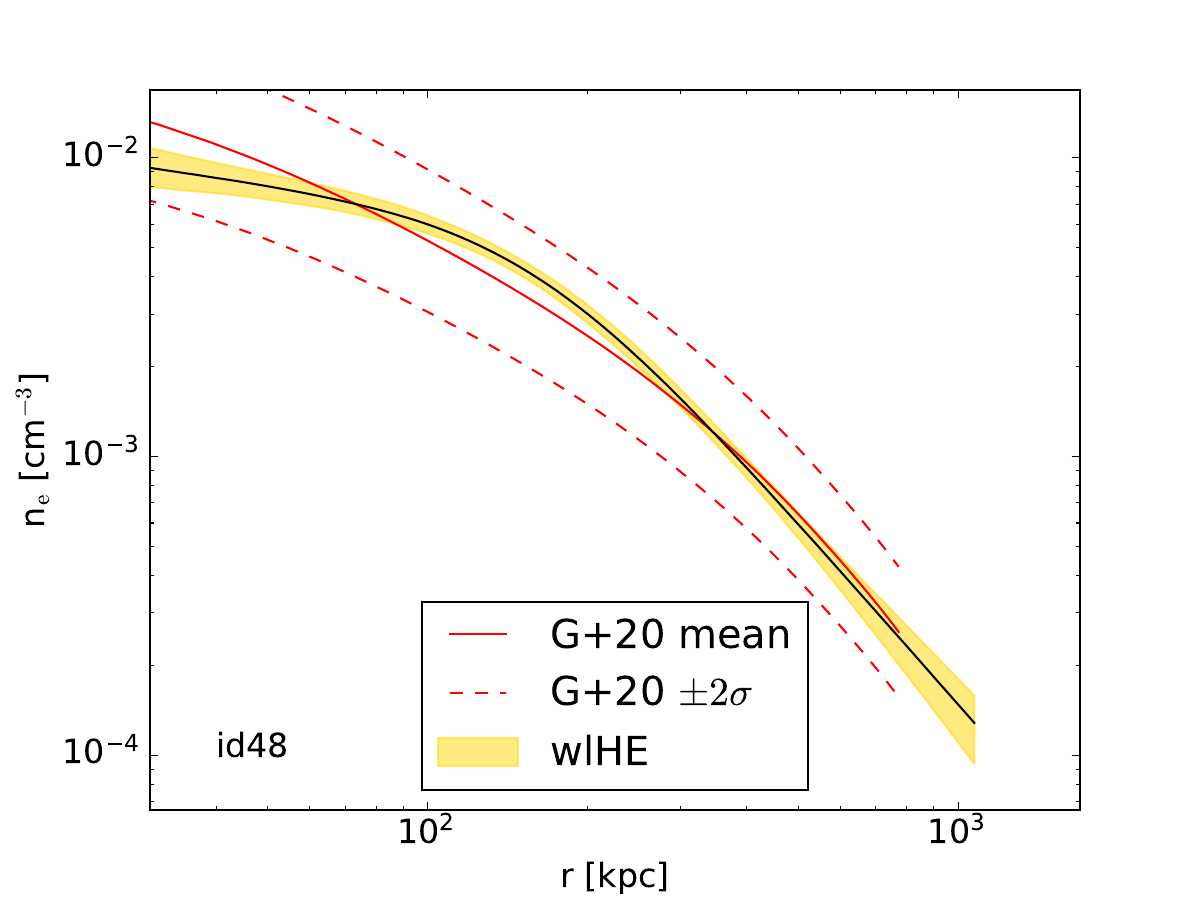}}
\caption[h]{Electron density profile of the four shear-selected clusters (posterior mean value with 68\% uncertainty shaded) and of SZ-selected clusters (Ghirardini et al. 2020) of the same mass (red line with dashed corridor marking the $\pm 2 \sigma$ range). The inset indicates the type of performed fit: a joint X-ray and weak lensing fit assuming hydrostatic equilibrium (``wlHE"), a disjoint X-ray fit, where  hydrostatic equilibrium is not assumed (``nHEmin"). 
The whole profile of id5 and the central value of electron density of id5, id17 and id34 
are near or outside the $\pm2\sigma$ range defined by SZ-selected samples. 
}
\label{fig:ne_prof_others}
\end{figure*}

Fig.~\ref{fig:M_rich} compares $M_{200}$ independently estimated from the cluster richness and from the shear data (the addition of the X-ray data in id17 and id 48 is irrelevant, since it does not constrain the mass at those large radii). The figure compares our shear-selected sample to the X-ray selected sample in Andreon (2015, 2016) which uses masses based on the caustic technique and identical measurements of richness. Errorbars account for intrinsic scatter between observable (richness, tangential shear profile, and escape velocity) and mass, unlike many literature works. In particular, they include cluster elongation uncertainties. There is a good agreement between the two mass estimates for three shear-selected clusters, whereas id5 is $>3\sigma$ away from the mean relation. The weak-lensing mass of this cluster, however, agrees with the caustic estimate (sec~\ref{richness_section}). The shear-based masses of our shear-selected sample calculated without the correction for Eddington bias would deviate from the mean relation (would be too large). On an individual base, shear- and richness-based mass estimates have very similar errors (see Table~\ref{tab1}) in spite of the fact that the former requires data of much better quality (the shear of background galaxies is a weak, and statistical only, signal).

As tested in Sec.~\ref{richness_section}, the id5 richness is robust to the choice of the used photometry, adopted code, and to whether we use a statistical or spectroscopic background subtraction. In terms of number of galaxies, id5 has 28 red massive galaxies within the richness-estimated $r_{200}$ (36 galaxies including the background), or 38 (statistical) members within the shear-estimated $r_{200}$, while it should have about 100 galaxies to obey the plotted relation. Since there are 60 galaxies in total in the area, including background galaxies, id5 richness is way too low for its mass. We are not aware of any example of so extreme outliers among over 100 clusters studied thus far (Andreon \& Hurn 2013; Andreon \& Congdon 2014, Andreon 2015) other than an unique outlier with a dubious (caustic) mass estimate, Abell 1068.

\begin{figure*}
\centerline{\includegraphics[width=8truecm]{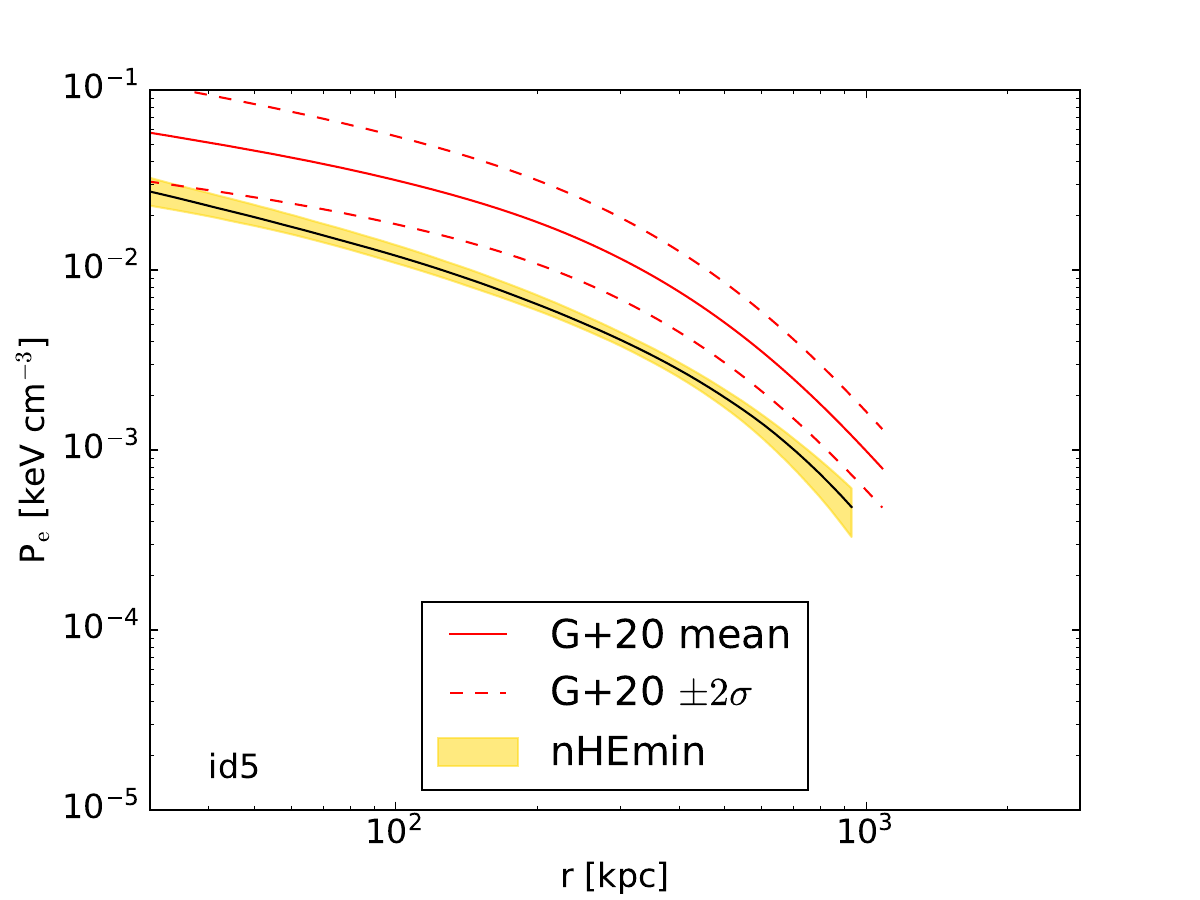}\includegraphics[width=8truecm]{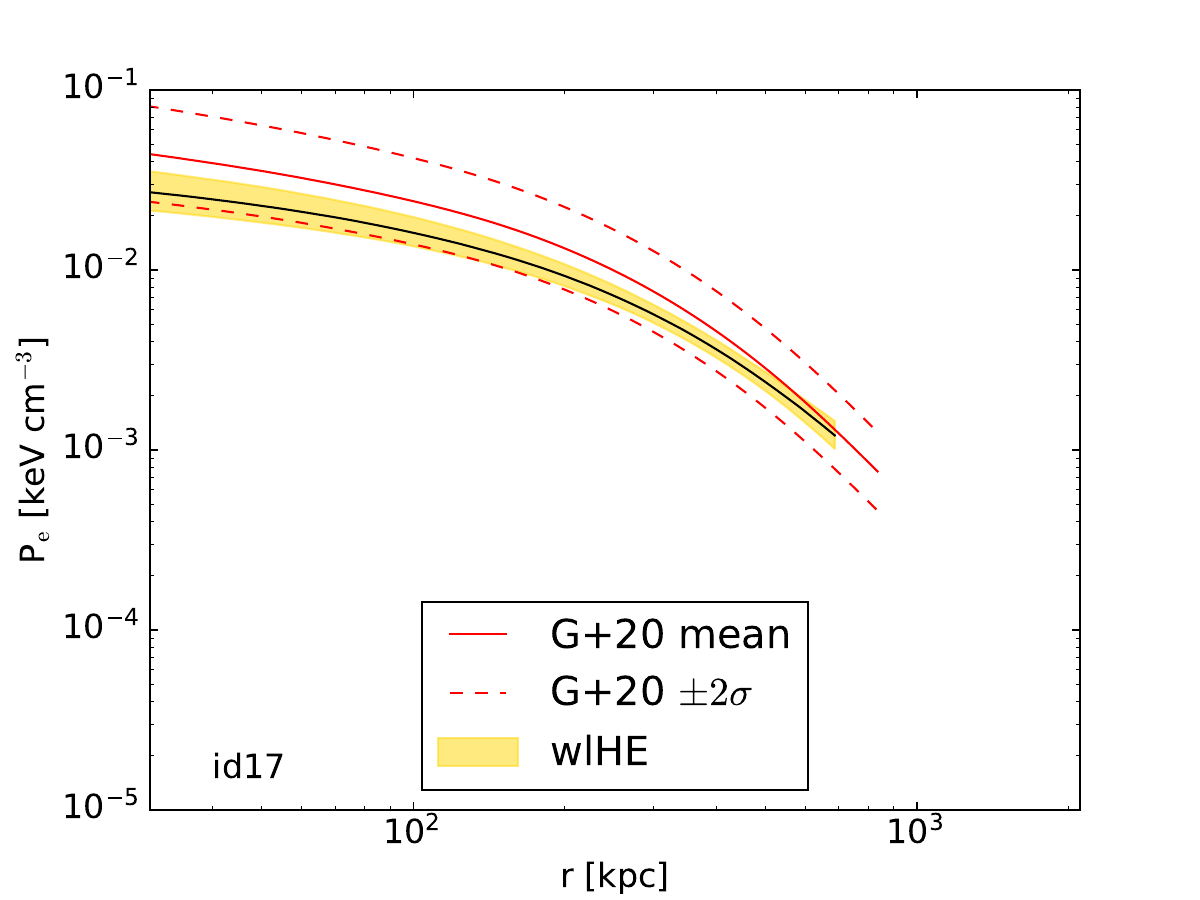}}
\centerline{\includegraphics[width=8truecm]{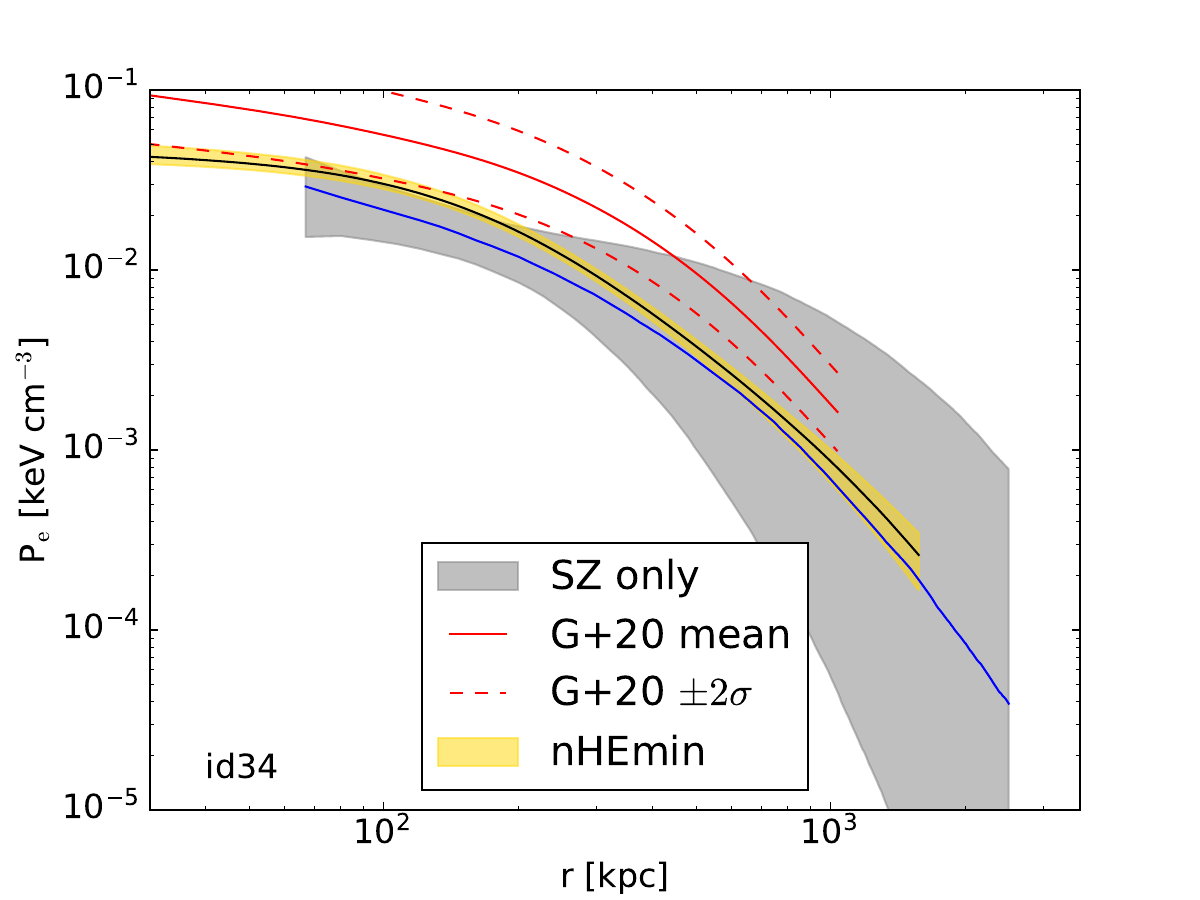}\includegraphics[width=8truecm]{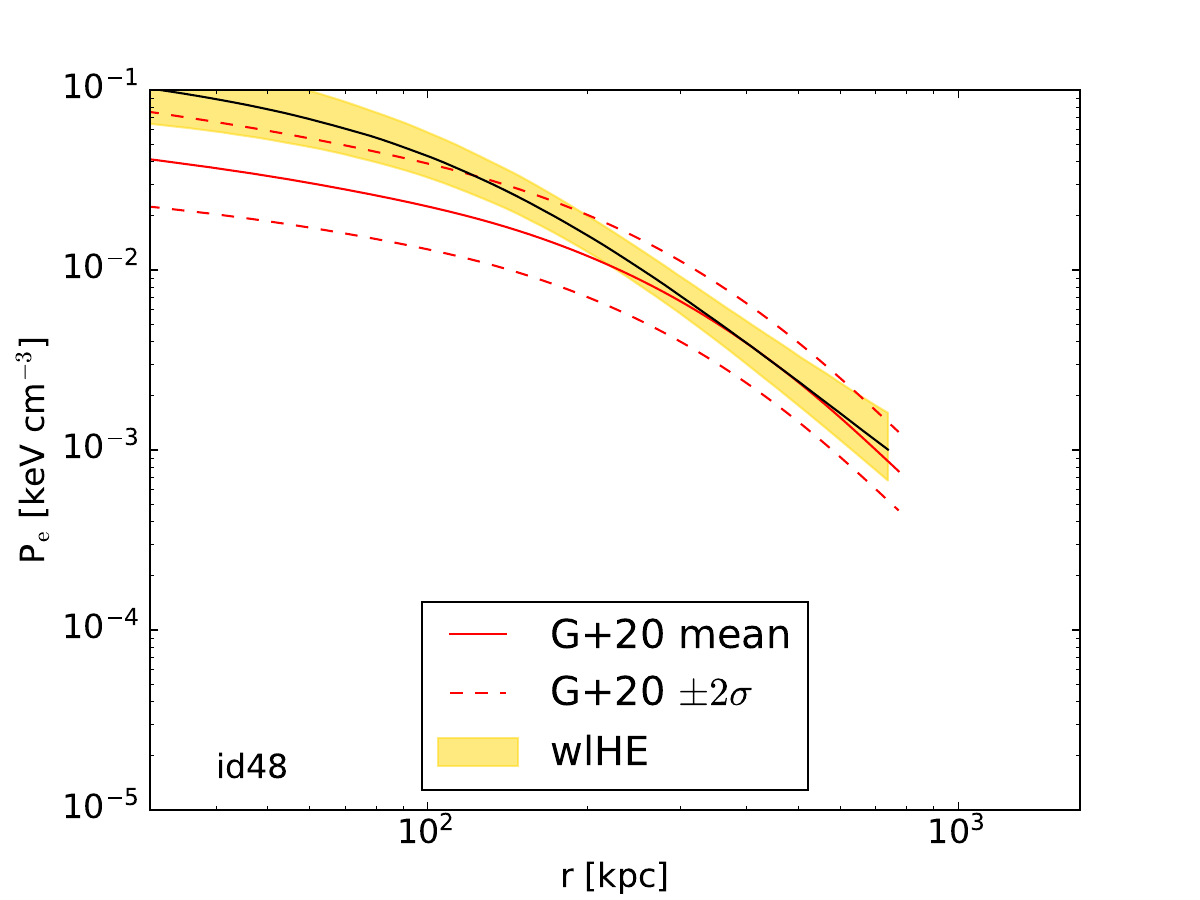}}
\caption[h]{Pressure density profile of the four shear-selected clusters (posterior mean value with 68\% uncertainty shaded) and of SZ-selected clusters (Ghirardini et al. 2020) of the same mass (red line with dashed corridor marking the $\pm 2 \sigma$ range). The inset indicates the type of performed fit: a joint X-ray and weak lensing fit assuming hydrostatic equilibrium (``wlHE"), a disjoint X-ray fit, where  hydrostatic equilibrium is not assumed(``nHEmin"), SZ-only fit (``SZ only"). 
The whole profile of id5 and of id34 and the central value of pressure density of 
id5 and id34 
are near or outside the $\pm2\sigma$ range defined by the SZ-selected samples. 
}
\label{fig:Pe_prof_others}
\end{figure*}

Fig.~\ref{fig:Lx_M} shows the [0.5-2] keV core-excised ($0.15<r/r_{500}<1$) luminosity vs mass of our sample and of the SZ-selected sample in Pratt et al. (2022). Masses in the latter sample are based on $Y_X$, which shows strong covariance at fixed mass with the X-ray luminosity.
Due to the strong covariance, the comparison sample mostly scatters along the mean relation and this induces an underestimate of the (vertical) intrinsic scatter around it (Andreon et al. 2016, 2017; Pratt et al. 2022), by a factor presumed to be about 1.7 (Pratt et al. 2022).
This problem does not affect our sample because our masses are based on the shear.

All clusters but id5 have a core-excised luminosity close to, or brighter than, the average, for their mass (the average is indicated by the green solid line). Using a larger sample that includes our four clusters, Miyakasi et al. (2018) found that, as a whole, shear-selected clusters are instead X-ray faint. However, as already mentioned, Miyakasi et al. (2018) masses are overestimated because they are not accounting for the Eddington (1913) bias.

Id5 is 0.45 dex below the mean core-excised X-ray ray luminosity at its mass, $6\sigma_{\rm intr}$ below the mean relation or $3.6\sigma_{\rm intr}$ after correcting for the intrinsic scatter underestimation. The same is true using the Pratt et al. (2022) subsample of X-ray selected clusters. 
If we instead consider a sample for which the probability of inclusion in the sample is independent of the quantity being investigated at fixed mass, the X-ray Unbiased Cluster sample (XUCS hereafter, plotted in red in Fig.~\ref{fig:Lx_M} see Andreon et al. 2016, 2017a, 2017b, 2024), id5 is closer to the mean (0.2 dex away). Id5 is also close to the mean relation in a relative sense: it is $0.4\sigma_{\rm intr}$ away from the XUCS mean. Therefore, in terms of its core-excised X-ray luminosity for its mass, id5 is atypical when compared to (biased) SZ or X-ray -selected samples, but not atypical when compared to a sample X-ray unbiased for its mass.

The log central (within 300 kpc) brightness of our four clusters is above 44 erg s$^{-1}$ Mpc$^{-2}$, much higher than the threshold of the objects of low surface brightness (43.35 erg s$^{-1}$ Mpc$^{-2}$) defined in Andreon et al. (2024).  Therefore they cannot be considered part of the population of clusters that are usually missed in X-ray and SZ catalogs.

Figs.~\ref{fig:ne_prof_others} and \ref{fig:Pe_prof_others} compare the electron density and pressure radial profiles of our shear-selected sample 
to those of an SZ-selected sample
(Ghirardini et al. 2020). The latter is formed by clusters at $0<z<1.8$
selected by their SZ signal. 
The electron density profile of id5 is below the $-2\sigma$ boundary of the range probed by the SZ-selected sample. The 
pressure profiles of id5 and id34 are at, or below, the $-2\sigma$. Neither of the two is as extreme as the very depressed CL2015 cluster (Andreon et al. 2019), nevertheless they are examples of clusters uncommon ($2\sigma$) in SZ-selected sample. 
Finally, three clusters have a low central electron density (id5, id17 and id34) and two have a low central pressure value (id5 and id34) compared to the SZ-selected sample. 
Qualitatively similar results are obtained using pressure profiles from Sayers et al. (2022),
that studied an X-ray selected sample.

\subsection{SZ}
\label{SZ}

\subsubsection{NIKA2}

As detailed in Appendix~\ref{NIKA2}, we observed id34 and id48,the two most massive clusters at the time of the selection, with NIKA2 for about 14h each. We detected id34 and we obtained a pressure profile of lower signal-to-noise, but different systematics, compared to the one derived from the X-ray data.  In our 150 GHz map of id48, there is a ($\sim 3 \sigma$) blip at the location of it, consistent with its much lower mass compared to id34. 

\begin{figure}
\centerline{\includegraphics[width=9truecm]{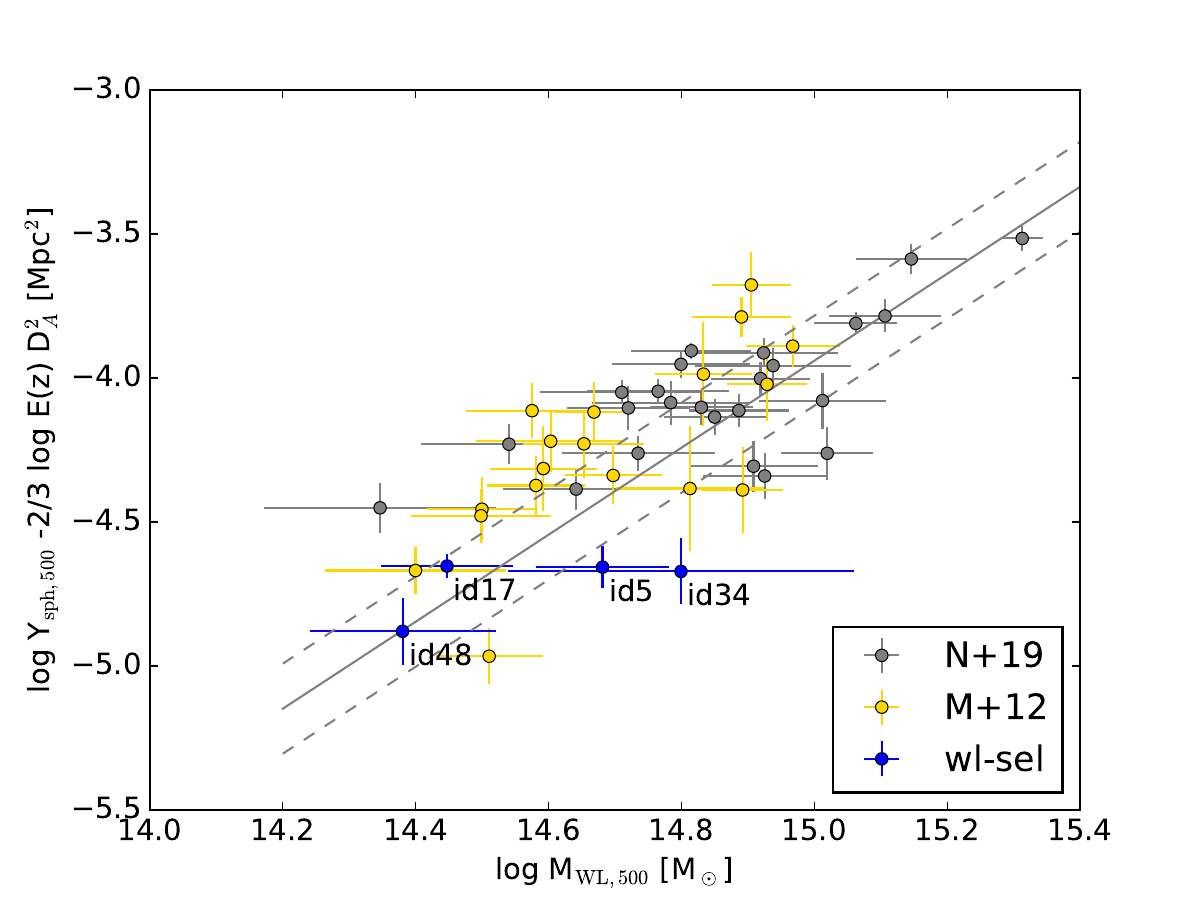}}
\caption[h]{$Y_{{\rm sph},500}-M$ plot 
for our shear-selected sample and for the biased sample formed by X-ray selected clusters at $z < 0.55$ (from Marrone
et al., 2012 and Nagarajan et al., 2019). The solid line indicates the fit to Nagarajan et al.(2019) data, that accounting for sample selection effects, while the dashed corridor is the mean model plus and minus the
intrinsic scatter (as derived by Nagarajan et al. 2019). 
Id5 and id34 clusters fall at about the location where the Nagarajan et al. (2019) analysis postulates
the existence of clusters (near and below the dashed line at $\log M/M_\odot<14.8$), although absent 
is in their sample.}
\label{fig:Y_M}
\end{figure}

\subsubsection{ACT}

Compton maps of one third of the sky have been released (Coulton et al. 2024) while we were writing this paper. All four shear-selected clusters stands out in those maps, including id5 that is instead missing in the list of detected clusters in earlier shallower maps (i.e., in Hilton et al. 2021). We measured the cylindrical Compton Y, $Y_{{\rm cyl},500}$, in aperture of $r_{500,{\rm wl}}$ radii. Errors are derived from the scatter of measurements with random centers spread around each cluster. The spherical Compton Y, $Y_{{\rm sph},500}$, is derived from $Y_{{\rm cyl},500}$ by dividing the latter by 1.203 as in Nagarajan et al. (2019). To check for systematics differences with this work, we also computed $Y_{{\rm sph},500}$ of the 19 Nagarajan et al. (2019) clusters in the footprint of the Coulton et al. (2024) map. By comparing our and their estimates,  we found that our estimates are biased high by 0.06 dex compared to theirs (with a scatter of 0.13, part of which due to errors). We therefore reduced our $Y_{{\rm sph},500}$ by that amount for the comparison (but values in Tab~\ref{tab1} are as measured).

Fig.~\ref{fig:Y_M} shows $Y_{{\rm sph},500}$ mass scaling for our sample and for two comparison samples with $z<0.55$. Both comparison samples are expected to be biased, because they are X-ray selected (i.e., include only X-ray bright clusters, $L_X\gtrsim 10^{45}$ erg s$^{-1}$ in the [0.1-2.4] band for the Nagarajan et al. 2019 sample) and Compton Y is covariant with the X-ray luminosity. Nagarajan et al. (2019) account for the biased sample selection, which is the reason why their mean model (solid line) tend to be at lower $Y_{{\rm sph},500}$ than the data at $\log M_{500}/M_\odot<14.9$.  Two of our shear-selected clusters (id5 and id34) are in empty parts of plot and in particular are 
0.5 dex fainter (i.e. a factor of 3) in $Y_{{\rm sph},500}$ than the typical cluster with $\log M =14.7$ present in the observational sample. 
These two clusters are in places where the authors assume the existence of clusters. Therefore, our sample confirms the population of objects postulated to exist in the Nagarajan et al. (2019) analysis but absent in both Marrone et al. (2012) and Nagarajan et al. (2019) observational samples. 
The low values of Compton parameters of id5 and id34 (plotted in Fig.~\ref{fig:Y_M}) are in line with the low pressure profiles for their mass derived in 
Sec.\ref{results}.

\begin{figure}
\centerline{\includegraphics[width=9truecm]{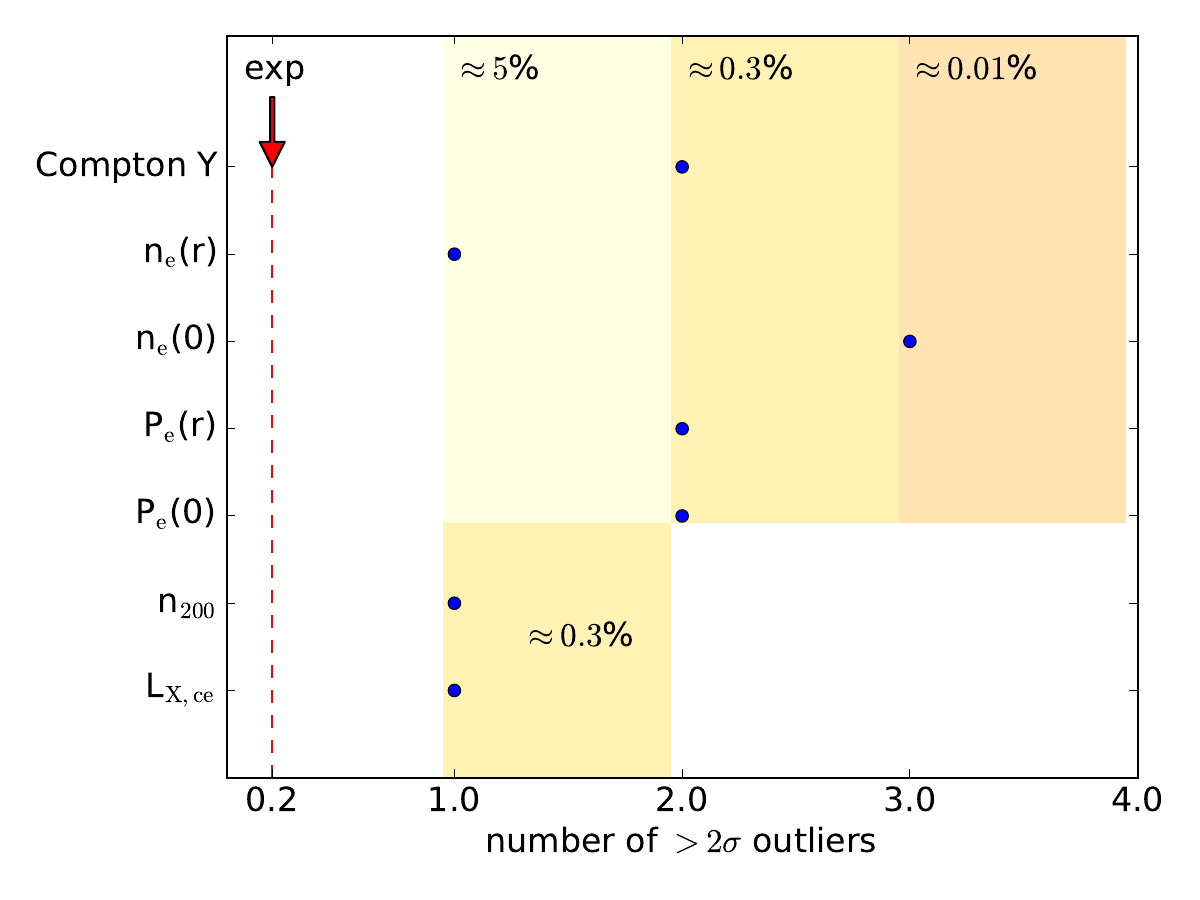}}
\caption[h]{Number of observed ($>2\sigma$) outliers for each of seven features studied. Associated probabilities (p-values) are indicated with numbers and shadings. The vertical red line indicates the expected number of outliers. We observe far too many outliers in our small sample of four objects for the expectation, based on ICM-selected samples, to be correct.}
\label{fig:summary}
\end{figure}

\section{Discussion}

\subsection{The atypical behaviour of the overall sample: a sign of a bias of the baryon selected samples?}

We studied only 7 (non-indepedent) features of a small sample of four objects selected differently than in almost all previous studies, as we selected clusters without relying on their baryons. Comparing these features of such a small sample with ICM-selected samples, we found (a summary can be found in Fig.~\ref{fig:summary}): a) two $>3\sigma$ features: one cluster, id5, has a low X-ray luminosity for its mass compared to X-ray or SZ-selected samples and a low richness compared to an X-ray selected sample; and b) 10 $2\sigma$, though already seen, features in ICM-selected samples, specifically two clusters with a low pressure profile (id5 and id34), one cluster with a low electron density profile (id5), three clusters with a low central electron density (id5, id17, and id34), two clusters with a low central pressure value (id5 and id34), and two clusters with a low Compton Y for their mass (id5 and id34). In a sample of 4 objects, we expect to see by chance 1.4 $2\sigma$ outliers every 7 independent features. We observed 12 outliers among our 7 non-independent features. Focusing on just one ICM-based feature to avoid dealing with the non-independence of the features, we observe on average 2 rare occurrences when the expected number is 0.2 (Fig.~\ref{fig:summary}). Such a large number of rare or just atypical features in a small sample of four objects is at least suspect and indicates a bias in our knowledge of the thermodynamic properties of clusters derived from ICM-selected samples. 

Some of the atypical behaviours
disappear when the bias induced by the ICM selection is corrected for: clusters with low Compton Y for their mass (id5 and id34) are actually assumed to exist in the $Y-M$ scaling Nagarajan et al. (2019), although objects of this type are absent in observational samples. These clusters will likely have low pressure profiles given that the Compton parameter is the integral of the pressure profile. Therefore, the low pressure profile of id5 and id34 compared to the ICM-selected sample in Ghirardini et al. (2019), in addition to the low Compton Y, could be explained by the bias of the ICM selection. The excessively low X-ray luminosity of one of the clusters, id5, when compared to X-ray or SZ-selected samples (Pratt et al. 2009, 2022) disappears when an X-ray unbiased sample is used for comparison. 
This seems to confirm that the bias in our knowledge of the thermodynamic properties of clusters derived from ICM-selected samples has not been completely corrected in all analyses (e.g., the X-ray vs mass scaling, Fig.~\ref{fig:Lx_M}).

It is worth noting that the two clusters with the most atypical behaviour, id5 and id34, are both dynamically complex with groups falling onto them and this could partially explain their properties. At the same time, we are confident that our analysis is robust to the derivation of their features because it is tailored to their complexity as detailed in Sect.\ref{fit}.

\begin{figure}
\centerline{\includegraphics[width=9truecm]{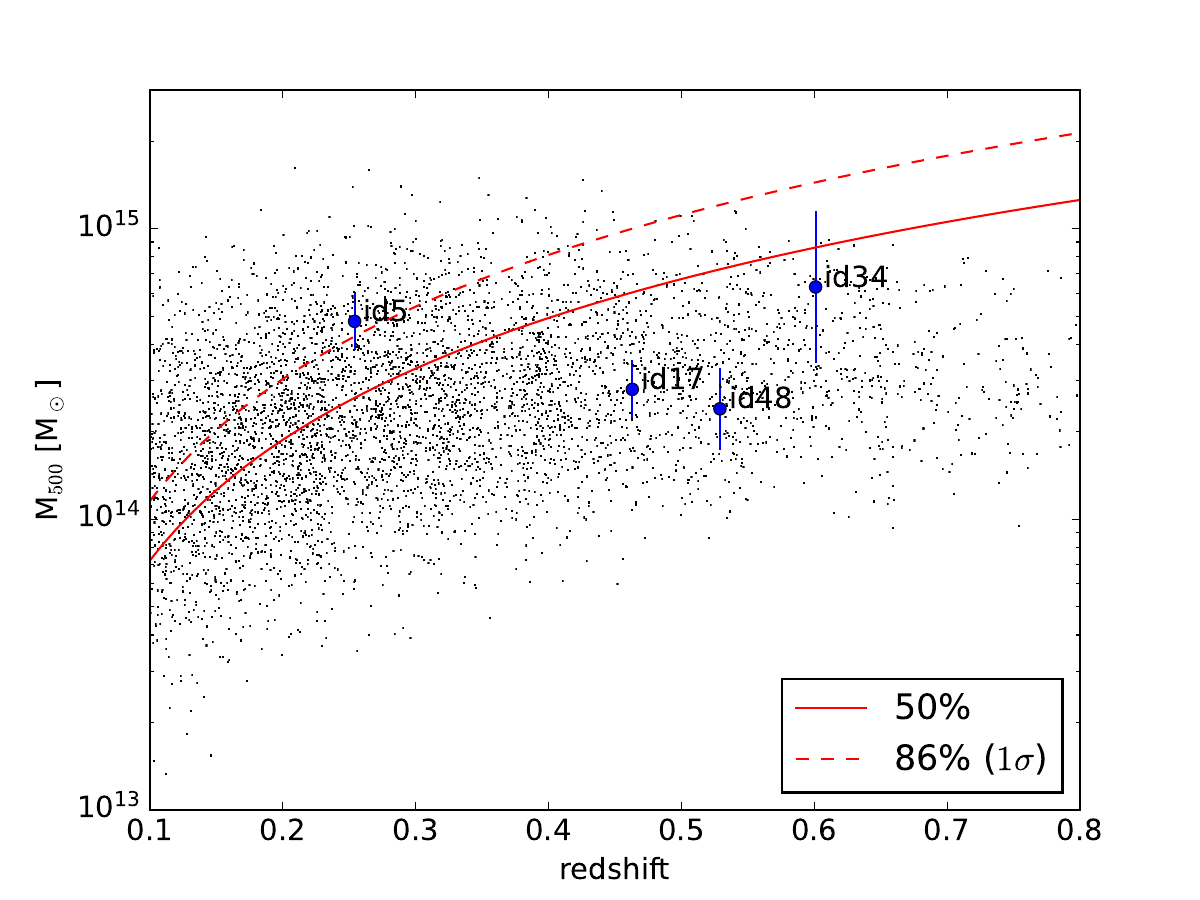}}
\caption[h]{Mass-redshift plot of the eROSITA cosmological sample (black dots) and of our weak-lensing selected sample (blue circles with error bars). eROSITA masses are, in practice, count-rates converted to mass assuming that clusters have exactly the average X-ray luminosity for their mass. The solid and dashed lines delimit the part of the plot having 
50\% and 86\% mass completeness. Only a minor part of the mass-redshift plane has a $>86$\% completeness and indeed id5 is currently undetected and absent from the eROSITA cosmological sample.}
\label{fig:Mz_erosita}
\end{figure}

\subsection{Id5}

Id5 is a very massive cluster ($\log M_{500}/M_\odot=14.7$) and one of the largest signal-to-noise shear-detected clusters in the HSC survey ($5^{\rm th}$ overall, $1^{\rm st}$ among massive clusters). According to 
Pillepich et al. (2012),
clusters ten times less massive than id5 should have 50 detected photons in the full eROSITA survey at the id5 redshift ($z=0.256$). In the eROSITA data release 1 catalog (Merloni et al. 2025, covering 1/8 of the full survey exposure time), there is no detection of extended X-ray emission at the cluster location. There is just one X-ray source within 1 arcmin of it with five net X-ray photons, in agreement with expectations based on the flux we measured from our much deeper X-ray data.  

Fig.~\ref{fig:Mz_erosita} shows the mass-redshift plot for the cosmological sample of the eROSITA data release 1 survey (Ghirardini et al. 2024). The redshift of id5 is well within the range considered in eROSITA cosmological analyses.
The lack of id5 in the eROSITA cosmological sample (and in the cluster catalog as well), despite its large mass, confirms that X-ray selected samples are not complete in mass, not even at low redshifts and for the most massive clusters. eROSITA masses are, in practice, count-rates converted to mass assuming that clusters have exactly the average X-ray luminosity for their mass (count rate at fixed mass and redshift, strictly speaking). The solid line indicates the detection threshold derived assuming
the first quartile count rate of detected clusters in the cosmological sample (barely 27 photons at the median exposure time of the survey). Because of the scatter between X-ray luminosity and mass (and the fact that below the median lies 50\% of the population) this line marks a $\sim50$\% completeness threshold. To have 86\% of the cluster population (down to $-1\sigma$), i.e. a 86\% complete sample, more massive clusters should be considered, as shown by the dashed line. Only a small fraction of the sample falls within the region of the mass-redshift plane with greater than 86\% completeness, aligning with Bulbul et al. (2024)'s more precise estimate of about 30\% when averaged over the entire cosmological sample.
The large 0.4 dex scatter in X-ray luminosity at fixed mass, as derived by Ghirardini et al. (2024) and used above,
is in line with the 0.5 dex reported by Andreon \& Moretti (2011) and Andreon et al. (2016). The scatter is significantly larger than those reported in analyses of SZ or X-ray selected samples that claim to account for sample selection effects (e.g., 0.06-0.17 dex in Pratt et al. 2009, 0.06-0.11 dex in Pratt et al. 2022, 0.01-0.03 dex in Lovisari et al. 2020, etc.). 
The substantial scatter in X-ray luminosity at fixed mass explains the non-detection of massive clusters in X-ray selected samples such as id5: clusters on the dashed line and $1\sigma$ faint would have 27 photons at the median depth of the survey, and therefore easily absent in the cosmological sample. Our other three clusters have comparable or lower masses (see, e.g., Fig~\ref{fig:Mz_erosita}) but are much brighter (Fig.~\ref{fig:Lx_M}). These clusters are  indeed in the current eROSITA cosmological sample, illustrating the expectation that at every location of the plot most of the points are indeed clusters with larger-than-average X-ray luminosity for their mass because of their easiness to be part of a sample and their large volume visibility.

Given that id5 brightness is too large to be a low surface brightness cluster, and that there are at most 5 photons in the eROSITA data release 1 in the id5 area, clusters of low surface brightness will go undetected, even if they are as massive and at as low a redshift ($z=0.25$) as id5.

Quite surprisingly, id5 also has a low richness for its mass. Its low richness does not preclude detection, and indeed the cluster was detected by Gal et al. (2009) using single-band 50-year-old photographic plates, by Wen et al. (2012) and Rozo et al. (2015) using shallow SDSS photometry, by Radovich et al. (2017) using KiDS photometry, and by Wen and Han (2015) using HSC photometry. The low richness hides this massive cluster among the plethora of low richness clusters, most of which have a low mass. The same situation will likely occur in X-ray surveys when they achieve the depth necessary to detect clusters with id5's luminosity.

Kinematically, id5 shows strong activity with two groups infalling on it, one of which is also obvious in X-ray, whereas the other is projected on the center, and therefore not distinguishable with our data.

The astrophysical reasons why id5 has a very low richness and electron density and pressure at the bottom of the range are worth investigating. Id5 (and id34 as well) may have a low gas fraction resulting from a very active AGN at the time of cluster formation (Ragagnin, Andreon \& Puddu 2022) or
gas properties influenced by the kinematic activity occurring, possibly temporarily pushing the gas to radii larger than $r_{500}$. Its low richness has no plausible explanation: reducing the richness inside $r_{200}$ would require the pre-existence of 98  massive quiescent galaxies that have merged into the 38 currently present (if they were star-forming, the merging would lead to a scattered red sequence), an extreme hypothesis never heard before. Moving those 60 additional galaxies outside $r_{200}$ is unfeasible given that they are collisionless.

The id5 Compton signal is atypical compared to studied observational samples, but is within the range where we expect clusters of this mass to be (Fig.\ref{fig:Y_M}). Id5 is not detected in early ACT maps (Hilton et al. 2021), in line with the depth of the map they used, but it will certainly be in catalogs based on the deeper data that we used (Sec.\ref{SZ}) given that the object stands out.

\section{Conclusions}

Previous studies of the thermodynamic properties of the ICM almost always focused on samples selected via the ICM. First, this is a  minor component in clusters, and second, selection via a quantity that is being investigated is obviously prone to introduce selection effects. Weak gravitational lensing allows us to select clusters independently of the baryon content. For this paper, we selected four clusters drawn from the weak-lensing survey of Miyazaki et al. (2018) and followed them up in X-ray and SZ with the aim of performing a first study of the thermodynamic radial profiles
in a sample free of baryon selection bias.

We derived core-excised X-ray luminosities, richness-based masses, Compton parameters, and profiles of mass, pressure, and electron density of the four clusters. These quantities are derived from shear data, the ACT Compton map, and our own X-ray and SZ follow-ups.

One of the clusters, id5, also has abundant DESI spectroscopy. The analysis of its caustic diagram indicates a lot of cluster substructures with at least two groups falling on it at $\Delta v \sim 1000$ km/s: one very close in projection to the cluster center and one 1.5 Mpc from it. The second group is in the cluster outskirts, is clearly visible as an obvious extended X-ray emission, and is associated with a galaxy group also visible in the optical. The cluster has been deeply scrutinized, and the data exclude the possibility of another structure along the line of sight boosting the shear signal, or that we might have missed galaxies due to corrupted photometry or their blue color. 
Our caustic analysis confirms the shear-based mass. In X-ray, id5 is unremarkable, showing no signs of activity. A second shear-selected cluster, id34, is a bimodal cluster, with a second group at 1 Mpc from the center, evident both in optical and X-rays. 

Because both id5 and id34 are out of equilibrium, these two clusters cannot be analyzed assuming hydrostatic equilibrium. We therefore performed disjoint analyses of the shear (from HSC) and X-ray data, and joint analyses for the other two clusters. Our analysis accounts for the Eddington bias 
using the cluster mass function. This correction turned out to be unnecessary for id5, given its strong shear signal, and for two more clusters because our X-ray data alone are very informative about the cluster mass. The Eddington bias-corrected masses agree well with the richness-based masses for three out of four clusters. id5 shear mass is instead much larger than inferred from its low richness and agrees with the caustic mass. After the inclusion of the Eddington bias, the four clusters originally selected to have $M_{500}\gtrsim 5 \times 10^{14}$ M$_\odot$ have $M_{500}\lesssim 5 \times 10^{14}$ M$_\odot$.

Comparing 7 non-independent properties derived for such a small sample with those of ICM-selected samples, we found: a) two rare features: namely one cluster, id5, has a low X-ray luminosity for its mass compared to X-ray or SZ-selected samples and a low richness compared to an X-ray selected sample; and b) 10 uncommon, although already seen, features in ICM-selected samples, namely two clusters with a low-pressure profile (id5 and id34), one cluster with a low electron density profile (id5), three clusters with a low central electron density (id5, id17, and id34), two clusters with a low central pressure value (id5 and id34), and two clusters with a low Compton Y for their mass (id5 and id34). Such a large amount of rare or just atypical features in a small sample of four objects is anomalous and it indicates a bias in our knowledge of the thermodynamic properties of clusters derived from ICM-selected samples.

Some of the atypical behaviours disappear when the bias induced by the ICM selection of the compared sample is corrected for. Their disappearance seems to confirm that the bias in our knowledge of the thermodynamic properties of clusters derived from ICM-selected samples has not been completely corrected in all analyses. Given the atypical, or rare, properties seen in such a small sample of four clusters, it would be useful to extend this analysis to a larger sample of baryon-free selected clusters to quantify the frequency of objects that are unusual in ICM-selected samples.

\section*{Acknowledgements}

We acknowledge INAF grant  
``Characterizing the newly  discovered clusters of low surface  brightness" and PRIN-MIUR grant
20228B938N ``Mass and selection biases of galaxy clusters: a multi-probe approach".
This work has been partially supported by the ASI-INAF program I/004/11/4.  
Based on observations carried out under projects number 222-19, 075-20, and 182-20 with the
IRAM 30-meter telescope. IRAM is supported by INSU/CNRS (France), MPG (Germany) and IGN (Spain).

\section*{Data Availability}
X-ray data used are publicly available at HEASARC\footnote{https://heasarc.gsfc.nasa.gov/docs/archive.html},
weak-lensing data and HSC photometry are publicly available at the HSC archive, https://hsc.mtk.nao.ac.jp/ssp/data-release/, SDSS photometry is available
at the SDSS archive, https://www.sdss.org/, DESI spectroscopic catalog is available at 
https://data.desi.lbl.gov/doc/releases/edr/vac/zcat/ .

{}

\appendix

\section{Optical images of the clusters and color-magnitude diagrams of id5}

\begin{figure*}
\centerline{\includegraphics[width=8truecm]{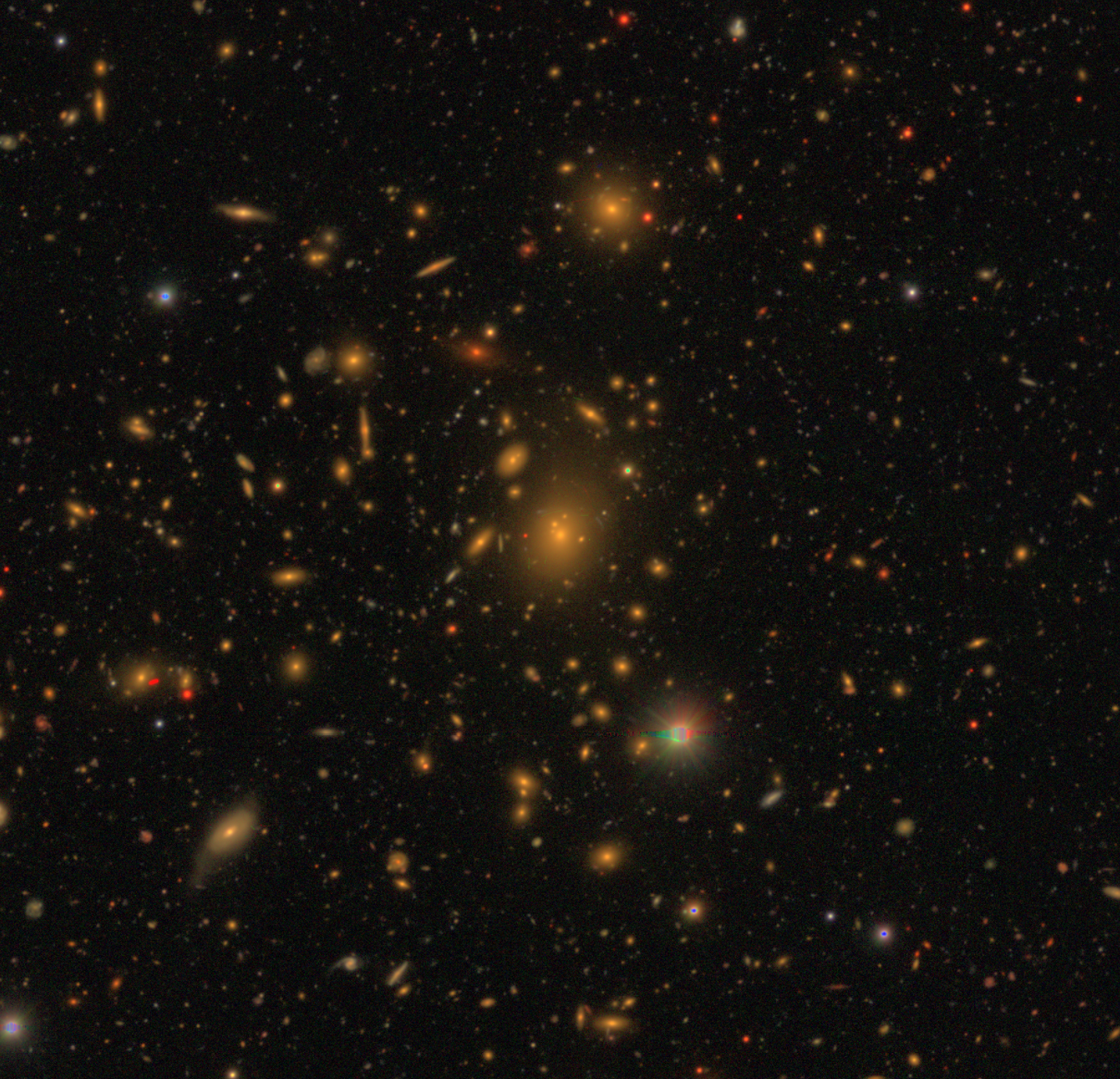}\,\includegraphics[width=8truecm]{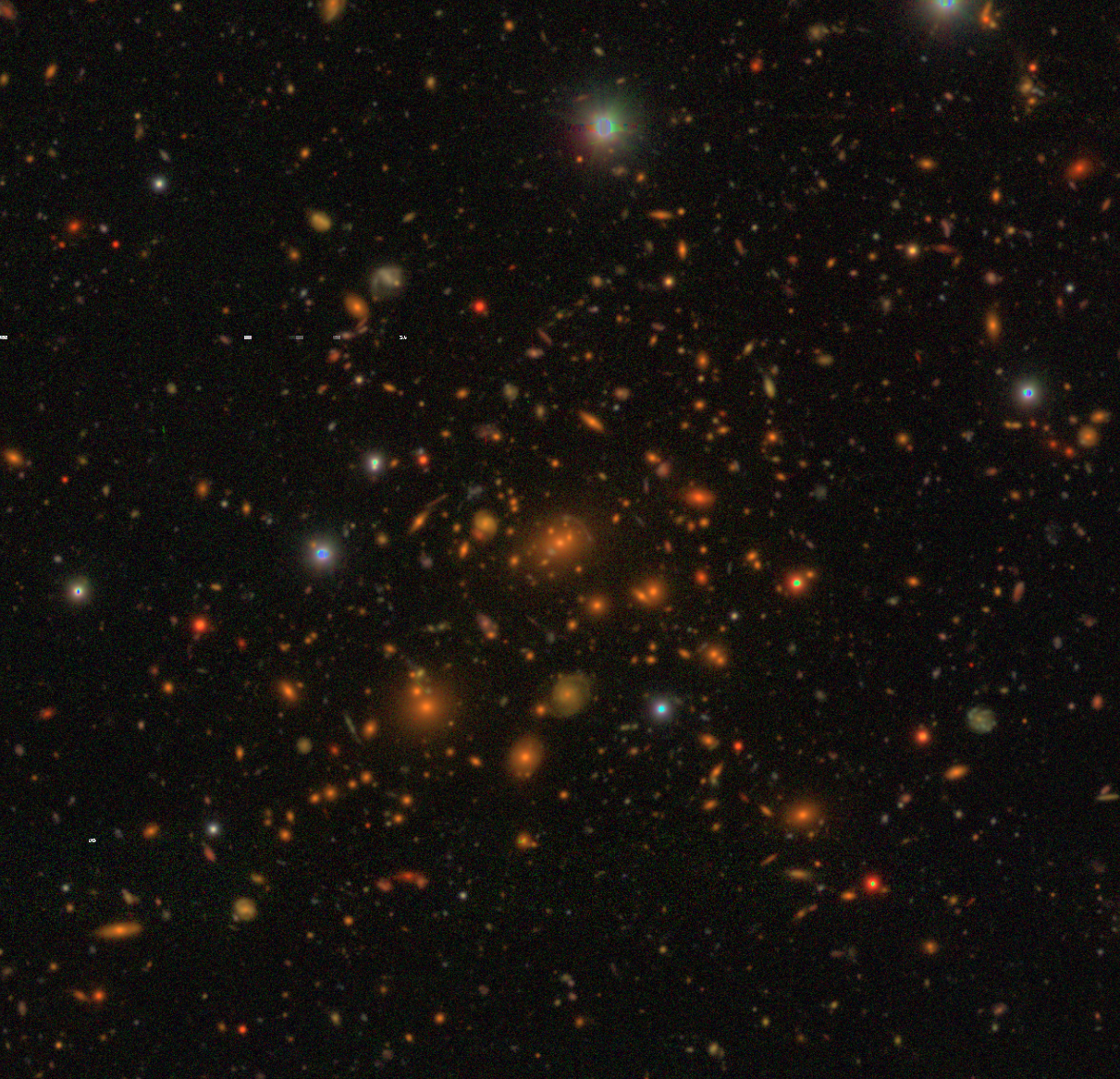}}
\centerline{\includegraphics[width=8truecm]{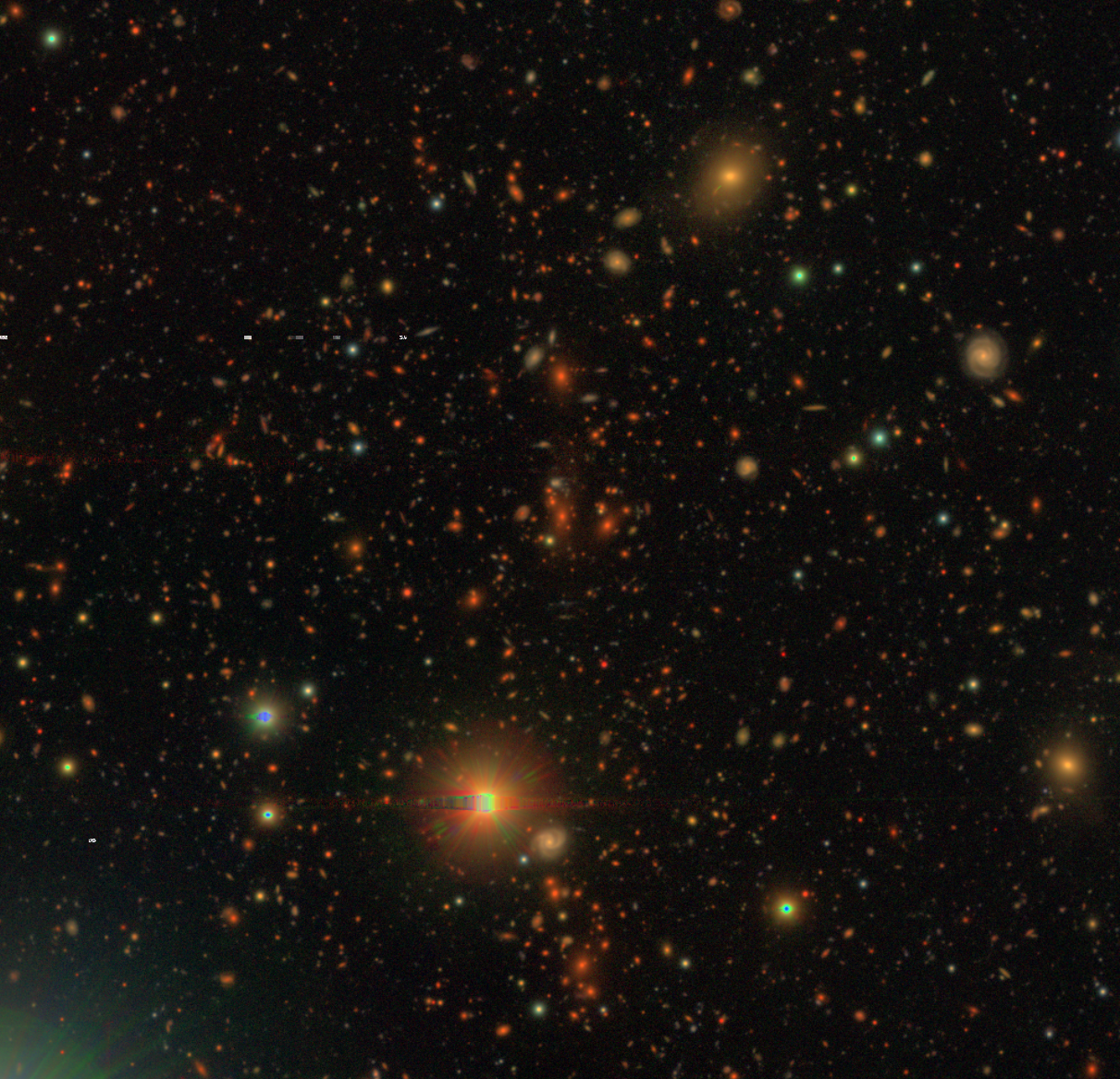}\,\includegraphics[width=8truecm]{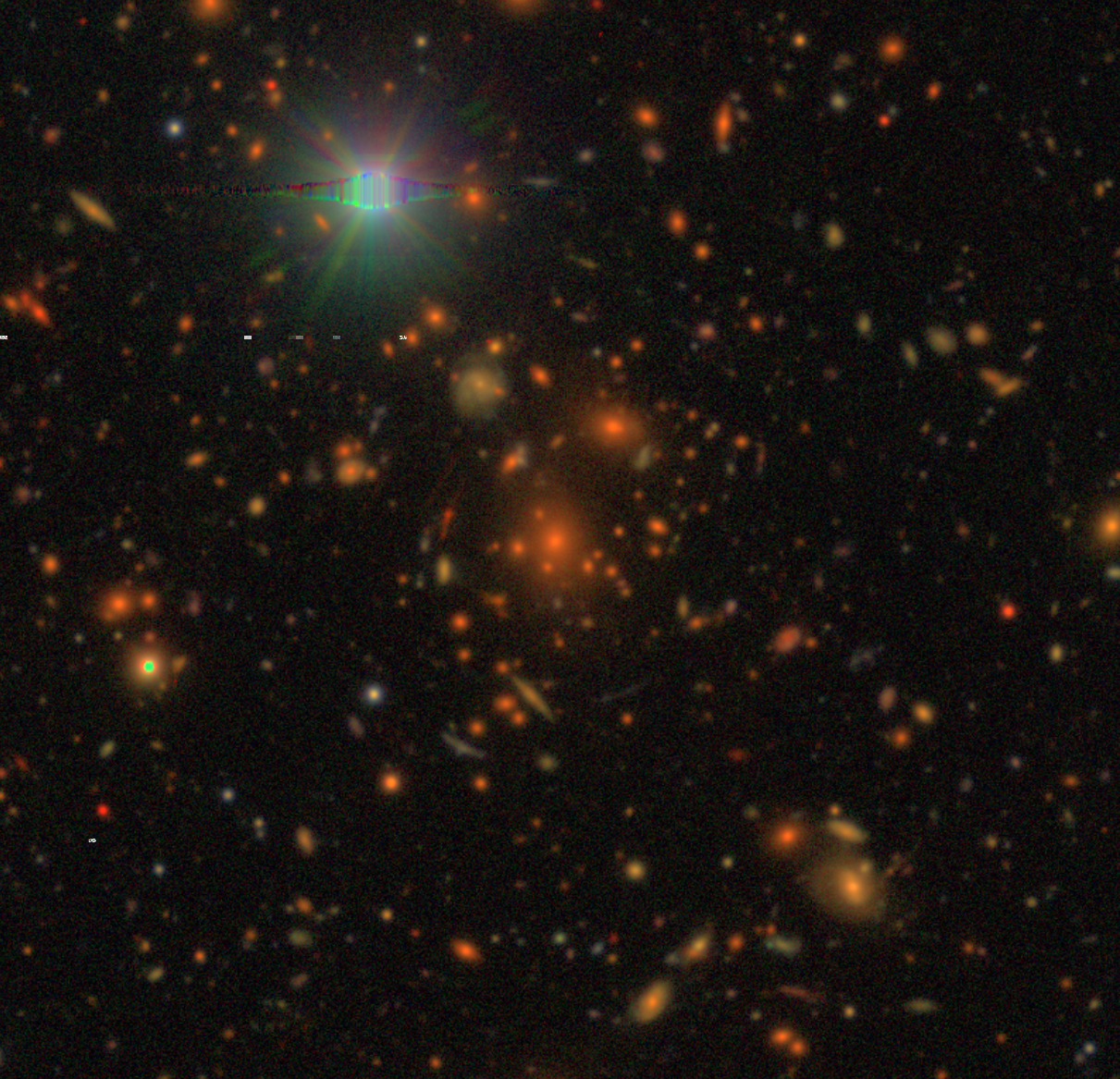}}
\caption[h]{True color images of id5 (top-left), id17 (top-right), id34 (bottom-left), and id48 (bottom-right). North is at the top, East to the left. The fields of views shown are square with sides of 5, 3.5, 5.5, and 2.2 arcmin, respectively. Id17 and id48 show gravitational arcs, id34 is clearly bimodal (the second peak is well visible to the South). Id5 is dynamically complex with two groups infalling on it (one is formed by many of the galaxies in the Eastern half of the image; the second group is outside the field of view). 
}
\label{fig:optima}
\end{figure*}

\begin{figure}
\centerline{\includegraphics[width=9truecm]{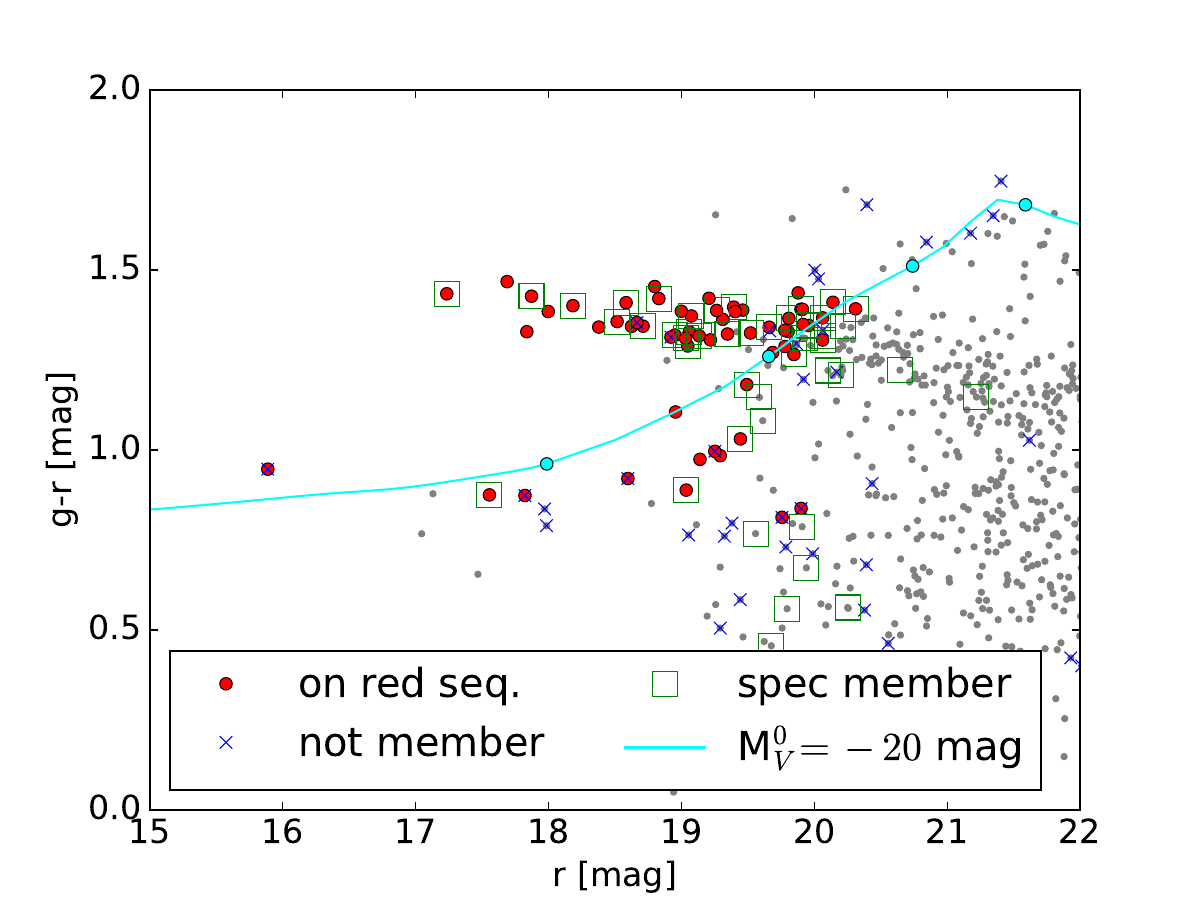}} 
\centerline{\includegraphics[width=9truecm]{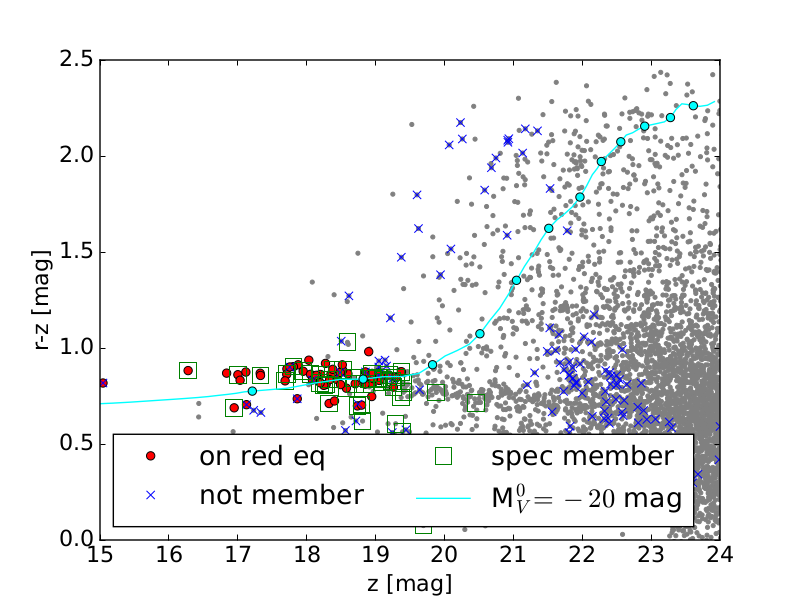}} 
\caption[h]{Color-magnitude plots of all galaxies within $r_{200,\rm wl}$ of the id5 center.
The two plotted color indexes are chosen to illustrate two different redshift regimes ($z<0.4$ in the top panel,
$0.3<z<1.2$ in the bottom panel), whereas the actual analysis uses the rest-frame $g-r$ at every redshift.
The solid line indicates the intercept of the color-magnitude relation at different redshifts (values multiple of 0.1 are marked with circles starting from $z=0.1$). 
Basically, massive galaxies are in the top left part of the diagram delimited by the cyan curve and 
massive clusters  
can be identified as a red sequence extending to the left of the curve.
Only one red sequence is present, at the color appropriate for the id5 redshift. 
The top panel also qualitatively 
illustrates the lack of a population of massive blue members 
(there are only few non-red points below the red sequence and at the left or near the cyan curve). Both panels show the lack of
other (contaminating) massive clusters along the line of sight (more than 60 additional massive galaxies are needed to bring id5 mass to $\log M_{200}/M_\odot = 14.7$, see Sec.~\ref{results}).
}
\label{fig:id5CM}
\end{figure}

\section{Shear technical details}
\label{appendix_shear}

As in \citet{Mandelbaum2018}, we define the calibration bias $\hat{m}$ and the responsivity factor $\mathcal{R}$ as:
\begin{equation}
\begin{aligned}
\hat{m}&=\frac{\sum_s {w_{ls} m_s}}{\sum_s{w_{ls}}} \\
\mathcal{R} &= 1-\frac{\sum_s{w_{ls}e^2_{rms,s} }}{\sum_s{w_{ls}}},
\end{aligned}
\end{equation}
where $\mathcal{R}\sim0.84$, $m_s$ is the catalog estimate of the calibration bias for each source, and the weights $w_{ls}$ are defined as
\begin{equation}
    w_{ls}=(<\Sigma^{-1}_{\rm cr,ls}>)^2\frac{1}{\sigma_{e,s}^2+e_{rms,s}^2},
\end{equation}
where: $\sigma_{e,s}$ and $e_{rms,s}$ are the catalog  shape measurement uncertainty and rms ellipticity respectively and 
\begin{equation}
  <\Sigma^{-1}_{\rm cr,ls}>= 
  \frac{\int_0^{\infty} P_s(z) dz}{\int_0^{\infty} P_s(z) \Sigma_{\rm cr}(z_l,z) dz}
\end{equation}.

As in \citet{Umetsu2020}, in each radial bin $R_i$, we computed the tangential shear as:
\begin{equation}
    \Delta\Sigma_+(R_i) = \frac{1}{2\mathcal{R}} \frac{\sum_s{w_{ls}\ e_{+}\ [<\Sigma^{-1}_{\rm cr,ls}>]^{-1} } }{[1+\hat{m}(R_i)] \sum_s{w_{ls}}},
\end{equation}
where $e_{+}$ is the tangential ellipticity $e_+=-\cos(2\phi) e_1 - \sin(2\phi) e_2$ where $e_1$, $e_2$, $\phi$ are the measured ellipticities components and the angle between the source and lens position,  in sky coordinates. 

In our tests of the id5 tangential shear profile we used KiDS ellipticities. For them, $m=0$ \citep[$|m|<10^{-2}$,][]{Giblin2021} and $2\mathcal{R}=1$ are appropriate and used. About 88\% of KiDS shape catalog have matches in the HSC-selected sample within 0.5".

\section{X-ray images of the clusters}

\begin{figure*}
\centerline{\includegraphics[width=6truecm]{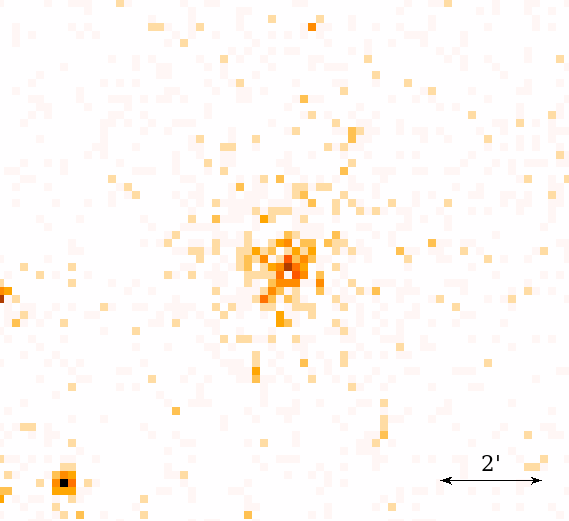} 
\includegraphics[width=6truecm]{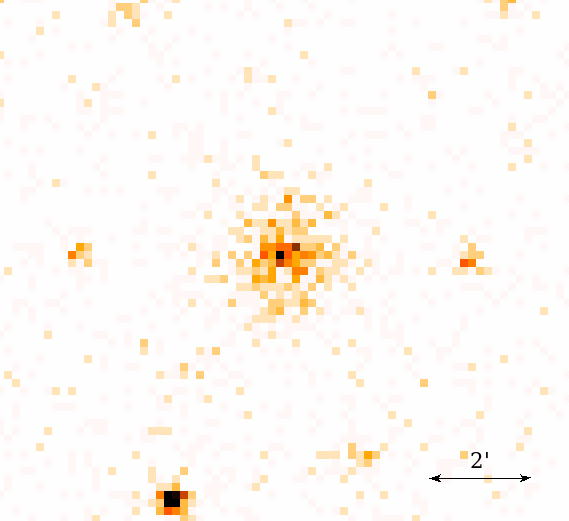}}
\centerline{\includegraphics[width=6truecm]{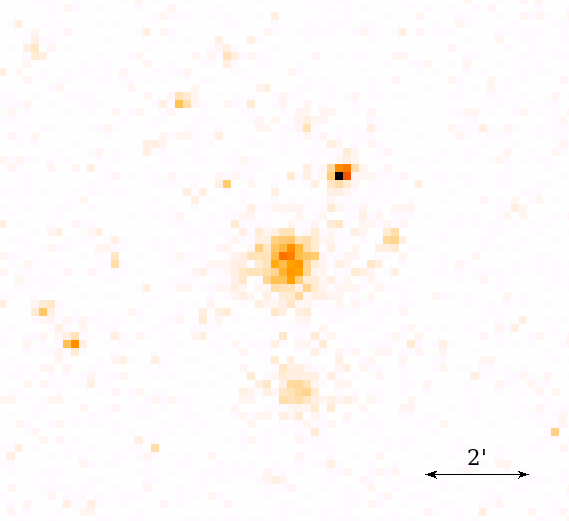} 
\includegraphics[width=6truecm]{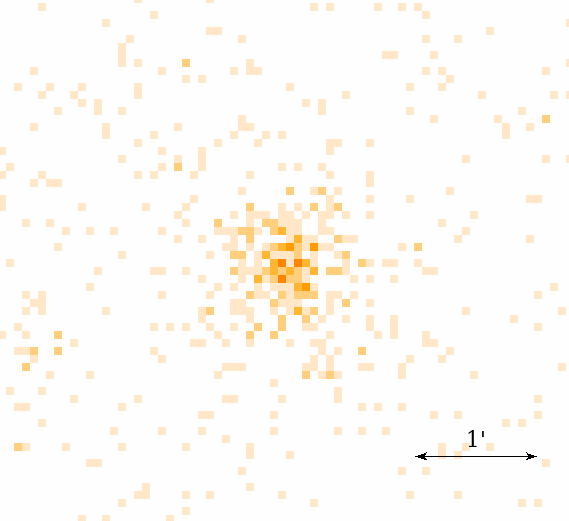}}
\caption[h]{[0.5-2] keV images of id5 (upper-left), id17 (upper-right), id34 (bottom-left) from Swift, and id48 (bottom-right) from Chandra. Id34 is a bimodal cluster (the second component is visible about 2 arcmin South).}
\label{fig:image}
\end{figure*}

\section{X-ray radial profiles of the other clusters}
\label{other_profs}

\begin{figure*}
\centerline{\includegraphics[trim={5 203 30 30}, clip,height=3.5truecm]{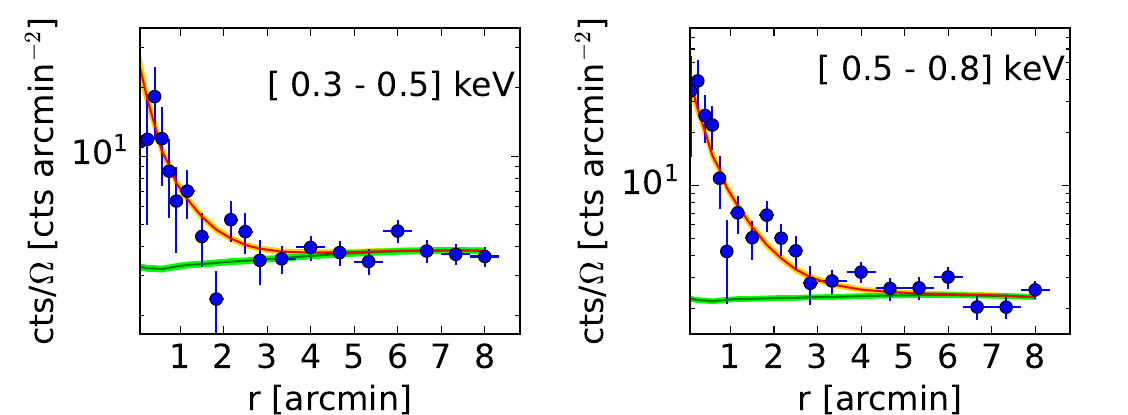}%
\includegraphics[trim={5 5 310 230}, clip,height=3.5truecm]{fitwithmod_1_nHEmin_id5.pdf}}
\centerline{\includegraphics[trim={255 5 40 230}, clip,height=3.5truecm]{fitwithmod_1_nHEmin_id5.pdf}%
\includegraphics[trim={5 203 30 30}, clip,height=3.5truecm]{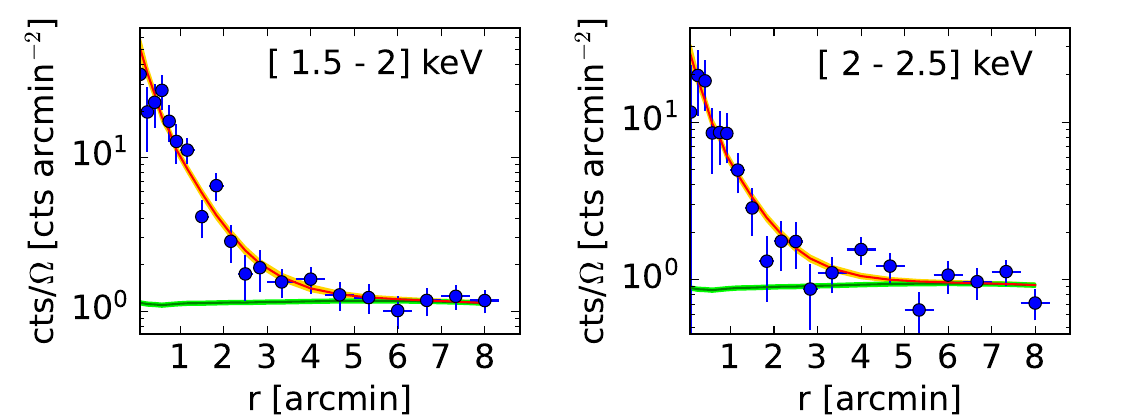}}
\centerline{\includegraphics[trim={5 5 30 230}, clip,height=3.5truecm]{fitwithmod_2_nHEmin_id5.pdf}%
\includegraphics[trim={5 203 310 30}, clip,height=3.5truecm]{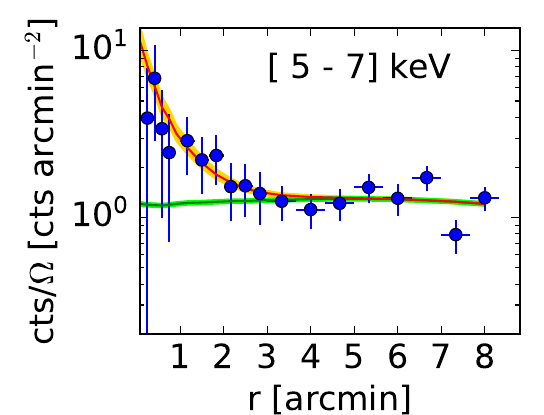}}
\caption[h]{Id5 surface brightness profiles (points with error bars) in the X-ray bands with 68\% uncertainties
on the fitted model (red line and yellow shading). 
The green line with lime shading (barely visible) is the background radial
profile and its 68\% uncertainty. The analysis
uses the Poisson likelihood, not the plotted $\sqrt{n}$ 
errorbars. The assumed X-ray cluster model (red line)
represents well
the observed X-ray profiles at all energies. 
}
\label{fig:x-ray_id5}
\end{figure*}

\begin{figure*}
\centerline{%
\includegraphics[trim={5 203 30 30}, clip,height=3.5truecm]{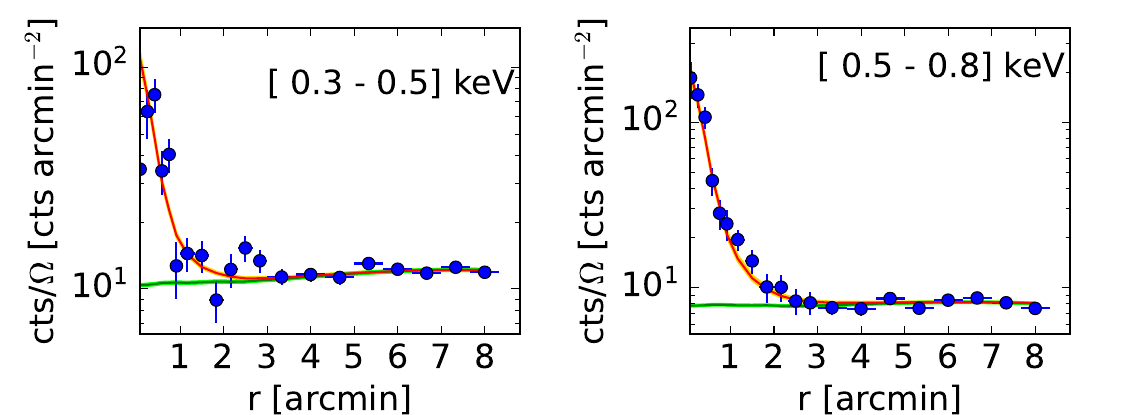}%
\includegraphics[trim={5 5 310 230}, clip,height=3.5truecm]{fitwithmod_1_nHEmin_id34.pdf}}
\centerline{\includegraphics[trim={255 5 40 230}, clip,height=3.5truecm]{fitwithmod_1_nHEmin_id34.pdf}%
\includegraphics[trim={5 203 30 30}, clip,height=3.5truecm]{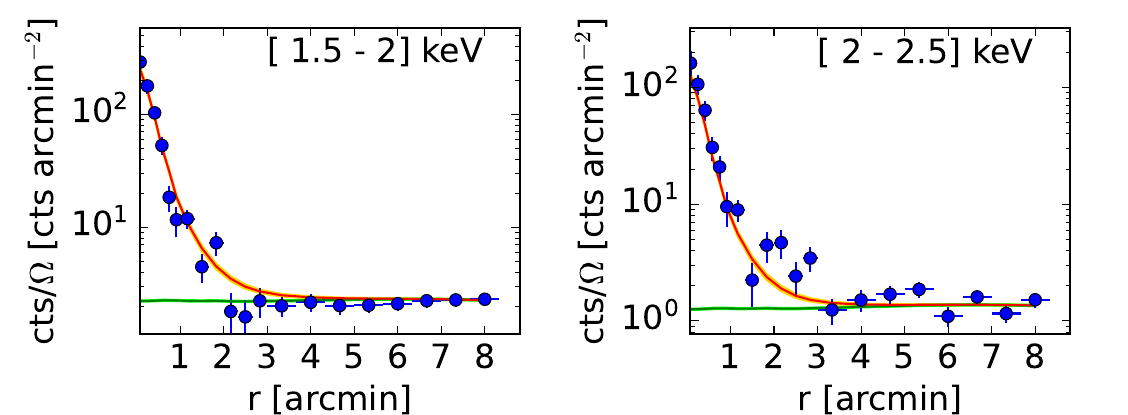}}
\centerline{\includegraphics[trim={5 5 30 230}, clip,height=3.5truecm]{fitwithmod_2_nHEmin_id34.pdf}%
\includegraphics[trim={5 203 310 30}, clip,height=3.5truecm]{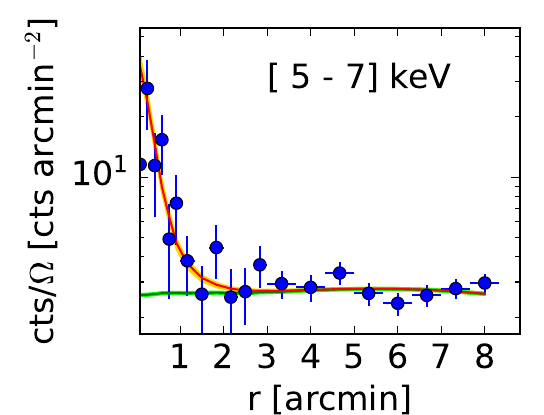}}
\caption[h]{Id34 surface brightness profiles (points with error bars) in the X-ray bands with 68\% uncertainties
on the fitted model (red line and yellow shading). 
The green line with lime shading (barely visible) is the background radial
profile and its 68\% uncertainty. The analysis
uses the Poisson likelihood, not the plotted  $\sqrt{n}$
errorbars. The assumed X-ray cluster model (red line)
represents well
the observed X-ray profiles at all energies. 
}
\label{fig:x-ray_id34}
\end{figure*}

\begin{figure*}
\centerline{
\includegraphics[trim={5 203 30 30}, clip,height=3.5truecm]{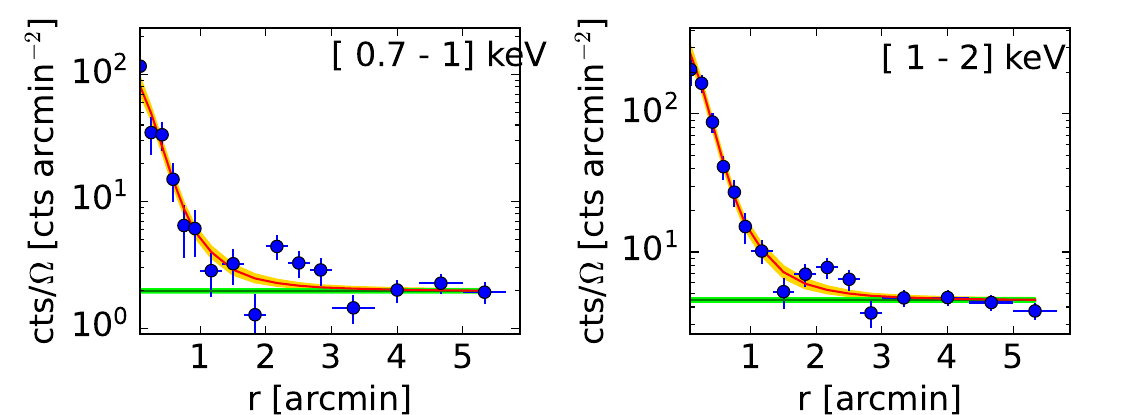}%
\includegraphics[trim={5 5 300 230}, clip,height=3.5truecm]{fitwithmod_1_wlHE_id48.pdf}}
\centerline{\includegraphics[trim={270 5 35 230}, clip,height=3.5truecm]{fitwithmod_1_wlHE_id48.pdf}%
\includegraphics[trim={5 203 30 30}, clip,height=3.5truecm]{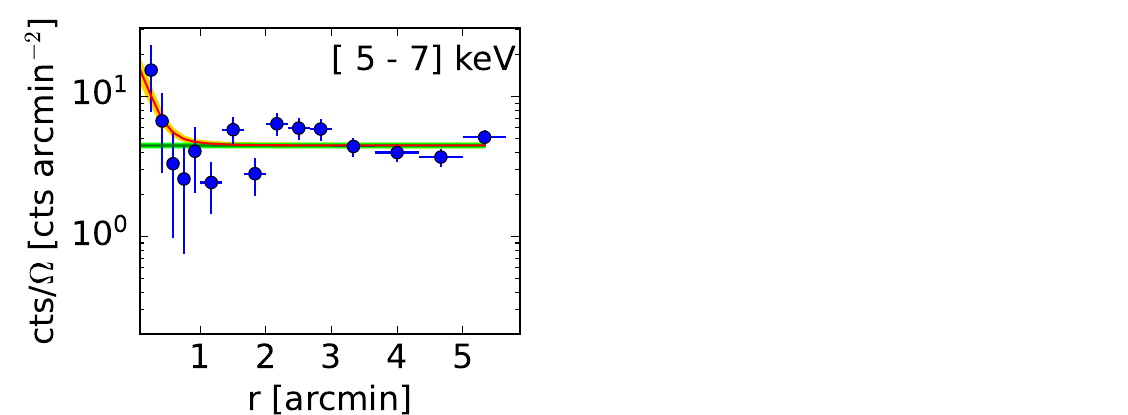}}
\caption[h]{Id48 surface brightness profiles (points with error bars) in the X-ray bands with 68\% uncertainties
on the fitted model (red line and yellow shading). 
The green line with lime shading (barely visible) is the background radial
profile and its 68\% uncertainty. The analysis
uses the Poisson likelihood, not the plotted $\sqrt{n}$
errorbars. The assumed X-ray cluster model (red line)
represents well the observed X-ray profiles at all energies. 
}
\label{fig:x-ray_id48}
\end{figure*}

\section{NIKA2 observations}
\label{NIKA2}

\begin{figure}
\centerline{\includegraphics[width=9truecm]{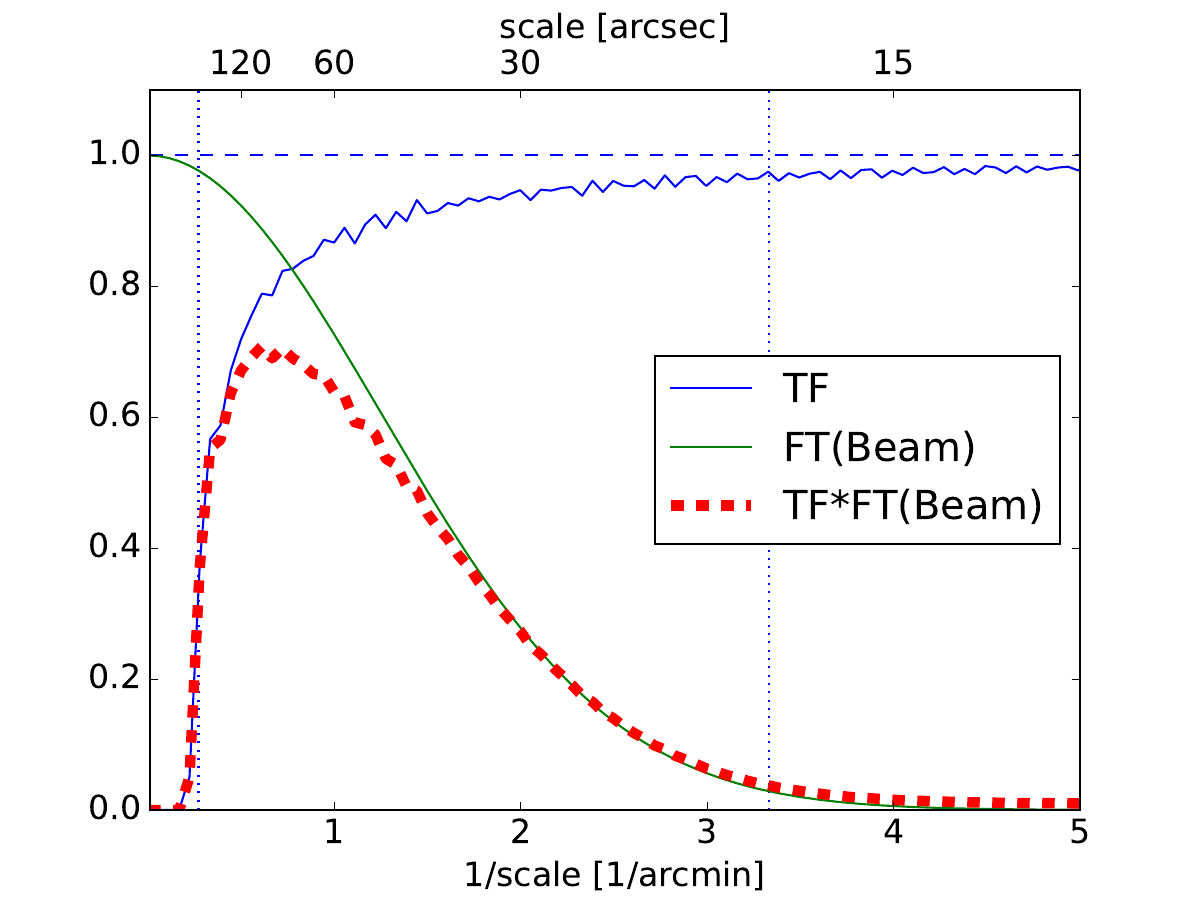}}
\caption[h]{Transfer function of the SZ reduction of the 150 GHz data of id34. The Fourier transform of the beam and the combined beam plus transfer function filtering, i.e., the effective filtering of the data, are also shown. Small scales (on the right) are washed out by the beam. Scales larger the maximal radius enclosed in the NIKA2 camera, 3.7 arcmin (the left dashed vertical line), are lost by the data reduction. }
\label{fig:TF}
\end{figure}

\begin{figure*}
\centerline{\includegraphics[width=8truecm]{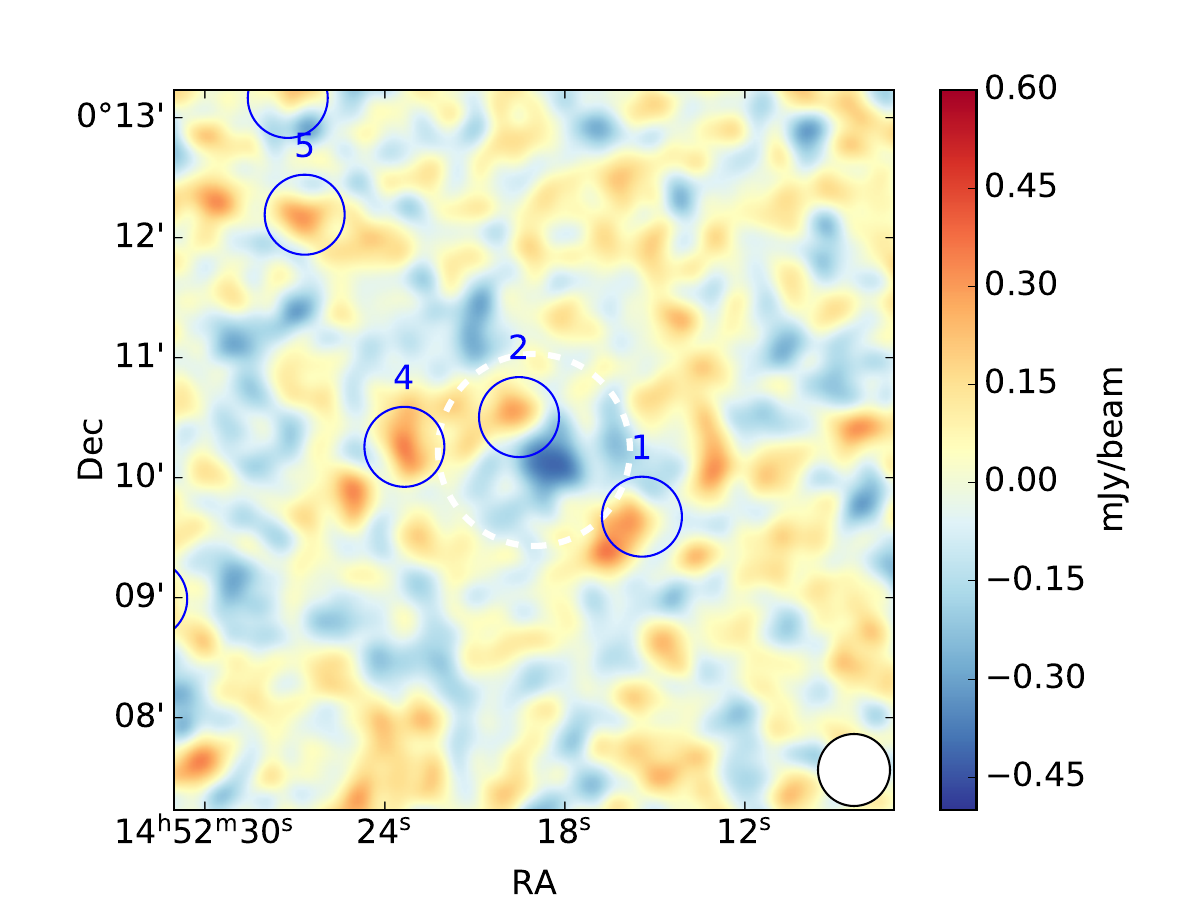}
\includegraphics[width=8truecm]{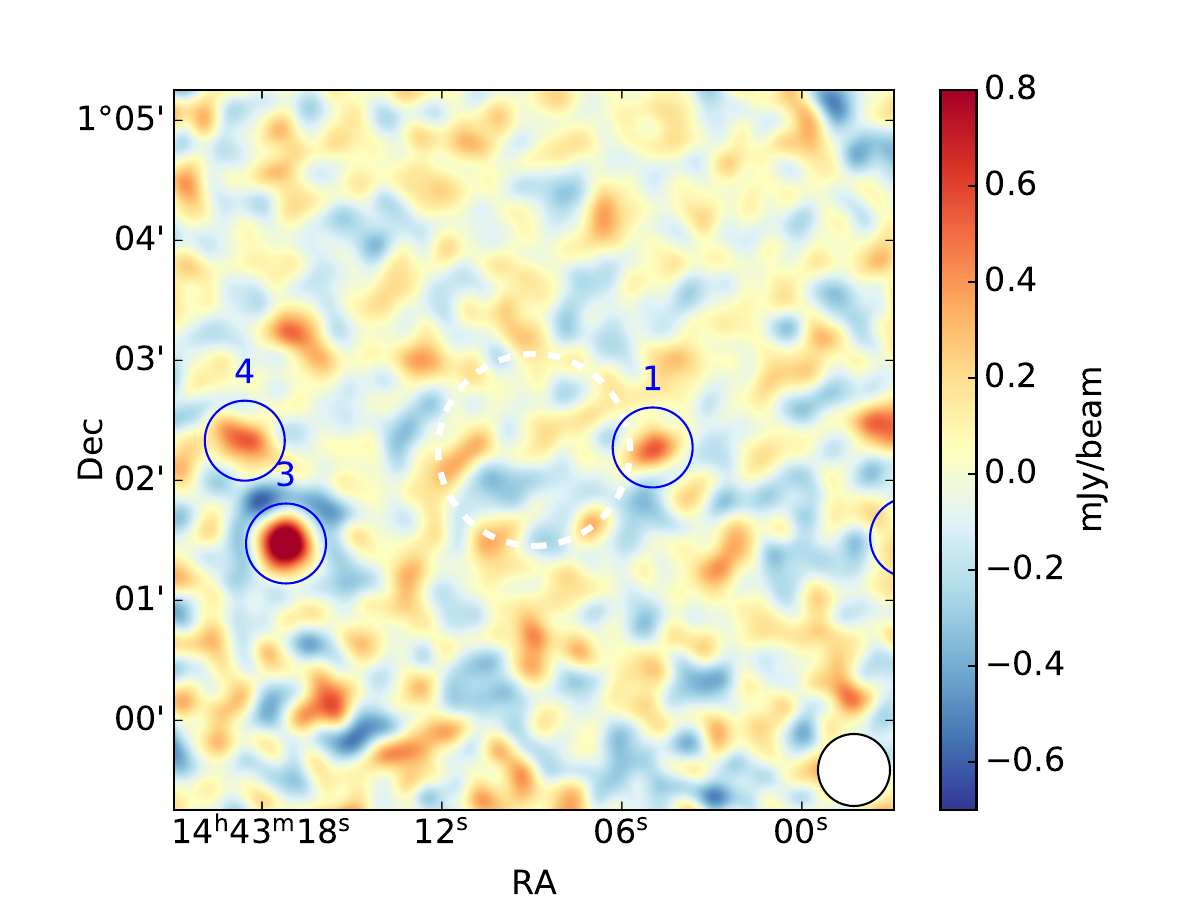}}
\caption[h]{NIKA2 images of id34 (left panel, SZ reduction) and of id48 (right panel, point source reduction) at 150 GHz with marked point sources with $S/N>4$ at 150 or 240 GHz. The cluster id34 is detected at 150 GHz at $4 \sigma$. The cluster id48 is undetected at 150 GHz.}
\label{fig:SZima}
\end{figure*}

\begin{figure}
\centerline{\includegraphics[width=9truecm]{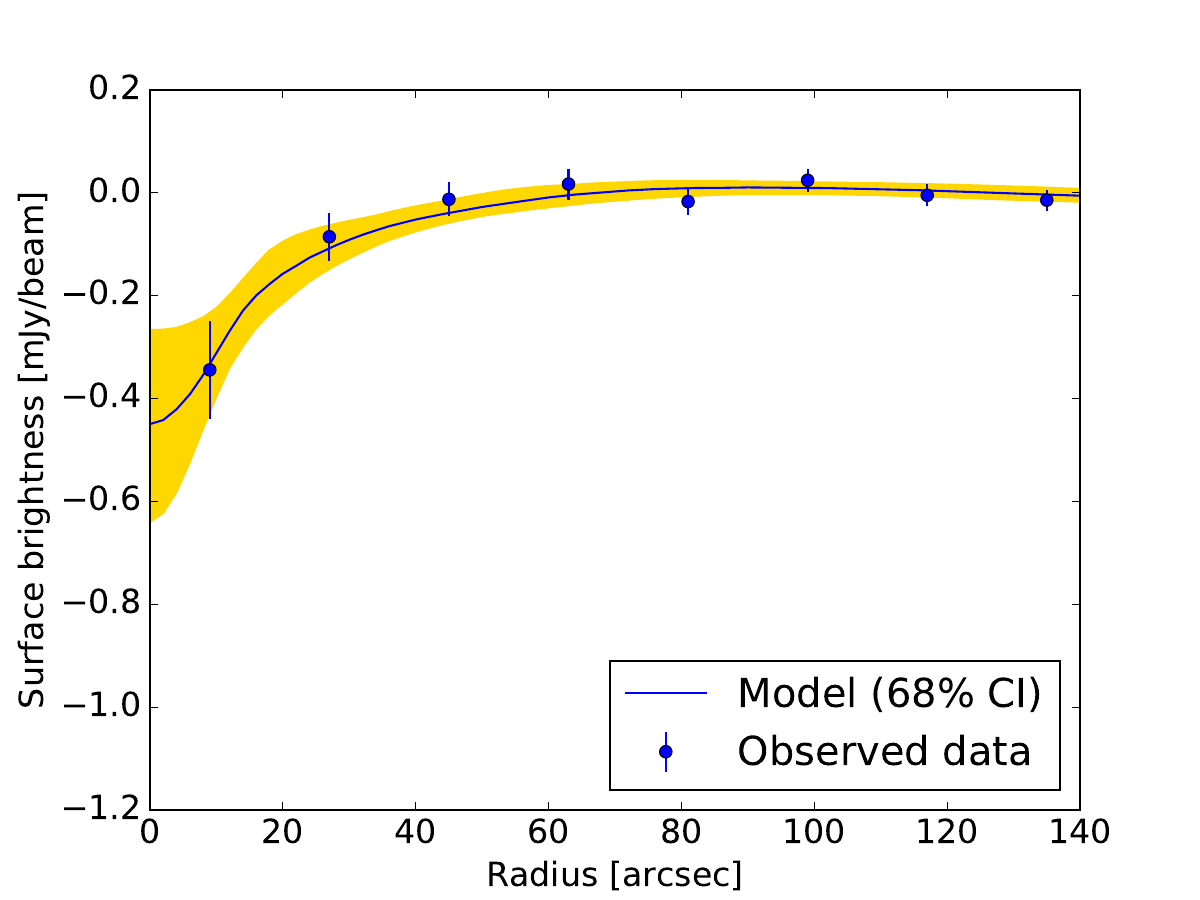}}
\caption[h]{Id34 observed SZ surface brightness profile (points with error bars) and mean fitted model (solid line) with 68 percent uncertainties. The adopted spline for the pressure profile is able to well describe the observed data.}
\label{fig:SZprof_id34}
\end{figure}

Id34 and id48 were observed with NIKA2 (Adam et al. 2018) between March and December 2020. NIKA2 is a three-array instrument with a 6.5 arcmin diameter field of view simultaneously observing at 150 and 260 GHz at the IRAM 30m telescope. The beam FWHM of the instrument is about 18 arcsec at 150 GHz and 11 arcsec at 260 GHz. The clusters were observed with a large number of raster scans of 8x4 or 10x5 arcmin with four different angles (at 0, 45, 90, and 135 degrees), centered on the center listed in Miyazaki et al. (2018). We spent 13.5 and 14.6 h on id34 and id48 scans, respectively, which lead to a maximal
effective exposure time at the scan center of 5.8h for both clusters. 

The raw NIKA2 data of each KID are processed with a modified version of the  pipeline described in Perotto et al. (2020). In particular, KIDs are calibrated using observations of Uranus and accounting for line-of-sight opacity absorption. They are then corrected from correlated atmospheric and electronic noise at the sub-scan level. A number of low-frequency sine-cosine templates are also used in the de-correlation effectively high-passing the signal above a 3-seconds period. That period was found as a trade-off between filtering the residual low-frequency excess noise and loss of signal power. Since we need information both on point and extended sources, two different data reduction settings were adopted, a more aggressive one that removes the atmosphere/electronics better, but also some of the extended emission (the SZ effect), and a second one, better at preserving the extended emission, but less efficient against atmospheric effect. The time-ordered signal of all KIDs of a same array is then projected into a common map using optimal weighing. Flux absolute calibration has a 10 percent uncertainty (Adam et al. 2018).

Point-sources are detected in the maps optimal for point-sources after the images are convolved with a fixed-width Gaussian kernel (with a FWHM of 12.5 and 18.5 arcsec at 150 and 260 GHz). The noise maps are consistently used in the process. Only detections above a signal-to-noise ratio (SNR) of 4 at 150 or at 260 GHz are considered. We make sure that the histogram of the SNR map is a normalized Gaussian in the process. Positional accuracy is of the order of 3 arcsec. 

As mentioned, the data reduction successfully removes non-astronomical signals, such as atmosphere variations, but also filters the astronomical signal. This filtering needs to be accounted for in the data analysis with a transfer function. The latter is computed analytically on-the-fly for each scan and coadded to give a final transfer function. The 150 GHz transfer function of the SZ reduction of id34 is shown in Fig.~\ref{fig:TF}. Scales larger than half the field of view diameter are lost by the data reduction, whereas small scales are washed out by the beam.

The final SZ-reduced maps have an RMS noise of 0.09 mJy/beam at 150 GHz within the central few arcmin radius. Fig.~\ref{fig:SZima} shows the SZ-reduced imageof id34 (left panel) at 150 GHz and the point-source reduced image of id48 (right panel) at 260 GHz. 

For the profile analysis, we extracted the radial profile of id34 and id48 at 150 GHz from the X-ray center in circular annuli of 18 arcsec width (the beam FWHM), accounting for flagged sources, positional variations in noise, and for data covariance, the latter estimated from a portion of the image free from contamination from sources. Figure~\ref{fig:SZprof_id34} shows the id34 SZ (150 GHz) radial surface brightness profile. The cluster is quite faint, with a peak of $\sim -0.4$ mJy/beam, more than two times  fainter than the lowest mass cluster in the NIKA2 large program  (K{\'e}ruzor{\'e} et al. 2020). Based on the two innermost data points, id34 is detected at $4\sigma$. The second id34 peak, about 2.6 arcmin South of the main clump, that is visible both in the optical and in X-ray images, is instead undetected in the NIKA2 data, mostly because of its faintness compared to the main clump. Id48 is even fainter than id34, and is undetected at 150 GHz (about $3\sigma$ away from the null, background only, model). Of course, both id48 and id34 clusters are undetected at 260 GHz because this frequency is close to the SZ null and the clusters are quite faint. 

The three dimensional pressure profile of id34 (no reliable profile can be derived for id48) is derived fitting the SZ data with a modified version of \texttt{PreProFit} (Castagna \& Andreon 2019), accounting for the transfer function, point spread function (beam), and pedestal (zero) level, following Andreon et al. (2021, 2023). 
The conversion from Jy/beam to Compton $y$ is taken to be $-11.34$ (Adam et al. 2017). 

Following Andreon et al. (2021, 2023), we adopt a pressure profile given by a cubic spline in log-log space with knots at radii of $r=15,30,60$ and $90$ arcsec. By adopting a cubic spline we allow the shape of the pressure profile to vary almost arbitrarily while keeping it continuous and doubly-differentiable. By defining the spline in log quantities (log pressure vs log radius), we naturally exclude non-physical (negative) values of pressure and radius and we can approximate a large variety of profiles and their derivatives. Our model then has 5 variables: the pressures at the four radii, $P_0,P_1,P_2,$ and $P_3$, and the pedestal (zero-level) value of the SZ surface brightness. Our analysis assumes spherical symmetry, uniform priors, zeroed for unphysical values (e.g., pressure cannot be negative). In particular, since the total mass of the cluster is finite, the logarithmic slope of the pressure should be steeper than $-4$ at large radii (Romero et al. 2018). We added a margin to this number and therefore adopted a logarithmic slope $<-2$ at $r=1$ Mpc as prior. This prior could alternatively be expressed as a maximal value for the pedestal level. This prior is primarily necessary to distinguish a radially-flat background from a radially-flat cluster signal.

The faintness of the cluster makes the $P_1,P_2,$ and $P_3$ parameters poorly determined, and with large covariance. Fig.~\ref{fig:SZprof_id34} shows the fitted model on the data. Fig.~\ref{fig:Pe_prof_others} shows the derived pressure profile of id34 and compares it with the one derived from the X-ray data. The SZ-derived pressure profile has larger errors than achieved from X-ray data, but different sensitivity to systematics such as the energy calibration of the XRT or the clumpiness of the ICM. Since the SZ data of id34 offer at most little constraining power, they are not jointly fitted with the X-ray and shear data.

\bsp	
\label{lastpage}
\end{document}